\documentclass[apj,tighten,iop]{emulateapj}
\usepackage{color}
\usepackage{graphicx}
\usepackage{natbib}
\usepackage{amssymb}
\usepackage{amsmath}
\usepackage{rotating}
\usepackage{url}

\shorttitle{On the Progenitors of Local Group Novae - I.}
\shortauthors{Williams et~al.}

\begin{document}

\title{On the Progenitors of Local Group Novae - I. The M31 Catalog}

%\author{S.~C. Williams, M.~J. Darnley, M.~F. Bode, \and
 % A. Keen\altaffilmark{1}}
%\affil{Astrophysics Research Institute, Liverpool John Moores
 % University, IC2 Liverpool Science Park IC2, Liverpool, L3~5RF, UK}
%\email{scw@astro.livjm.ac.uk}
%\altaffiltext{1}{Department of Physics, The University of Liverpool,
 % Liverpool, L69~7ZE, UK}
%
%and
%
%\author{A.~W. Shafter}
%\affil{Department of Astronomy, San Diego State University, San Diego,
 % CA 92182, USA}

\author{S.~C. Williams$^{1}$, M.~J. Darnley$^{1}$, M.~F. Bode$^{1}$,
  A. Keen$^{1,2}$, and A.~W. Shafter$^{3}$}
\affil{$^{1}$Astrophysics Research Institute, Liverpool John Moores
  University, Liverpool, L3~5RF, UK\\
$^{2}$Department of Physics, The University of Liverpool,
  Liverpool, L69~7ZE, UK\\
$^{3}$Department of Astronomy, San Diego State University, San Diego,
  CA 92182, USA}

\begin{abstract}
We report the results of a survey of M31 novae in quiescence. This is
the first catalog of extragalactic systems in quiescence to be
published, and contains data for 38 spectroscopically confirmed novae
from 2006 to 2012. We used Liverpool Telescope (LT) images of each
nova during eruption to define an accurate position for each
system. These positions were then matched to archival {\it Hubble
  Space Telescope} ({\it HST}) images and we performed photometry on
any resolved objects that were coincident with the eruption
positions. The survey aimed to detect quiescent systems with red giant
secondaries, as only these, along with a few systems with bright
sub-giant secondaries, will be resolvable in the {\it HST}
images. There are only a few confirmed examples of such red giant
novae in our Galaxy, the majority of which are recurrent novae. However, we find a relatively high percentage of
the nova eruptions in M31 may occur in systems containing red giant
secondaries.  Of the 38 systems in this catalog, 11
have a progenitor candidate whose probability of being a coincidental
alignment is less than $5\%$.  We show that, at the $3\sigma$ limit,
up to only two of these eleven systems may be due to chance
alignments, leading to an estimate of the M31 nova population with
evolved secondaries of up to $24\%$, compared to the $\sim3\%$
seen Galactically. Such an elevated proportion of nova systems with evolved
secondaries may imply the presence of a much larger population of
recurrent novae than previously thought.  This would have considerable impact, particularly with regards
their potential as Type Ia supernova progenitors. 

Additionally, for several novae, serendipitous {\it HST} images had
been taken when the nova was still fading; this allowed us to produce
light curves that go fainter than is usually achievable for most
extragalactic systems. Finally, as this survey is astrometric in
nature, we also update the position of each nova in the catalog. 
\end{abstract}


\section{Introduction}

The eruptions of classical novae (CNe) are only surpassed in
luminosity by a handfull of other phenomena, such as gamma-ray bursts,
supernovae (SNe), and some luminous blue variable stars, although in any given galaxy, nova eruptions are far
more commonplace that these other transients.  Novae are a sub-class of the cataclysmic variables,
their canonical model consisting of a white dwarf (WD; the primary)
which accretes material from a typically late-type, main-sequence star
(the secondary).  Mass is lost from the secondary usually via Roche
Lobe overflow and is typically transferred to the WD by means of an
accretion disk around the primary, although in a number of systems the
magnetic field of the WD may significantly affect the accretion
process.  CN eruptions are powered by a thermonuclear runaway within
the accreted material on the surface of the WD and are predicted to
recur on time-scales $<10^{6}$~years \citep{1986ApJ...308..721T}.  CNe
are typically classified based on the properties of the eruptions;
either by the rate of decline of their optical light curve (the speed
class; \citealp{1957gano.book.....G}) or via their spectroscopic
behaviour (into the sub-types of \ion{Fe}{2}\ and He/N novae;
\citealp{1992AJ....104..725W}). 

Closely related to CNe are the recurrent novae (RNe), which have been
seen to have undergone more than one eruption, with observed
recurrence time-scales of approximately $1-100$~years \citep{2014A&A...563L...9D}. Observed RNe
have typically been divided into three sub-types, based not only on
the properties of their eruption but also on the nature of the
progenitor system. The three subclasses are RS~Oph/T~CrB, U~Sco and
T~Pyx type, which harbor red giant, sub-giant and main-sequence
secondaries respectively \citep[see][for a
review]{2008ASPC..401...31A}.

\citet{2012ApJ...746...61D} recently proposed a nova classification
system that unifies CNe and RNe and is based only on the properties of
each system at quiescence; i.e.\ the evolutionary state of the
secondary star.  The majority of CN systems harbor a main-sequence
secondary star and define the MS-nova class; this class also includes
the T~Pyx sub-class of RNe.  The U~Sco class of RNe and also CNe such
as GK~Persei all have sub-giant secondaries and make up the SG-nova
class.  The third class, the RG-nova class (also commonly referred to
as symbiotic novae), contain nova systems with red giant secondaries,
including the RS~Oph/T~CrB class of RNe and systems such as EU~Scuti.

Amongst the hundreds of known Galactic CNe, there are only ten firmly
established RNe.  Traditionally, RNe have only been confirmed by the
observation of more than one eruption from the system.  Using multiple
eruptions as the sole discriminator between RNe and CNe is subject to
a number of significant selection effects, mainly the decreasing
completeness and deepness of sky coverage as one goes back in time.
As such, a Galactic population of ten RNe is a lower limit and clearly
far from the true picture.  Hence, there is approximately only one
confirmed Galactic RN eruption each year, compared to an overall
Galactic nova rate of $\sim35$~year$^{-1}$
\citep{1997ApJ...487..226S,2006MNRAS.369..257D}.  However, more
indirect methods have recently been employed to introduce a small
number of new candidate RN systems that have only one previous
recorded eruption.  These systems include V2491~Cygni
\citep{2010MNRAS.401..121P,2011A&A...530A..70D}, KT~Eridani \citep{2012A&A...537A..34J} and the Andromeda
Galaxy (M31) nova M31N~2007-12b \citep{2009ApJ...705.1056B}.  These
analyses include other parameters of the system, such as ejection
velocity, accretion rate, WD mass, super-soft source (SSS) behaviour,
orbital period and secondary type.

From the Sun's position within the Milky Way, our ability to study the
entire nova population of the Galaxy is hindered by the obscuring
effect of dust in the Galactic plane and large uncertainties when
determining distances and extinction.  As such, we have turned to the
nearby galaxy M31, which, whilst far from ideal, provides us with the
best opportunity to observe the nova population of a whole galaxy.
Novae in M31 can all be assumed to be at the same distance, and whilst
there are still dust extinction effects, these are generally much
smaller than Galactic extinction.  With a nova rate of
$65^{+16}_{-15}\;\mathrm{yr}^{-1}$ \citep{2006MNRAS.369..257D}, M31
also presents us with a much larger sample size to study, compared to
the $\sim10$ Galactic novae that are practically observable each
year.

The ability to successfully detect RNe in M31 using the traditional
method -- the observation of more than one eruption from the same
system -- is severely hampered in a number of ways.  These include;
our inability to observe M31 for several months each year; large gaps
in the observations of M31, time between surveys, etc.; the
astrometric accuracy required to match two eruptions at the distance
of M31; the nova distribution in M31 -- bulge-centric; all of these
selection effects are worse the older the data source.  Nonetheless,
there have been a number of attempts to explore the recurrent nova
population of M31 by searching for multiple eruptions at similar
positions (e.g.\
\citealp{1973A&AS....9..347R,1996ApJ...473..240D,2004A&A...421..509A,2008A&A...477...67H,2012A&A...537A..43L};
Shafter et~al.\ in preparation). However, these have all been severely
limited by the large positional uncertainties of the old novae, and
have only yielded a handful of RN candidates.  Additionally,
astrometric errors can give rise to incorrect association between
multiple eruptions, \citep[see Figure~4 of][for
example]{2009ApJ...705.1056B}.  This technique is further hampered by
the misidentification of long period Mira variables as M31 novae
\citep[see, for
example,][]{2004MNRAS.353..571D,2008ATel.1834....1S,2008ATel.1851....1S}.

Some of these problems can be overcome by using a spectroscopically
confirmed sample of novae which all have well determined astrometric
positions. But this essentially limits us only to M31 novae that have produced eruptions since
around 2006, when spectroscopy of large samples of M31 novae in
eruption became viable \citep[see][]{2011ApJ...734...12S}.  Based on the
recurrence time-scales of their Galactic counterparts, such a short
baseline is not long enough to recover a good sample of RNe using
multiple eruptions alone.

Here we propose a technique that can be used to recover the progenitor
systems of novae belonging to the RG-nova class (which is dominated by
confirmed and candidate RNe of the RS~Oph/T~CrB sub-class) from a
spectroscopically confirmed sample of M31 novae.  In some
(exceptional) cases this technique may also be able to recover
SG-novae in M31 (a class dominated by RNe of the U~Sco sub-class) as
was recently achieved for the (albeit much closer) Large Magellanic
Cloud (LMC) RN LMCN~2009-02a (Bode et~al.\ in preparation). 

The technique presented in this paper was also recently used to locate
the progenitor system of a RN in M31 with a $\sim1$~year
recurrence time-scale (\citealp{2014A&A...563L...9D}, see also
\citealp{2013ATel.5611....1W}).  The first optical eruption of this
system, M31N~2008-12a, was
recorded in 2008\footnote{\url{http://www.cbat.eps.harvard.edu/CBAT\_M31.html\#2008\mbox{-}12a}},
with coincident eruptions also being reported in
2011\footnote{\url{http://www.cbat.eps.harvard.edu/unconf/followups/J00452885+4154094.html}},
2012\footnote{\url{http://www.cbat.eps.harvard.edu/unconf/followups/J00452884+4154095.html}} and 2013 \citep{2013ATel.5607....1T}. A
spectrum taken after the 2012 eruption confirmed the transient to be a
He/N nova in M31
\citep{2012ATel.4503....1S}. \citet{2013ATel.5611....1W}
found a coincident source, with only a 2.5\% probability of being a
chance alignment, making this system the first confirmed RN in M31 with a
resolved progenitor. Work by \citet{2014A&A...563L...9D}, \citet{2014A&A...563L...8H} and \citet{2014ApJ...786...61T}
following the 2013 eruption has revealed
that the nova system likely contains a high-mass WD and has a
high accretion rate. Their work also indicated the nova may have a
$\sim1$-year recurrence time, with further eruptions being found in
1992, 1993 \citep{1995ApJ...445L.125W}, 2001
\citep{2004ApJ...609..735W} and 2009
\citep{2014ApJ...786...61T}. M31N~2008-12a was not included in the survey presented here, as we did not
have any eruption photometry taken prior to the end of the survey
(February 2013), with the nova having faded beyond detection by the
time observations were taken after the 2008 and 2012 eruptions and no
observations being made after the 2011 eruption.

A number of space-based and the larger ground-based optical telescopes
are capable of resolving red giant stars within M31.  As such, the
progenitor systems of RG-novae can in principle be directly imaged.
Similar work has been used to directly image the lensed system for M31
microlensing events \citep[see, for example][]{2001ApJ...553L.137A}
and indeed for the RN candidate M31N~2007-12b \citep[see][on which this work is
largely based]{2009ApJ...705.1056B}, as well as the RN M31N~2008-12a. \citet{2011ApJ...735...94K} have
also recently discussed such an approach, as have
\citet{2013arXiv1303.1980W}, who presented a summary of the initial
results of this survey.

\section{Type Ia Supernovae}

Novae are candidates for the progenitors of Type Ia supernovae (SNe
Ia). It is now widely believed that SNe Ia are the result of the
thermonuclear explosion of a carbon-oxygen WD, but their progenitors
are still uncertain. The two proposed pathways for SN Ia progenitors
are the single-degenerate and the double-degenerate models.

The single-degenerate model involves a WD accreting matter from a less
evolved secondary and increasing in mass (e.g.\
\citealp{1973ApJ...186.1007W}). Eventually the WD mass approaches the
Chandrasekhar mass and a SN explosion occurs when carbon ignites. One
of the leading single-degenerate progenitor candidates are novae,
particularly those classified as RNe, due to their high WD masses and
low ejecta masses. The main problem with the single-degenerate model
concerns the occurrence rate of SNe Ia versus the population of likely
progenitors (e.g.\ \citealp{1996ApJ...473..240D}). As most WDs in RN
systems are generally thought to be increasing in mass with each
eruption (e.g.\ \citealp{2000ApJ...528L..97H,2007ApJ...659L.153H,2011ApJ...727..124O}) they are prime candidates for the progenitors of SNe
Ia. However as most CNe are thought to be decreasing in mass as they
undergo eruption, they are unlikely candidates for progenitors of SNe
Ia. One of the reasons for this difference may be related to mixing
between the WD material and the accreted envelope (see e.g.\
\citealp{2012BaltA..21...76S,2013arXiv1303.3642N} for recent
discussions).

The double-degenerate model involves the merger of two WDs, with their
combined mass exceeding the Chandrasekhar limit. There seems to be
evidence supporting both of these channels and it may be the case that
SNe Ia have more than one progenitor pathway. For example
\citet{2012Natur.481..164S} could find no resolvable companion star in
the LMC SN Ia remnant SNR 0509$-$67.5, implying this SN at least was
produced by a double-degenerate system. On the other hand
\citet{2012Sci...337..942D} suggest that a single-degenerate system
with a red giant companion was the progenitor of SN
PTF~11kx. \citet{2011Natur.480..344N} suggested that SN 2011fe was
caused by a single-degenerate system with a main-sequence companion.
\citet{2011Natur.480..348L} showed that a system very similar to the
RN RS~Ophiuchi could not be the progenitor of the same SN; however
\citet{2013arXiv1303.2711D} pointed out that this analysis did not
exclude a less (optically) luminous RG-nova being the SN~Ia progenitor system.

The first nova progenitor system discovered in M31 was that of
M31N~2007-12b \citep{2009ApJ...705.1056B}. The current paper builds on
this work, using Liverpool Telescope (LT) data to search archival {\it
  Hubble Space Telescope} ({\it HST}) images for 38 quiescent M31
novae. The light curves for some M31 novae published by
\citet{2011ApJ...734...12S} are also extended to include {\it HST}
data taken whilst the nova was still fading. This allowed us to
produce light curves that go deeper than is usually possible for
extragalactic systems. Additionally, from LT data, we produced a few
light curves for novae that underwent eruptions in 2010 as well as
updating the positions of all 38 systems.

\section{Method}

\begin{figure}
\includegraphics[width=\columnwidth]{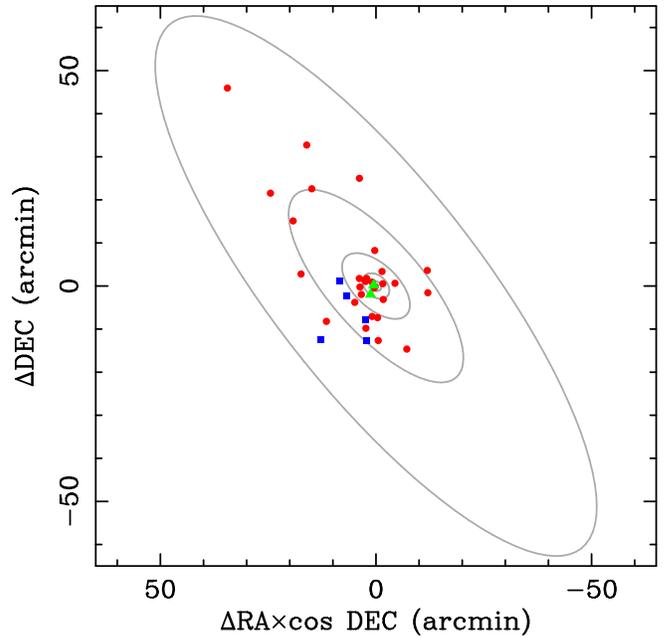}
\caption{The spatial distribution of the 38 M31 novae with known
  spectroscopic classes examined in this study.  The \ion{Fe}{2}\ novae are
  indicated by filled (red) circles, the He/N novae by filled (blue)
  squares and the hybrid types by filled (green) triangles
  \citep[see][for a summary of each pre-2010 nova spectroscopic type
  determination]{2011ApJ...734...12S}. The gray ellipses represent
  isophotes from the surface photometry of
  \citet{1987AJ.....94..306K}.  (A color version of this figure is available in the online journal.)\label{spatialplot}}
\end{figure}

The technique employed in this paper to attempt to recover the
progenitor system for each nova is that outlined in
\citet{2009ApJ...705.1056B}.  This method relies upon accurate
registration between images containing the nova in eruption and deeper
(typically archival) high-resolution images where the system is likely
to be in quiescence.  For this paper we use a catalog of Local Group
novae selected by virtue of their spectroscopic confirmation
\citep[see][]{2011ApJ...734...12S} and astrometric precision of the
eruption. The input catalog of 38 novae is presented in
Table~\ref{inputcatalogue} and their spatial distribution is shown in
Figure~\ref{spatialplot}.

\begin{deluxetable*}{llllllll}
\tablecaption{The input nova catalog: spectroscopically confirmed
  sample with LT/FTN eruption astrometry. The availability of LT and
  FTN eruption data along with {\it HST} WFPC2 and ACS/WFC quiescent
  data are shown for each nova. We also list the spectroscopic type
  and $t_2$ of each eruption. For novae that have a $t_2$ value
  published in more than one filter, we note the one in the filter
  closest to that used to search for the progenitor. The table also
  shows the proposal IDs of the {\it HST} data used in this paper. The
  first number is for the data used to determine the position of the
  system, with any subsequent IDs for data used in the light
  curves.\label{inputcatalogue}}
\tablewidth{0pt}
\tablehead{
\colhead{Nova} & \colhead{Spectral-} &\colhead{$t_2$ (band)} & \multicolumn{2}{c}{Astrometry}  & \multicolumn{2}{c}{{\it {HST}} archive}                    & \colhead{Proposal ID of}\\
\colhead{}    & \colhead{Class} &\colhead{in days} & \colhead{LT} & \colhead{FTN}    & \colhead{WFPC2}  & \colhead{ACS/WFC} & \colhead{images used}
}
\startdata
M31N~2006-09c &\ion{Fe}{2}\tablenotemark{1}     &$23.1\pm1.6$ ($R$)\tablenotemark{1}   &$\checkmark$  &$\times$      &$\checkmark$  &$\times$      &10273\\
M31N~2006-11a &\ion{Fe}{2}\tablenotemark{1}     &$28.7\pm2.6$ ($R$)\tablenotemark{1}   &$\checkmark$  &$\times$      &$\checkmark$  &$\checkmark$  &10273\\
M31N~2007-02b &\ion{Fe}{2}\tablenotemark{1,2}   &$34.1\pm3.6$ ($R$)\tablenotemark{1}   &$\checkmark$  &$\times$      &$\checkmark$  &$\checkmark$  &10260, 11218\\
M31N~2007-10a &He/Nn\tablenotemark{1,3}   &$7.9\pm0.4$ ($V$)\tablenotemark{1}    &$\checkmark$  &$\checkmark$  &$\times$      &$\checkmark$  &9719\\
M31N~2007-10b &He/Nn\tablenotemark{1,4}   &$3.1\pm0.4$ ($R$)\tablenotemark{1}    &$\checkmark$  &$\checkmark$  &$\times$      &$\checkmark$  &12058\\
M31N~2007-11b &He/Nn\tablenotemark{1,5}   &$74.4\pm16.7$ ($i'$)\tablenotemark{1} &$\checkmark$  &$\checkmark$  &$\checkmark$  &$\times$      &10273\\
M31N~2007-11c &\ion{Fe}{2}\tablenotemark{6}     &$11.7\pm0.9$ ($i'$)\tablenotemark{1}  &$\checkmark$  &$\checkmark$  &$\times$      &$\checkmark$  &12058\\
M31N~2007-11d &\ion{Fe}{2}\tablenotemark{7}     &$9.2\pm0.5$ ($i'$)\tablenotemark{1}   &$\checkmark$  &$\checkmark$  &$\times$      &$\checkmark$  &12057\\
M31N~2007-11e &\ion{Fe}{2}\tablenotemark{8}     &27 ($R$)\tablenotemark{9}             &$\checkmark$  &$\times$      &$\times$      &$\checkmark$  &12110\\
M31N~2007-12a &\ion{Fe}{2}\tablenotemark{1}     &$29.6\pm2.0$ ($i'$)\tablenotemark{1}  &$\checkmark$  &$\checkmark$  &$\checkmark$  &$\checkmark$  &12057\\
M31N~2007-12b &He/N\tablenotemark{10}     &$5.0\pm0.5$ ($R$)\tablenotemark{1}    &$\checkmark$  &$\times$      &$\checkmark$  &$\checkmark$  &12058\\
M31N~2008-10b &\ion{Fe}{2}\tablenotemark{11}    &$98.4\pm14.9$ ($B$)\tablenotemark{1}  &$\checkmark$  &$\checkmark$  &$\checkmark$  &$\checkmark$  &10006, 11833, 12058\\
M31N~2008-12b &\ion{Fe}{2}\tablenotemark{12}    &$24.7\pm3.6$ ($i'$)\tablenotemark{1}  &$\checkmark$  &$\times$      &$\times$      &$\checkmark$  &12109\\
M31N~2009-08a &\ion{Fe}{2}\tablenotemark{13}    &$36.7\pm4.1$ ($B$)\tablenotemark{1}   &$\checkmark$  &$\times$      &$\times$      &$\checkmark$  &10760, 12058\\
M31N~2009-08b &\ion{Fe}{2}\tablenotemark{14}    &$26.9\pm2.2$ ($i'$)\tablenotemark{1}  &$\checkmark$  &$\times$      &$\times$      &$\checkmark$  &12114\\
M31N~2009-08d &\ion{Fe}{2}\tablenotemark{15}    &$27.9\pm6.5$ ($B$)\tablenotemark{1}   &$\checkmark$  &$\times$      &$\checkmark$  &$\checkmark$  &10006\\
M31N~2009-10b &\ion{Fe}{2}\tablenotemark{16}    &$8.0\pm0.2$ ($B$)\tablenotemark{1}    &$\checkmark$  &$\times$      &$\times$      &$\checkmark$  &11013, 12058\\
M31N~2009-10c &\ion{Fe}{2}\tablenotemark{17}    &$14.9\pm0.8$ ($B$)\tablenotemark{1}   &$\checkmark$  &$\times$      &$\checkmark$  &$\checkmark$  &10006, 12058\\
M31N~2009-11a &\ion{Fe}{2}\tablenotemark{18}    &$21.7\pm1.2$ ($V$)\tablenotemark{1}   &$\checkmark$  &$\times$      &$\times$      &$\checkmark$  &10273\\
M31N~2009-11b &\ion{Fe}{2}\tablenotemark{19}    &$74.8\pm10.6$ ($V$)\tablenotemark{1}  &$\checkmark$  &$\times$      &$\checkmark$  &$\times$      &10273\\
M31N~2009-11c &\ion{Fe}{2}\tablenotemark{20}    &$32.5\pm2.4$ ($V$)\tablenotemark{1}   &$\checkmark$  &$\times$      &$\checkmark$  &$\checkmark$  &10273, 12058\\
M31N~2009-11d &\ion{Fe}{2}\tablenotemark{21}    &$11.2\pm0.4$ ($B$)\tablenotemark{1}   &$\checkmark$  &$\times$      &$\times$      &$\checkmark$  &12105\\
M31N~2009-11e &\ion{Fe}{2}\tablenotemark{22}    &$55.7\pm3.1$ ($R$)\tablenotemark{1}   &$\checkmark$  &$\times$      &$\checkmark$  &$\checkmark$  &5907, 12058\\
M31N~2010-01a &\ion{Fe}{2}\tablenotemark{23}    &$28\pm21$ ($B$)\tablenotemark{24}      &$\checkmark$  &$\times$      &$\times$      &$\checkmark$  &10760, 12058\\
M31N~2010-05a &\ion{Fe}{2}\tablenotemark{25}    &$39\pm17$ ($B$)\tablenotemark{24}      &$\checkmark$  &$\times$      &$\checkmark$  &$\checkmark$  &10006, 12058\\
M31N~2010-09b &\ion{Fe}{2}\tablenotemark{26,27} &$3.8\pm0.2$ ($B$)\tablenotemark{24}    &$\checkmark$  &$\times$      &$\times$      &$\checkmark$  &12073\\
M31N~2010-10a &\ion{Fe}{2}\tablenotemark{27}    &$<16\pm2$ ($B$)\tablenotemark{24}      &$\checkmark$  &$\times$      &$\times$      &$\times$      &12109\\
M31N~2010-10b &\ion{Fe}{2}\tablenotemark{28}    &$>41$\tablenotemark{29}                &$\checkmark$  &$\times$      &$\times$      &$\checkmark$  &10273\\
M31N~2010-10c &\ion{Fe}{2}\tablenotemark{30}    &20\tablenotemark{29}                   &$\checkmark$  &$\times$      &$\checkmark$  &$\checkmark$  &10407\\
M31N~2010-10d &\ion{Fe}{2}\tablenotemark{31}    &$23\pm7$ ($B$)\tablenotemark{24}       &$\checkmark$  &$\times$      &$\checkmark$  &$\checkmark$  &10006, 12058\\
M31N~2010-10e &He/N\tablenotemark{32}     &$>5$\tablenotemark{29}                 &$\checkmark$  &$\times$      &$\checkmark$  &$\checkmark$  &10273\\
M31N~2011-10a &\ion{Fe}{2}\tablenotemark{33,34} & \nodata &$\checkmark$  &$\times$      &$\times$      &$\checkmark$  &12058\\
M31N~2011-10d &\ion{Fe}{2}\tablenotemark{35,36} & \nodata &$\checkmark$  &$\times$      &$\checkmark$  &$\checkmark$  &12058\\
M31N~2011-12a &\ion{Fe}{2}\tablenotemark{37}    & \nodata &$\checkmark$  &$\times$      &$\checkmark$  &$\times$      &10273\\
M31N~2012-01a &\ion{Fe}{2}\tablenotemark{38}    & \nodata &$\checkmark$  &$\times$      &$\times$      &$\checkmark$  &11647\\
M31N~2012-09a &\ion{Fe}{2}b\tablenotemark{39}   & \nodata &$\checkmark$  &$\times$      &$\checkmark$  &$\checkmark$  &12058\\
M31N~2012-09b &\ion{Fe}{2}b\tablenotemark{40}   & \nodata &$\checkmark$  &$\times$      &$\checkmark$  &$\checkmark$  &12058\\
M31N~2012-12a &\ion{Fe}{2}\tablenotemark{41}    & \nodata &$\checkmark$  &$\times$      &$\checkmark$  &$\checkmark$  &12058
\enddata
\tablerefs{(1)~\citet{2011ApJ...734...12S},
  (2)~\citet{2007ATel.1009....1P}, (3)~\citet{2007ATel.1236....1G},
  (4)~\citet{2007ATel.1242....1R}, (5)~\citet{2007ATel.1276....1R},
  (6)~\citet{2007ATel.1292....1C}, (7)~\citet{2009ApJ...690.1148S},
  (8)~\citet{2007ATel.1325....1D}, (9)~\citet{2011ApJ...727...50S},
  (10)~\citet{2009ApJ...705.1056B}, (11)~\citet{2010AN....331..197D},
  (12)~\citet{2009ATel.1886....1K}, (13)~\citet{2009ATel.2208....1V},
  (14)~\citet{2009ATel.2166....1R}, (15)~\citet{2009ATel.2171....1D},
  (16)~\citet{2009ATel.2251....1B}, (17)~\citet{2009ATel.2240....1F},
  (18)~\citet{2009CBET.2062....1H}, (19)~\citep{2009CBET.2015....3K},
  (20)~\citet{2009CBET.2057....3H}, (21)~\citet{2009CBET.2058....3H},
  (22)~\citet{2009CBET.2061....5H}, (23)~\citet{2010CBET.2127....1H},
  (24)~this work (see Section~\ref{sec:lc}),
  (25)~\citet{2010CBET.2319....1H}, (26)~\citet{2010ATel.2898....1S},
  (27)~\citet{2010ATel.2909....1S}, (28)~\citet{2010ATel.3039....1S},
  (29)~\citet{2012ApJ...752..133C}, (30)~\citet{2010ATel.2949....1S},
  (31)~\citet{2010ATel.2987....1S}, (32)~\citet{2010ATel.3006....1S},
  (33)~\citet{2011ATel.3701....1C}, (34)~\citet{2011ATel.3702....1C}
  (35)~\citet{2011ATel.3699....1S}, (36)~\citet{2011ATel.3727....1S},
  (37)~\citet{2011ATel.3825....1S}, (38)~\citet{2012ATel.3850....1S},
  (39)~\citet{2012ATel.4368....1S}, (40)~\citet{2012ATel.4391....1S},
  (41)~\citet{2012ATel.4658....1S}}
\end{deluxetable*}

LT and Faulkes Telescope North (FTN) data were taken as part of a
monitoring survey of Local Group novae post-eruption.  Observations
were generally taken through $B$, $V$, $r'$ and $i'$ filters, with a
typical exposure time per filter of 180s (each comprising
$3\times60$s images combined by taking the median).  For most of these
observations the cameras were operated in a $2\times2$ binning mode,
giving an effective pixel size of $0^{\prime\prime}\!\!.279$.  Towards
the end of this survey, the observations in the $r'$ and $i'$ filters
were dropped in favour of first a higher cadence through the $B$ and
$V$ filters and more recently, greater exposure times.

The accuracy of the registration between the LT data and the {\it HST}
data is dependent upon a number of factors, including: the size of the
common overlap region; the stellar density in the overlap; the seeing
of the ground-based data; the luminosity of the nova, and the {\it
  HST} instrument available.  Typically, it is the seeing in the
ground-based data that has the largest detrimental effect, but some
compromise must be reached between the ground-based seeing and the
luminosity of the nova for each eruption.  In order to minimise these
effects we employ a number of approaches.  We preferentially used
ground-based data in $i'$ and $r'$ filters, as these are where the LT is most sensitive, yielding superior position measurements
on fainter objects and usually more numerous objects for registration.
Ideally we would compare the ground-based data to {\it HST} data taken
using the Wide-Field Camera 3 - UVIS Channel (WFC3/UVIS), however
there is to-date little data taken within the M31 field in suitable
filters and indeed none coincident with the novae in this
survey. Therefore we preferentially choose Advanced Camera for Surveys
- Wide-Field Channel (ACS/WFC) data over Wide-Field Planetary Camera 2
(WFPC2) data. This is due to the greater spatial resolution and
quantum efficiency improvement and hence better sampling of the PSF.
As the ideal ground based data are generally ``red'', we chose the
most appropriate {\it HST} filter if possible, typically F814W.

For this study, we use data taken by the LT
\citep{2004SPIE.5489..679S} on La Palma, backed-up by one of its
sister telescopes, the FTN in Hawaii (the FTN data were not required
to complete this survey), to determine the eruption position of the
nova and archival {\it HST} data for the identification and photometry
of the progenitor candidate. The ground-based data were processed and
analysed using a combination of
Starlink\footnote{\url{http://starlink.jach.hawaii.edi}} and
IRAF \citep{1986SPIE..627..733T,
1993ASPC...52..173T} software.  We used imaging data from
WFPC2, ACS/WFC and WFC3/UVIS on-board {\it HST}, all three of which
provide very good overlap with the
$4^{\prime}\!.6\times4^{\prime}\!.6$ LT RATCam fields.  The {\it HST}
data are again processed using Starlink and IRAF software, the {\it
  HST} photometry is performed using HSTphot
\citep[v1.1;][]{2000PASP..112.1383D} for the WFPC2 data and
DOLPHOT (v1.1)\footnote{\url{http://purcell.as.arizona.edu/dolphot}}, a
photometry package based on HSTphot, for the ACS/WFC data. 

\section{Observations}

\subsection{Eruption Photometry}
The LT data were reduced using a combination of IRAF and Starlink
software, calibrated using standard stars from
\citet{1992A&AS...96..379M}, \citet{1994A&A...286..725H} and
\citet{2006AJ....131.2478M}. The time (in days) it takes a nova to
fade by two magnitudes ($t_{2}$) was calculated using different methods,
depending on the data available.  Typically, for well populated light
curves, this involved a simple linear extrapolation of the data around
maximum and around 2 magnitudes below the peak.  Where different
approaches were employed this is described for the respective novae in
Section~\ref{sec:lc}.

\subsection{LT Image Selection}
We used LT images taken using the  RATCam instrument in the Sloan-like
$i'$ and $r'$ filters along with $B$ and $V$ Bessel filters. We
selected the images that would produce the most reliable position for
each nova, taking into consideration both the seeing and brightness of
the nova.  In this survey, no ground-based data with FWHM seeing
  poorer than $2^{\prime\prime}\!\!.1$ were used.  We also matched the filters with those used for the {\it
  HST} images where possible. The position of each nova in the catalog
with respect to 2MASS \citep{2006AJ....131.1163S}
is shown in Table~\ref{tab:pos}.

\subsection{{\it HST} Image Selection}
\begin{figure*}[ht]
\begin{center}
\includegraphics[scale=0.8]{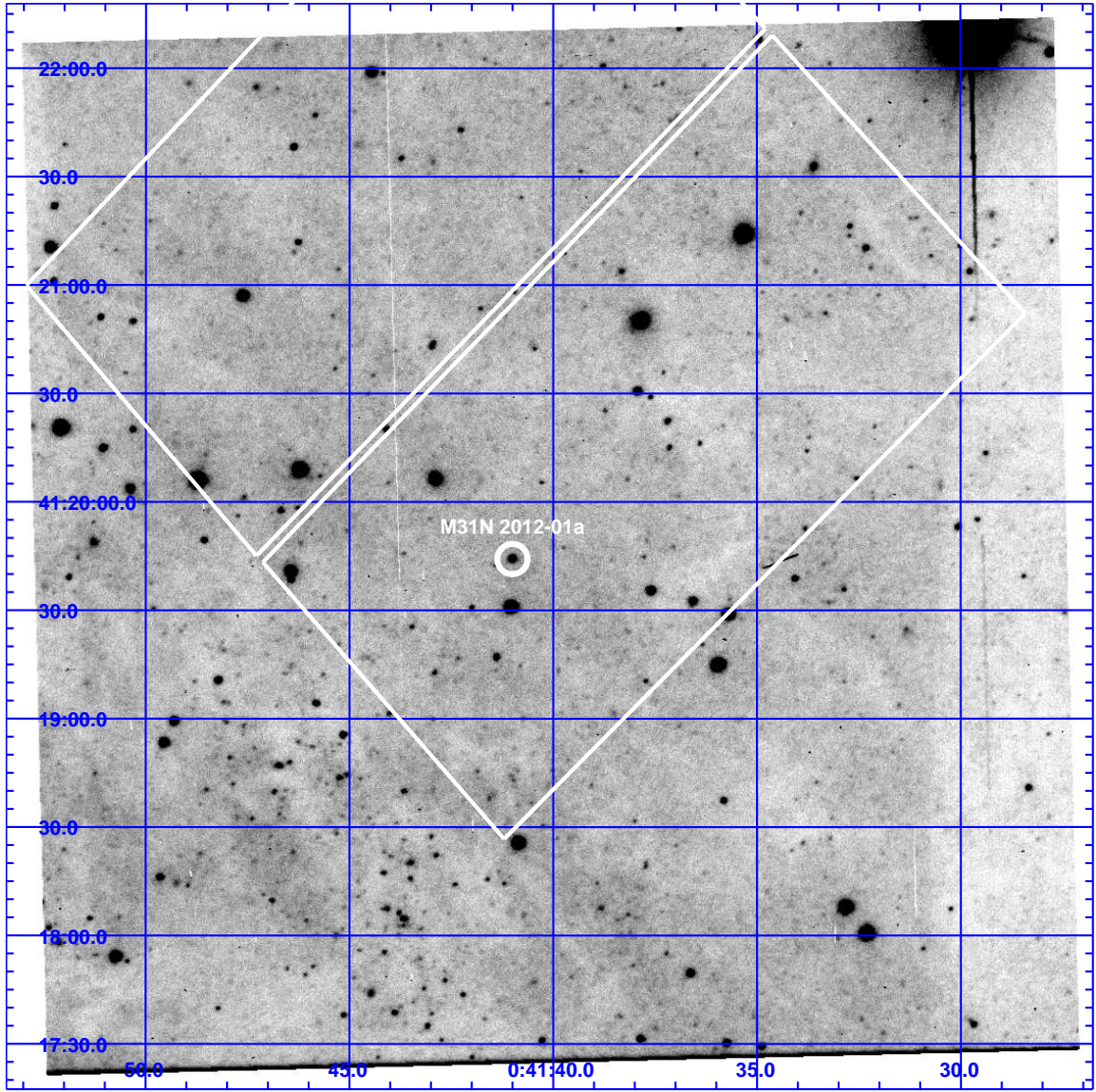}
\caption{M31N~2012-01a during eruption in a $V$-band LT image taken on
  2012 January 10.83, with the ``footprint'' of the coincident ACS/WFC
  F555W {\it HST} image overlaid as white boxes. M31N~2012-01a is
  shown with the white circle (the nova being the visible source at
  the center of the circle).{\label{fig:grid}}}
\end{center}
\end{figure*}

For this work, we used archival {\it HST} images taken with ACS/WFC
and WFPC2 using the F435W, F475W, F555W, F606W, F625W, and F814W
filters. For some novae, pre-eruption archival {\it HST} images were
not available, but images taken long after the eruption were. In such
cases we have ensured that the system is likely to be at quiescence
using the speed of the nova and the images themselves. For example, a
nova with a $t_{2}$ of just a few days will clearly have faded to
quiescence if the images were taken some years after the
eruption. Additionally, for novae which we could not be sure were at
quiescence using the previous method, we looked at multiple {\it HST}
images. A system was assumed to be at quiescence if it has not faded
between one {\it HST} image and the next when they were taken several
months apart. For the individual novae with only post-eruption data,
we have noted why the system is believed to be at quiescence in
Section~\ref{sec:prog}. A typical example of how LT images of novae in
eruption coincide with archival {\it HST} data is shown in
Figure~\ref{fig:grid}.

\subsection{Astrometry} \label{sec:ast}
We combined the best images for a given nova, taking into account the
seeing and the brightness of the nova itself, in order to produce the
most precise position possible for the system. Typically, around
$25-30$ stars were selected and matched up with the corresponding
stars in the other images. The necessary translation, rotation,
magnification and distortion needed for the images to align were then
calculated and executed using standard routines in IRAF. We then
combined the newly aligned images using the median. Finally
the position of the nova within the combined image was measured.

The positions of about $15-25$ stars in the combined image were then
recorded, as were their respective positions in the chosen {\it HST}
image. For the astrometric analysis we used the drizzled (drz) {\it
  HST} images. The drizzling process is an algorithm used on {\it HST}
data to correct for geometric distortion and cosmic rays by combining
pre-processed images. We then matched the two lists of coordinates and
transformed the LT coordinates using the same IRAF routines. The
position of the nova in the LT data was therefore transformed into a
position in the {\it HST} image.

Using the position determined from the LT image we also calculated the
coordinates of each nova using 2MASS \citep{2006AJ....131.1163S}. These were computed using the
same method as described above by utilising the positions of the stars
in 2MASS, rather than the {\it HST} images. The resulting
positions are more accurate than those previously published in the
majority of cases and are shown in Table~\ref{tab:pos}.

\begin{deluxetable*}{lclcll}[ht]
\tablecaption{A summary of the eruption images used to define the
  position of each nova, including the date the images were taken and
  the filter each was observed through. The coordinates of all the
  novae were calculated from the LT images using positions from 2MASS \citep{2006AJ....131.1163S}. \label{tab:pos}}
\tablewidth{0pt}
\tablehead{
\colhead{Nova} & \colhead{Telescope} &\colhead{Date (UT)} &
\colhead{Filter} & \colhead{Right Ascension (J2000)}  &
\colhead{Declination (J2000)}
}
\startdata
M31N~2006-09c &LT &2006 September 19.0 &$V$ &$0^{\mathrm{h}}42^{\mathrm{m}}42^{\mathrm{s}}\!.38\pm0^{\mathrm{s}}\!.02$ &$+41^{\circ}08^{\prime}45^{\prime\prime}\!\!.4\pm0^{\prime\prime}\!\!.2$\\
M31N~2006-11a &LT &2007 February 9.9 &$i'$ &$0^{\mathrm{h}}42^{\mathrm{m}}56^{\mathrm{s}}\!.800\pm0^{\mathrm{s}}\!.009$ &$+41^{\circ}06^{\prime}18^{\prime\prime}\!\!.3\pm0^{\prime\prime}\!\!.1$\\
M31N~2007-02b &LT &2007 February 14.9 &$i'$ &$0^{\mathrm{h}}41^{\mathrm{m}}40^{\mathrm{s}}\!.307\pm0^{\mathrm{s}}\!.009$ &$+41^{\circ}14^{\prime}33^{\prime\prime}\!\!.4\pm0^{\prime\prime}\!\!.1$\\
M31N~2007-10a &LT &2007 October 10.9 &$V$ &$0^{\mathrm{h}}42^{\mathrm{m}}55^{\mathrm{s}}\!.947\pm0^{\mathrm{s}}\!.007$ &$+41^{\circ}03^{\prime}21^{\prime\prime}\!\!.9\pm0^{\prime\prime}\!\!.1$\\
M31N~2007-10b &LT &2007 October 16.0 &$i'$ &$0^{\mathrm{h}}43^{\mathrm{m}}29^{\mathrm{s}}\!.47\pm0^{\mathrm{s}}\!.01$ &$+41^{\circ}17^{\prime}13^{\prime\prime}\!\!.9\pm0^{\prime\prime}\!\!.1$\\
M31N~2007-11b &LT &2007 November 16.9 &$i'$ &$0^{\mathrm{h}}43^{\mathrm{m}}52^{\mathrm{s}}\!.99\pm0^{\mathrm{s}}\!.01$ &$+41^{\circ}03^{\prime}35^{\prime\prime}\!\!.9\pm0^{\prime\prime}\!\!.1$\\
M31N~2007-11c &LT &2007 November 17.0 &$i'$ &$0^{\mathrm{h}}43^{\mathrm{m}}04^{\mathrm{s}}\!.16\pm0^{\mathrm{s}}\!.01$ &$+41^{\circ}15^{\prime}53^{\prime\prime}\!\!.93\pm0^{\prime\prime}\!\!.09$\\
M31N~2007-11d &LT &2007 November 28.0 &$i'$ &$0^{\mathrm{h}}44^{\mathrm{m}}54^{\mathrm{s}}\!.59\pm0^{\mathrm{s}}\!.01$ &$+41^{\circ}37^{\prime}39^{\prime\prime}\!\!.8\pm0^{\prime\prime}\!\!.1$\\
M31N~2007-11e &LT &2007 December 5.1 &$i'$ &$0^{\mathrm{h}}45^{\mathrm{m}}47^{\mathrm{s}}\!.76\pm0^{\mathrm{s}}\!.01$ &$+42^{\circ}02^{\prime}03^{\prime\prime}\!\!.7\pm0^{\prime\prime}\!\!.1$\\
M31N~2007-12a &LT &2007 Dec 16.9, 31.8, 2008 Jan 8.9 &$i'$ &$0^{\mathrm{h}}44^{\mathrm{m}}03^{\mathrm{s}}\!.52\pm0^{\mathrm{s}}\!.01$ &$+41^{\circ}38^{\prime}40^{\prime\prime}\!\!.9\pm0^{\prime\prime}\!\!.1$\\
M31N~2007-12b &LT &2007 December 14.9 &$i'$ &$0^{\mathrm{h}}43^{\mathrm{m}}19^{\mathrm{s}}\!.96\pm0^{\mathrm{s}}\!.02$ &$+41^{\circ}13^{\prime}46^{\prime\prime}\!\!.3\pm0^{\prime\prime}\!\!.1$\\
M31N~2008-10b &LT &2008 October 21.0 &$B$ &$0^{\mathrm{h}}43^{\mathrm{m}}02^{\mathrm{s}}\!.41\pm0^{\mathrm{s}}\!.02$ &$+41^{\circ}14^{\prime}09^{\prime\prime}\!\!.9\pm0^{\prime\prime}\!\!.2$\\
M31N~2008-12b &LT &2009 January 15.0 &$i'$ &$0^{\mathrm{h}}43^{\mathrm{m}}04^{\mathrm{s}}\!.85\pm0^{\mathrm{s}}\!.01$ &$+41^{\circ}17^{\prime}51^{\prime\prime}\!\!.6\pm0^{\prime\prime}\!\!.2$\\
M31N~2009-08a &LT &2009 August 27.1, September 4.1 &$i'$ &$0^{\mathrm{h}}42^{\mathrm{m}}58^{\mathrm{s}}\!.105\pm0^{\mathrm{s}}\!.007$ &$+41^{\circ}17^{\prime}29^{\prime\prime}\!\!.56\pm0^{\prime\prime}\!\!.06$\\
M31N~2009-08b &LT &2009 August 18.0 &$i'$ &$0^{\mathrm{h}}44^{\mathrm{m}}09^{\mathrm{s}}\!.89\pm0^{\mathrm{s}}\!.02$ &$+41^{\circ}48^{\prime}50^{\prime\prime}\!\!.7\pm0^{\prime\prime}\!\!.1$\\
M31N~2009-08d &LT &2009 August 20.1 &$B$ &$0^{\mathrm{h}}42^{\mathrm{m}}46^{\mathrm{s}}\!.74\pm0^{\mathrm{s}}\!.02$ &$+41^{\circ}15^{\prime}37^{\prime\prime}\!\!.4\pm0^{\prime\prime}\!\!.1$\\
M31N~2009-10b &LT &2009 October 15.0 &$B$ &$0^{\mathrm{h}}42^{\mathrm{m}}20^{\mathrm{s}}\!.83\pm0^{\mathrm{s}}\!.02$ &$+41^{\circ}16^{\prime}44^{\prime\prime}\!\!.3\pm0^{\prime\prime}\!\!.1$\\
M31N~2009-10c &LT &2009 October 15.0 &$B$ &$0^{\mathrm{h}}42^{\mathrm{m}}45^{\mathrm{s}}\!.72\pm0^{\mathrm{s}}\!.02$ &$+41^{\circ}15^{\prime}56^{\prime\prime}\!\!.99\pm0^{\prime\prime}\!\!.09$\\
M31N~2009-11a &LT &2009 November 13.9 &$V$ &$0^{\mathrm{h}}43^{\mathrm{m}}04^{\mathrm{s}}\!.789\pm0^{\mathrm{s}}\!.009$ &$+41^{\circ}41^{\prime}07^{\prime\prime}\!\!.79\pm0^{\prime\prime}\!\!.08$\\
M31N~2009-11b &LT &2009 December 6.8 &$V$ &$0^{\mathrm{h}}42^{\mathrm{m}}39^{\mathrm{s}}\!.596\pm0^{\mathrm{s}}\!.009$ &$+41^{\circ}09^{\prime}02^{\prime\prime}\!\!.9\pm0^{\prime\prime}\!\!.1$\\
M31N~2009-11c &LT &2009 November 12.9 &$V$ &$0^{\mathrm{h}}43^{\mathrm{m}}10^{\mathrm{s}}\!.46\pm0^{\mathrm{s}}\!.01$ &$+41^{\circ}12^{\prime}18^{\prime\prime}\!\!.5\pm0^{\prime\prime}\!\!.1$\\
M31N~2009-11d &LT &2009 November 24.0 &$V$ &$0^{\mathrm{h}}44^{\mathrm{m}}16^{\mathrm{s}}\!.866\pm0^{\mathrm{s}}\!.009$ &$+41^{\circ}18^{\prime}53^{\prime\prime}\!\!.6\pm0^{\prime\prime}\!\!.2$\\
M31N~2009-11e &LT &2009 November 27.0 &$V$ &$0^{\mathrm{h}}42^{\mathrm{m}}35^{\mathrm{s}}\!.33\pm0^{\mathrm{s}}\!.01$ &$+41^{\circ}12^{\prime}59^{\prime\prime}\!\!.4\pm0^{\prime\prime}\!\!.2$\\
M31N~2010-01a &LT &2010 January 15.9 &$B$ &$0^{\mathrm{h}}42^{\mathrm{m}}56^{\mathrm{s}}\!.70\pm0^{\mathrm{s}}\!.02$ &$+41^{\circ}17^{\prime}20^{\prime\prime}\!\!.2\pm0^{\prime\prime}\!\!.1$\\
M31N~2010-05a &LT &2010 June 17.2 &$B$ &$0^{\mathrm{h}}42^{\mathrm{m}}35^{\mathrm{s}}\!.899\pm0^{\mathrm{s}}\!.008$ &$+41^{\circ}16^{\prime}38^{\prime\prime}\!\!.24\pm0^{\prime\prime}\!\!.04$\\
M31N~2010-09b &LT &2010 October 5.1 &$B$ &$0^{\mathrm{h}}43^{\mathrm{m}}45^{\mathrm{s}}\!.545\pm0^{\mathrm{s}}\!.008$ &$+41^{\circ}07^{\prime}54^{\prime\prime}\!\!.5\pm0^{\prime\prime}\!\!.1$\\
M31N~2010-10a &LT &2010 October 10.1 &$V$ &$0^{\mathrm{h}}42^{\mathrm{m}}45^{\mathrm{s}}\!.82\pm0^{\mathrm{s}}\!.03$ &$+41^{\circ}24^{\prime}22^{\prime\prime}\!\!.0\pm0^{\prime\prime}\!\!.1$\\
M31N~2010-10b &LT &2010 October 11.1 &$V$ &$0^{\mathrm{h}}42^{\mathrm{m}}41^{\mathrm{s}}\!.55\pm0^{\mathrm{s}}\!.02$ &$+41^{\circ}03^{\prime}27^{\prime\prime}\!\!.7\pm0^{\prime\prime}\!\!.1$\\
M31N~2010-10c &LT &2010 October 22.1 &$B$ &$0^{\mathrm{h}}44^{\mathrm{m}}26^{\mathrm{s}}\!.575\pm0^{\mathrm{s}}\!.008$ &$+41^{\circ}31^{\prime}13^{\prime\prime}\!\!.6\pm0^{\prime\prime}\!\!.1$\\
M31N~2010-10d &LT &2010 October 30.0 &$B$ &$0^{\mathrm{h}}42^{\mathrm{m}}36^{\mathrm{s}}\!.914\pm0^{\mathrm{s}}\!.008$ &$+41^{\circ}19^{\prime}28^{\prime\prime}\!\!.9\pm0^{\prime\prime}\!\!.1$\\
M31N~2010-10e &LT &2010 November 7.0 &$V$ &$0^{\mathrm{h}}42^{\mathrm{m}}57^{\mathrm{s}}\!.75\pm0^{\mathrm{s}}\!.01$ &$+41^{\circ}08^{\prime}12^{\prime\prime}\!\!.3\pm0^{\prime\prime}\!\!.1$\\
M31N~2011-10a &LT &2011 October 26.1 &$B$ &$0^{\mathrm{h}}42^{\mathrm{m}}57^{\mathrm{s}}\!.13\pm0^{\mathrm{s}}\!.01$ &$+41^{\circ}17^{\prime}10^{\prime\prime}\!\!.9\pm0^{\prime\prime}\!\!.1$\\
M31N~2011-10d &LT &2011 October 26.1 &$B$ &$0^{\mathrm{h}}42^{\mathrm{m}}55^{\mathrm{s}}\!.74\pm0^{\mathrm{s}}\!.01$ &$+41^{\circ}17^{\prime}52^{\prime\prime}\!\!.3\pm0^{\prime\prime}\!\!.1$\\
M31N~2011-12a &LT &2011 December 26.8 &$V$ &$0^{\mathrm{h}}42^{\mathrm{m}}06^{\mathrm{s}}\!.277\pm0^{\mathrm{s}}\!.009$ &$+41^{\circ}01^{\prime}28^{\prime\prime}\!\!.7\pm0^{\prime\prime}\!\!.1$\\
M31N~2012-01a &LT &2012 January 10.8 &$V$ &$0^{\mathrm{h}}41^{\mathrm{m}}41^{\mathrm{s}}\!.01\pm0^{\mathrm{s}}\!.01$ &$+41^{\circ}19^{\prime}44^{\prime\prime}\!\!.3\pm0^{\prime\prime}\!\!.1$\\
M31N~2012-09a &LT &2012 September 10.0 &$r'$ &$0^{\mathrm{h}}42^{\mathrm{m}}47^{\mathrm{s}}\!.16\pm0^{\mathrm{s}}\!.01$ &$+41^{\circ}16^{\prime}19^{\prime\prime}\!\!.63\pm0^{\prime\prime}\!\!.07$\\
M31N~2012-09b &LT &2012 September 18.2 &$r'$ &$0^{\mathrm{h}}42^{\mathrm{m}}50^{\mathrm{s}}\!.98\pm0^{\mathrm{s}}\!.02$ &$+41^{\circ}14^{\prime}09^{\prime\prime}\!\!.7\pm0^{\prime\prime}\!\!.2$\\
M31N~2012-12a &LT &2012 December 20.9 &$r'$ &$0^{\mathrm{h}}42^{\mathrm{m}}49^{\mathrm{s}}\!.13\pm0^{\mathrm{s}}\!.02$ &$+41^{\circ}17^{\prime}02^{\prime\prime}\!\!.5\pm0^{\prime\prime}\!\!.1$
\enddata
\end{deluxetable*}

\subsection{Photometry}

Crowded field, PSF-fitting photometry was performed on the available
{\it HST} data around the position of each nova using the specific
ACS/WFC modules for the DOLPHOT photometry software and the HSTphot
software for the WFPC2 data.  These software packages create catalogs
of all objects (above a certain detection threshold; $3\sigma$ above
the local background in this case) simultaneously across all available
filters for each data-set.  For the purpose of this work HSTphot and
DOLPHOT were essentially run as ``black-boxes'' following the standard
procedure and parameters given in the manuals/cookbooks.  We combined
photometry taken in the same filter from multiple observations (all
taken at one {\it HST} observation epoch; i.e.\ all the raw (flt) data
used to created one drz image) by using a weighted mean, with the
associated errors propagated with the weightings in mind.

The results of this photometry were first used to determine the
position of all objects in the vicinity of the nova eruption, and from
this the position of the most likely (statistically closest) matching
object is drawn. We also use the positional information to determine
the object density in the region surrounding each nova eruption.  The
object density is then used -- through a Monte Carlo technique -- to
calculate the probability that a positional offset, between the nova
and progenitor candidate, at least as small can occur through
chance. We use this probability to examine how significant the
detection of any progenitor candidates, which appear to be coincident
with the nova, may be.

The HSTphot/DOLPHOT software also computes photometry for each object
in the catalog.  This photometry is presented in the native {\it HST}
photometric/filter system, and where photometry in at least two
filters is available for a particular object we convert these data
from the {\it HST} flight system to the {\it UBVRI} system using the
relations from \citet{2009PASP..121..655D} and
\citet{2005PASP..117.1049S} for the WFPC2 and ACS/WFC data
respectively. Additionally, the extinction internal to M31 that the
novae may be subject to was calculated for each individual system
based on its apparent position within the galaxy. The extinction
estimates were based on the internal extinction map of M31 published
by \citet{2005PhDT.........2D}. This extinction map gives the $r'$-band extinction for an object positioned at the far side of
M31. We then used the extinction law from \citet{1989ApJ...345..245C}
to estimate the equivalent $B$, $V$, $R$ and $I$
extinction. Extrapolation was used for novae located in regions not
covered by the extinction map. We assumed that the actual
line-of-sight internal extinction will be distributed randomly
(linearly) between this maximum value and zero, and incorporated this
into the magnitude and color determinations of the quiescent systems.

\section{Progenitor Systems} \label{sec:prog}

Here we present detailed information regarding the outcome of the
progenitor candidate search for each of the 38 novae in the input
catalog.  The positional information for the most likely matching
objects for each nova, along with the probability of coincidence are
summarised in Table~\ref{tab:res}.

\begin{deluxetable*}{lccccc}
\tablecaption{The separation between the calculated nova eruption
  position and the nearby objects resolvable in archival {\it HST}
  data in arc-seconds and $\sigma$ (to nearest 0.05$\sigma$), along with
  the probability of the detection being a chance alignment. The table
  shows all resolved objects within 3$\sigma$ of the calculated
  eruption position or, if no source is within 3$\sigma$, the closest
  source. The progenitor candidates for the two novae at the bottom of
  the table were found using {\it HST} images taken during
  eruption. In each of the two systems, the progenitor candidate
  identified is the same source as that found using LT
  data.\label{tab:res}}
\tablewidth{0pt}
\tablehead{
\colhead{Nova} & \multicolumn{2}{c}{Distance from nova position} & \colhead{Significance} & \colhead{Coincidence} &\colhead{Limiting}\\
\colhead{}     & \colhead{(pixels)} & \colhead{(arcsec)}      & \colhead{($\sigma$)}   & \colhead{probability} &\colhead{Magnitude (filter)}
}
\startdata
M31N~2006-09c &1.688 &$0.169^{\prime\prime}$ &3.90 &0.285 &23.8 (F814W)\\
M31N~2006-11a &2.825 &$0.141^{\prime\prime}$ &3.95 &0.580 &26.1 (F814W)\\
M31N~2007-02b &0.325 &$0.016^{\prime\prime}$ &0.30 &0.011 &26.6 (F814W)\\
M31N~2007-10a &0.517 &$0.026^{\prime\prime}$ &1.00 &0.027 &26.3 (F814W)\\
M31N~2007-10b &1.529 &$0.076^{\prime\prime}$ &1.85 &0.212 &25.4 (F814W)\\
M31N~2007-11b &0.319 &$0.032^{\prime\prime}$ &1.75 &0.0005 &22.3 (F814W)\\
M31N~2007-11c &2.317 &$0.115^{\prime\prime}$ &2.80 &0.422 &24.7 (F814W)\\
M31N~2007-11d &0.181 &$0.009^{\prime\prime}$ &0.25 &0.003 &26.3 (F814W)\\
                            &2.018 &$0.101^{\prime\prime}$ &1.75 &0.344 &\\
M31N~2007-11e &0.786 &$0.039^{\prime\prime}$ &1.40 &0.042 &26.4 (F814W)\\
M31N~2007-12a &0.446 &$0.022^{\prime\prime}$ &1.30 &0.017 &26.1 (F814W)\\
M31N~2007-12b &0.684 &$0.034^{\prime\prime}$ &0.95 &0.043 &25.5 (F814W)\\
M31N~2008-10b &1.815 &$0.091^{\prime\prime}$ &2.95 &0.309 &26.4 (F435W)\\
M31N~2008-12b &1.431 &$0.072^{\prime\prime}$ &1.75 &0.180 &26.0 (F475W)\tablenotemark{a}\\
M31N~2009-08a &1.118 &$0.056^{\prime\prime}$ &1.40 &0.121 &26.0 (F435W)\\
M31N~2009-08b &2.830 &$0.142^{\prime\prime}$ &3.25 &0.571 &26.6 (F814W)\\
M31N~2009-08d &3.236 &$0.163^{\prime\prime}$ &2.95 &0.627 &25.9 (F435W)\\
M31N~2009-10b &1.803 &$0.091^{\prime\prime}$ &4.30 &0.291 &26.7 (F435W)\\
M31N~2009-10c &2.128 &$0.107^{\prime\prime}$ &2.40 &0.287 &25.9 (F435W)\\
M31N~2009-11a &3.999 &$0.201^{\prime\prime}$ &3.45 &0.683 &26.5 (F814W)\\
M31N~2009-11b &1.971 &$0.197^{\prime\prime}$ &4.15 &0.386 &23.8 (F814W)\\
M31N~2009-11c &2.975 &$0.150^{\prime\prime}$ &3.35 &0.550 &25.8 (F814W)\\
M31N~2009-11d &0.505 &$0.025^{\prime\prime}$ &0.60 &0.022 &26.4 (F814W)\\
M31N~2009-11e &1.104 &$0.110^{\prime\prime}$ &0.80 &0.099 &22.8 (F814W)\\
M31N~2010-01a &1.058 &$0.053^{\prime\prime}$ &1.45 &0.094 &26.1 (F435W)\\
M31N~2010-05a &1.939 &$0.098^{\prime\prime}$ &2.20 &0.369 &25.8 (F435W)\\
M31N~2010-09b &0.579 &$0.029^{\prime\prime}$ &0.55 &0.026 &26.3 (F814W)\\
        &2.264 &$0.114^{\prime\prime}$ &2.05 &0.363 &\\
M31N~2010-10a &1.181 &$0.059^{\prime\prime}$ &1.20 &0.112 &26.7 (F814W)\\
M31N~2010-10b &1.968 &$0.099^{\prime\prime}$ &3.60 &0.334 &26.1 (F814W)\\
M31N~2010-10c &1.083 &$0.055^{\prime\prime}$ &1.05 &0.124 &27.1 (F606W)\\
        &1.988 &$0.100^{\prime\prime}$ &1.90 &0.385 &\\
        &2.084 &$0.105^{\prime\prime}$ &2.00 &0.416 &\\
        &3.405 &$0.171^{\prime\prime}$ &3.00 &0.786 &\\
M31N~2010-10d &2.992 &$0.151^{\prime\prime}$ &4.25 &0.656 &26.3 (F435W)\\
M31N~2010-10e &1.216 &$0.061^{\prime\prime}$ &0.80 &0.156 &25.9 (F814W)\\
        &1.859 &$0.094^{\prime\prime}$ &1.45 &0.344 &\\
        &4.310 &$0.217^{\prime\prime}$ &2.75 &0.923 &\\
M31N~2011-10a &2.540 &$0.128^{\prime\prime}$ &4.80 &0.468 &26.1 (F475W)\tablenotemark{a}\\
M31N~2011-10d &1.400 &$0.070^{\prime\prime}$ &3.25 &0.164 &26.1 (F475W)\tablenotemark{a}\\
M31N~2011-12a &1.129 &$0.113^{\prime\prime}$ &1.15 &0.036 &24.2 (F814W)\\
M31N~2012-01a &1.712 &$0.086^{\prime\prime}$ &2.50 &0.220 &26.2 (F814W)\\
M31N~2012-09a &2.152 &$0.108^{\prime\prime}$ &3.65 &0.328 &25.8 (F475W)\tablenotemark{a}\\
M31N~2012-09b &3.659 &$0.184^{\prime\prime}$ &10.55 &0.784 &24.6 (F814W)\\
M31N~2012-12a &1.532 &$0.077^{\prime\prime}$ &2.00 &0.199 &25.6 (F475W)\tablenotemark{a}\\
\hline
M31N~2009-08a &0.516 &$0.026^{\prime\prime}$ &2.85 &0.024 &26.0 (F435W)\\
M31N~2010-01a &0.641 &$0.032^{\prime\prime}$ &2.65 &0.038 &26.1 (F435W)
\enddata
\tablenotetext{a}{Due to problems with the F814W photometry, short
  exposure images had to be used, leading to a relatively bright
  limiting magnitude. For F814W limiting magnitudes see text for each
  nova.}
\end{deluxetable*}

\begin{enumerate}
\item {\it M31N~2006-09c.}
Nova M31N~2006-09c was an \ion{Fe}{2}\ type nova with an $R$-band $t_{2}$ of
$23.1\pm1.6$~days \citep{2011ApJ...734...12S}. The LT eruption
detection images were taken using a $V$-band filter on 2006 September
19.0 and we calculated the position of the nova to be
$0^{\mathrm{h}}42^{\mathrm{m}}42^{\mathrm{s}}\!.38\pm0^{\mathrm{s}}\!.02$,
$+41^{\circ}08^{\prime}45^{\prime\prime}\!\!.4\pm0^{\prime\prime}\!\!.2$. The
{\it HST} images used were taken using WFPC2 with F814W and F555W
filters on 2004 August 22. There is no resolvable source within
3$\sigma$ of the calculated position, with the nearest object being
1.688 WFPC2 pixels, $0^{\prime\prime}\!\!.169$ or $3.90\sigma$ away
from the defined position. The local population density suggests there
is a 28.5\% probability of an object alignment this close occurring by
chance.  We note that due to the data being taken with WFPC2, which
typically cannot resolve sources as faint as those accessible to
ACS/WFC, there may be sources that would have been visible in ACS/WFC
data closer to the position of the nova than
$0^{\prime\prime}\!\!.169$. Therefore the 28.5\% coincidence
probability is valid for objects with an F814W magnitude brighter than
23.8. From the overall distribution of stars in the M31 field that are
detectable by ACS/WFC, we would expect a source to be at least as
close as $0^{\prime\prime}\!\!.169$ to a random point in a typical
ACS/WFC image about $69\%$ of the time. The location around the nova
eruption position is shown in Figure~\ref{progenitor-grid-1} (top
left).
\begin{figure*}
\begin{center}
\includegraphics[width=0.45\textwidth]{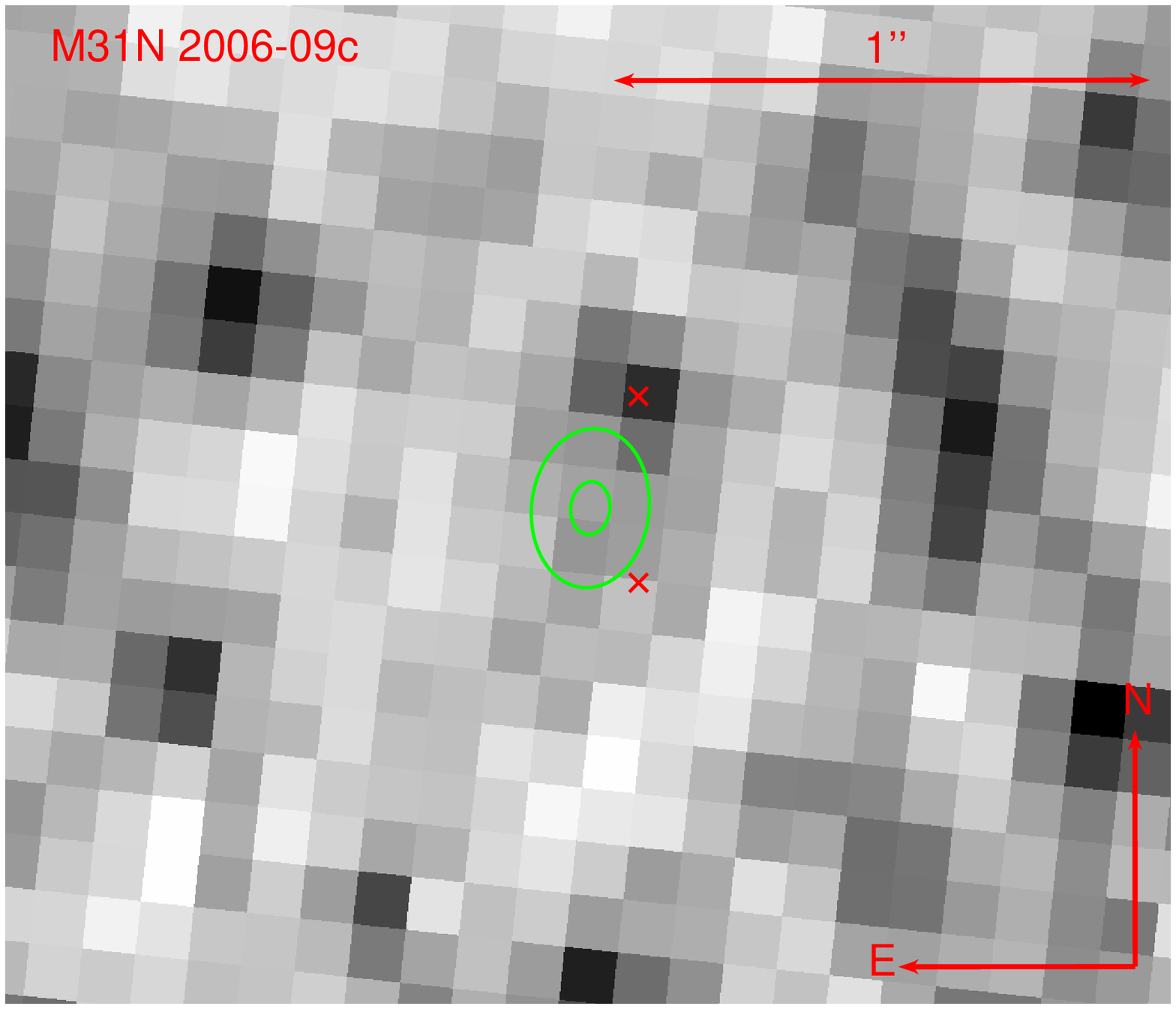}
\includegraphics[width=0.45\textwidth]{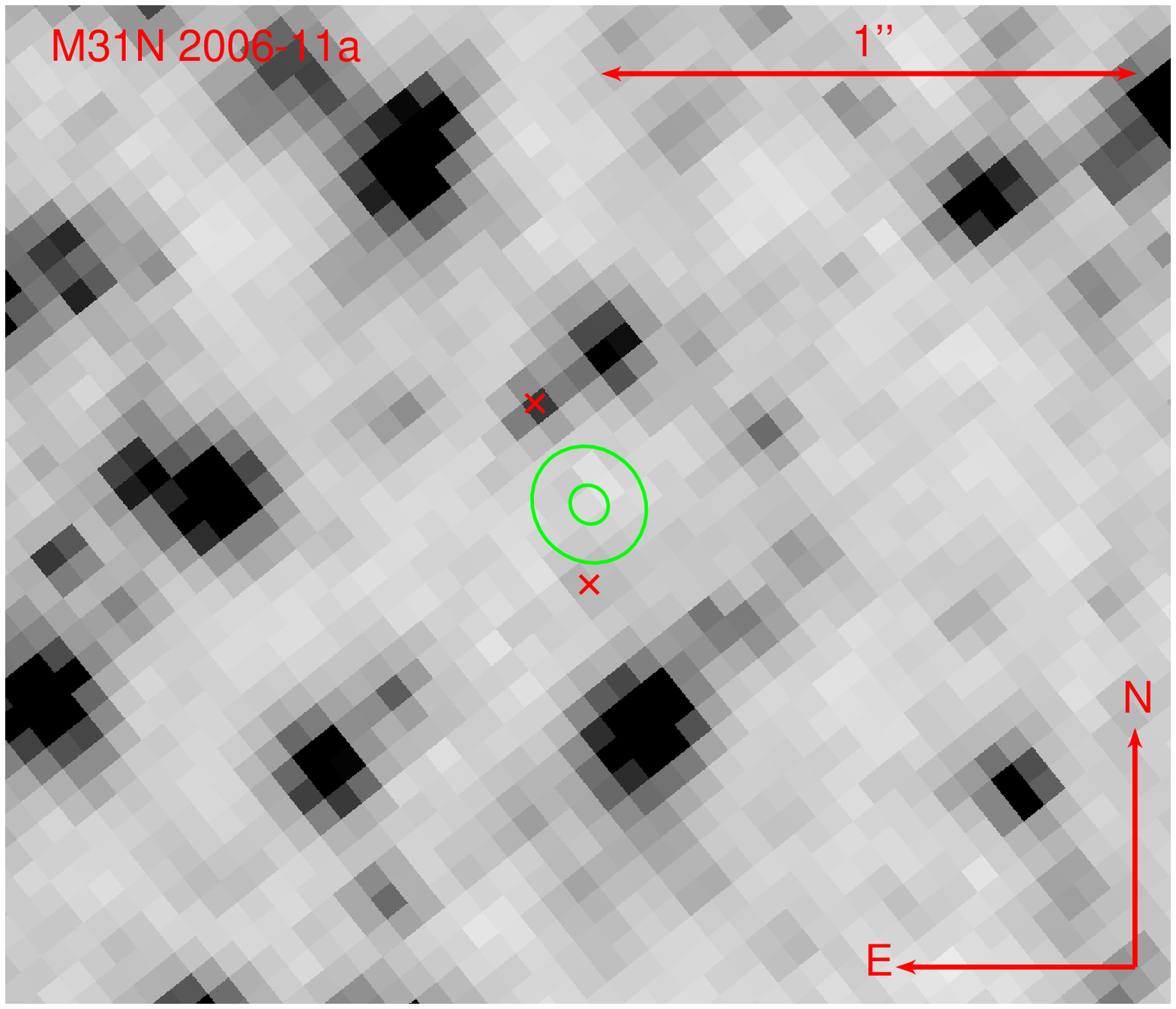}\\
\includegraphics[width=0.45\textwidth]{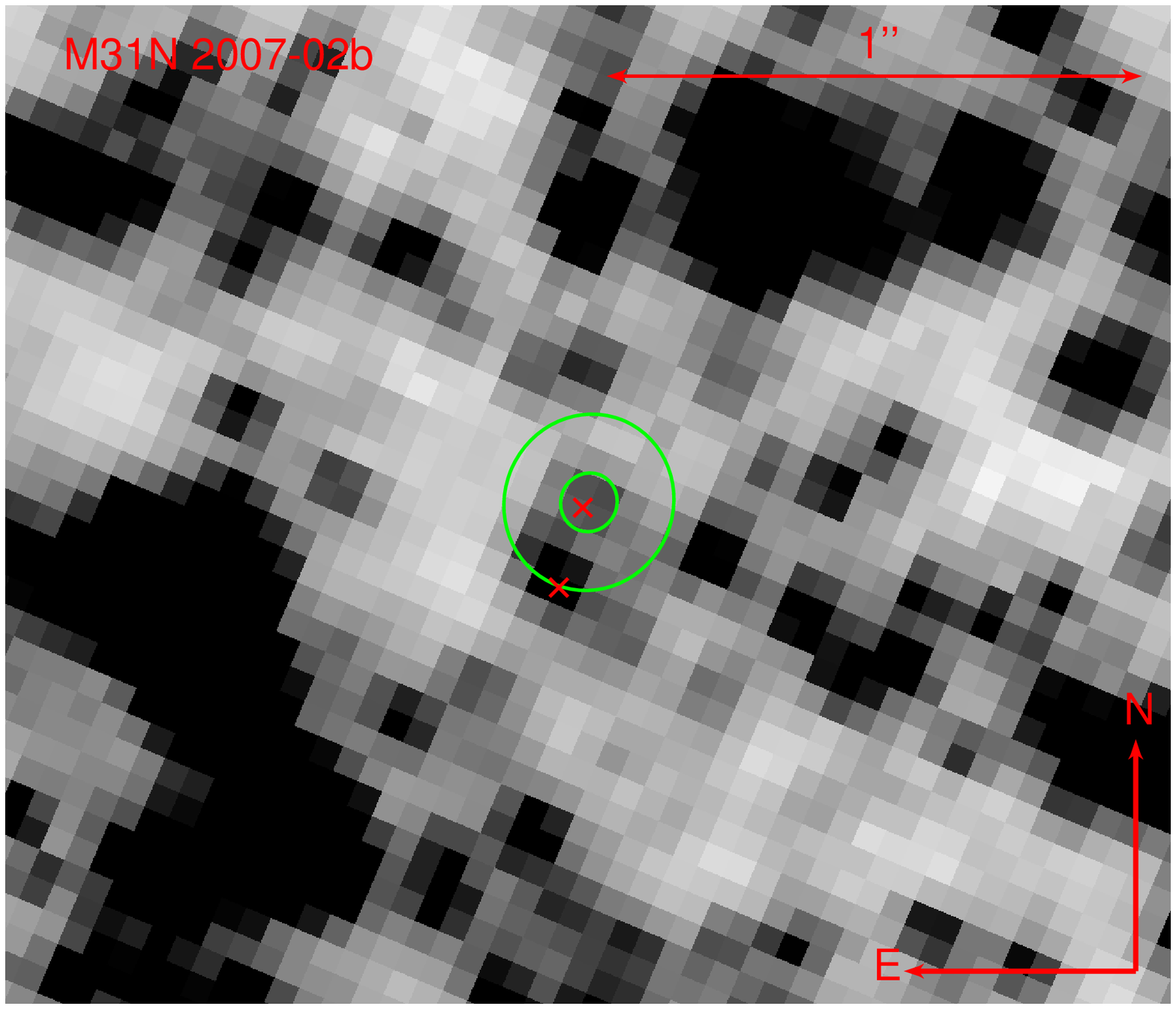}
\includegraphics[width=0.45\textwidth]{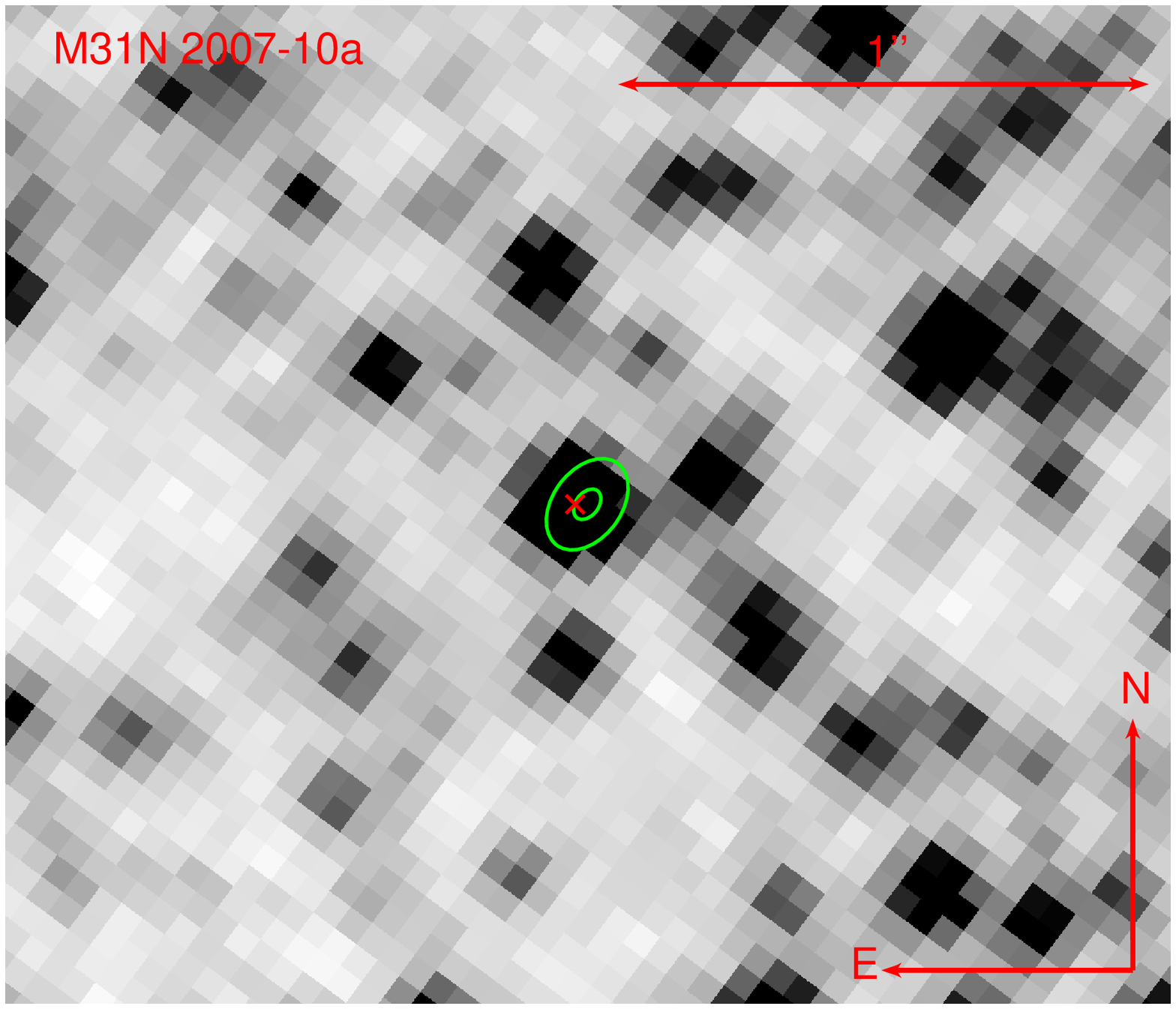}\\
\includegraphics[width=0.45\textwidth]{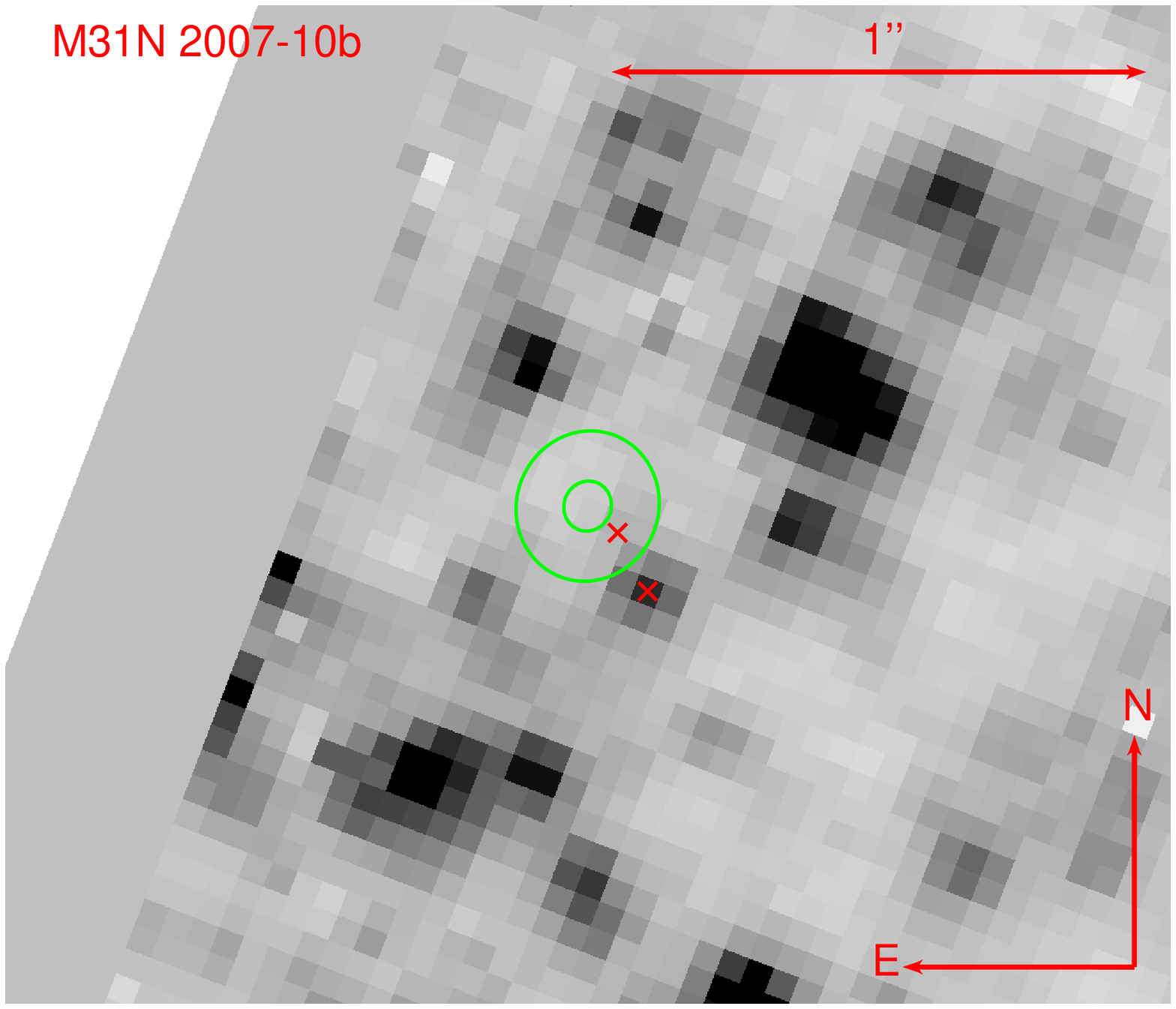}
\includegraphics[width=0.45\textwidth]{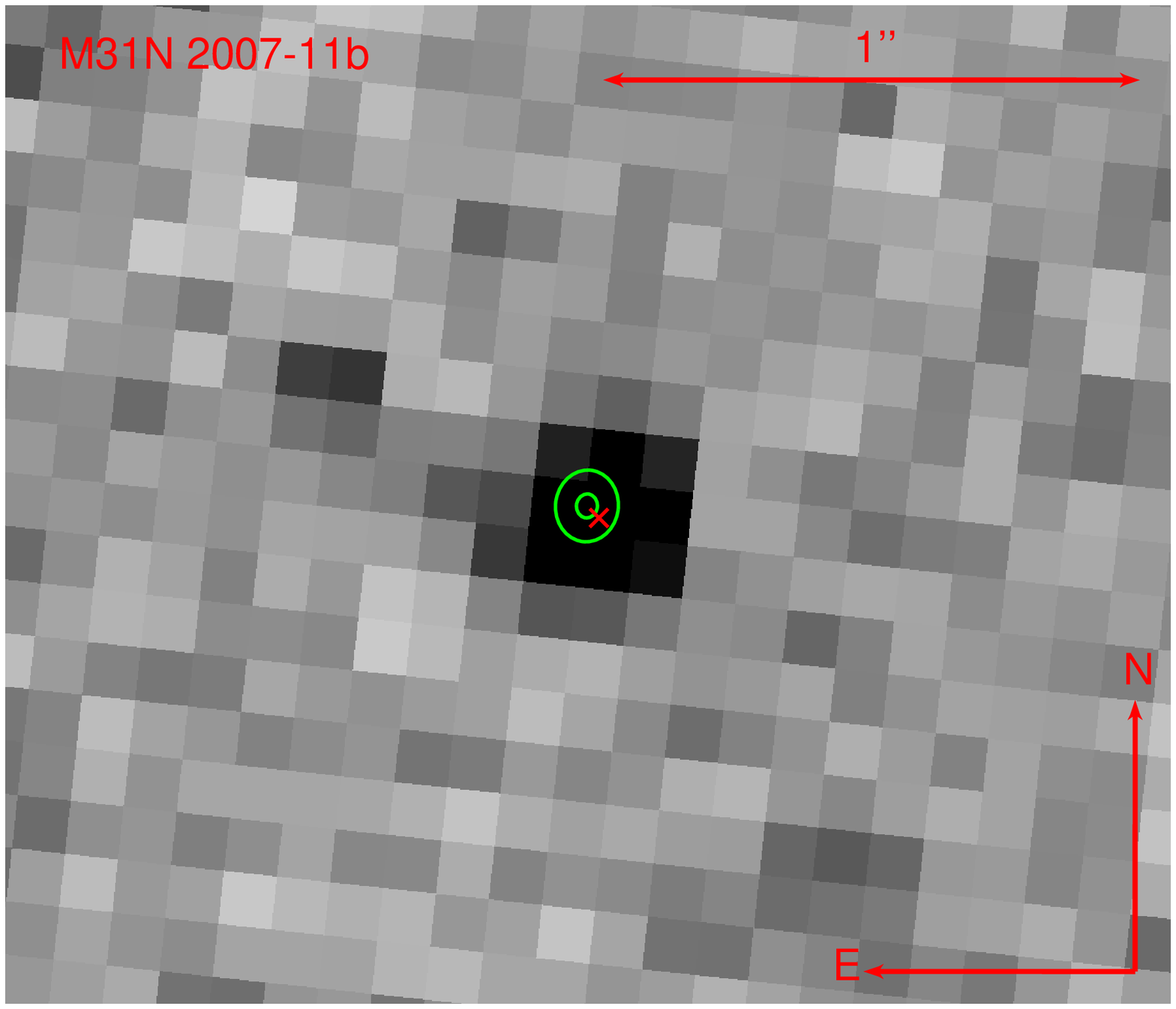}
\caption{{\it HST} images of the
  $\sim2.2^{\prime\prime}\times1.9^{\prime\prime}$ region surrounding each
  nova.  Top left: WFPC2 F555W image, M31N~2006-09c eruption position
  determined from LT $V$-band data.  Top right: ACS/WFC F814W image,
  M31N~2006-11a eruption position determined from LT $i'$-band data.
  Middle left: ACS/WFC F814W image, M31N~2007-02b eruption position
  determined from LT $i'$-band data.  Middle right: ACS/WFC F625W
  image, M31N~2007-10a eruption position determined from LT $V$-band
  data.  Bottom Left: ACS/WFC F814W image, M31N~2007-10b eruption
  position determined from LT $i'$-band data.  Bottom Right: WFPC2
  F814W image, M31N~2007-11b eruption position determined from LT
  $i'$-band data.  In each case, the inner green ellipse indicates the
  $1\sigma$ radius uncertainty of the eruption position, the outer
  ellipse the $3\sigma$ region.  The red $\times$ indicates the
  position of any nearby resolvable objects in the archival
  data. (A color version of this figure is available in the online journal.)\label{progenitor-grid-1}}
\end{center}
\end{figure*}
\item {\it M31N~2006-11a.}
The region where the progenitor system of nova M31N~2006-11a lies is
covered by {\it HST} images that were taken with ACS/WFC using F814W
and F555W filters on 2004 August 17. Again, there is no resolvable
source within $3\sigma$ of the calculated position, with the closest
resolvable source being 2.825 ACS/WFC pixels,
$0^{\prime\prime}\!\!.141$ or $3.95\sigma$ away from the defined
position. The local population density suggests that the coincidence
probability at this separation is $58.0\%$.  The limiting F814W
magnitude is 26.1. The region where this system lies is shown in
Figure~\ref{progenitor-grid-1} (top right).
\item {\it M31N~2007-02b.}
The archival {\it HST} images of the location of M31N~2007-02b were
taken with ACS/WFC using an F814W filter on 2004 July 4 and an F606W
filter on 2004 August 2. There is a resolvable source within $1\sigma$
of the calculated position and no other resolvable source within
$3\sigma$. The source is 0.325 ACS/WFC pixels,
$0^{\prime\prime}\!\!.016$ or $0.3\sigma$ away from the defined
position. The local population density, which is resolved down to an
F814W magnitude of 26.6, suggests there is only a 1.1\% chance of
coincidence at such a separation. The candidate had an F814W magnitude
of $24.82\pm0.03$ on 2004 July 4 and an F606W magnitude of
$25.95\pm0.05$ on 2004 August 2. If we simply assume the source remained at a constant brightness between the two dates, this gives a $V$-band
magnitude of $26.39\pm0.06$, $I$-band magnitude of $24.83\pm0.04$
and ($V-I$) color of $1.55\pm0.07$.  However, all novae show some
  variation in their luminosity during quiescence.  Such variation is typically
  related to the accretion disk, generally the greater the accretion
  rate the higher the amplitude of the variation.  RS~Oph shows some
  of the largest amplitude variation at quiescence, around 1 magnitude
\citep{2008ASPC..401..203D}.  We can see from
\citet{2005PhDT.........2D} that a nova at the apparent position of
M31N~2007-02b would be subject to $r'$-band extinction of
$A_{r'}=0.69$ if it were at the far side of the galaxy. Using the
extinction law from \citet{1989ApJ...345..245C} this equates to
maximum $I$-band extinction of $A_{I}=0.46$ and $V$-band
extinction of $A_{V}=0.80$. If we take this into account we derive a
$V$-band magnitude of $25.99\pm0.40$, $I$-band magnitude of
$24.60\pm0.23$ and ($V-I$) color of $1.39\pm0.18$. This system is
shown in Figure~\ref{progenitor-grid-1} (middle left). By chance,
M31N~2007-02b was also observed during its eruption with WFPC2 on 2007
September 13, which is described in Section~\ref{sec:lc}. As the nova
is relatively faint at this time any attempt to calculate the position
of the system in the quiescent {\it HST} image from the eruption {\it
  HST} image is dominated by the error on the position in the WFPC2
eruption image and does not aid the progenitor search.
\item {\it M31N~2007-10a.}
The region where M31N~2007-10a lies is covered by {\it HST} images
that were taken with ACS/WFC using an F625W filter on 2003 August
5. There is a resolvable source near $1\sigma$ of the calculated
position and no other resolvable source within $3\sigma$. The source
is 0.517 ACS/WFC pixels, $0^{\prime\prime}\!\!.026$ or $1.00\sigma$
away from the defined position. The local resolved stellar density
suggests that the coincidence probability at this separation is
2.7\%. The source had an F625W magnitude of $22.397\pm0.008$ at this
time. The {\it HST} images were only taken in one filter, so we were
unable to calculate the color of the source, although we estimate it
to have an $R$-band magnitude of $\sim22.3$ with internal
extinction in M31 taking it to $\sim22.0$. The location of this system
is shown in Figure~\ref{progenitor-grid-1} (middle right).
\item {\it M31N~2007-10b.}
The location around the position of nova M31N~2007-10b was observed by
{\it HST} with ACS/WFC using F814W and F475W filters on 2010 July
22. In the F475W data, there is one source within $3\sigma$ of the calculated position that is not detected in the F814W image, which has a limiting
magnitude of 25.4. The source is 1.529 ACS/WFC pixels,
$0^{\prime\prime}\!\!.076$ or $1.85\sigma$ away from the defined
position and the probability of such an alignment occurring by chance is 21.2\%. The nearest resolvable source in the F814W data is 3.842 ACS/WFC pixels,
$0^{\prime\prime}\!\!.193$ or 4.40$\sigma$ away from the defined
position. The local population density suggests that the coincidence
probability at this separation is 81.1\%.  The region where this nova
erupted is shown in Figure~\ref{progenitor-grid-1} (bottom left).
\item {\it M31N~2007-11b.}
Although the area of M31 where M31N~2007-11b lies has not yet been
observed with ACS/WFC, {\it HST} images were taken with WFPC2 using
F814W and F555W filters on 2005 February 18. There is a resolvable
source within $2\sigma$ of the calculated {\it HST} position and no
other resolvable source within 3$\sigma$. The source is 0.319 WFPC2
pixels, $0^{\prime\prime}\!\!.032$ or $1.75\sigma$ away from the
defined position, with the local object density suggesting a
coincidence probability at this separation of only 0.05\%. However
these images have an F814W limiting magnitude as bright as
22.3. Therefore the 0.05\% coincidence probability is valid only for
objects with an F814W magnitude $<22.3$. The candidate had an F555W
magnitude of $22.6\pm0.1$ and F814W magnitude of $20.44\pm0.06$ at
this time. This gives an {\it I}-band magnitude of $20.30\pm0.06$ and
{\it V}-band magnitude of $22.6\pm0.1$, with $(V-I)$ color of
$2.3\pm0.2$. \citet{2005PhDT.........2D} showed that a nova at the
apparent position of M31N~2007-11b would be subject to $r'$-band
extinction of approximately $A_{r'}=0.52$ if it were at the far side
of the galaxy. Therefore we derive a {\it V}-band magnitude of
$22.3\pm0.3$, {\it I}-band magnitude of $20.1\pm0.2$ and ($V-I$) color
of $2.1\pm0.2$. Notably, from the overall distribution of stars in the
M31 field in ACS/WFC, we would predict a source to be at least as
close as $0^{\prime\prime}\!\!.032$ in a typical ACS/WFC image only
about 4\% of the time. M31N~2007-11b is shown in
Figure~\ref{progenitor-grid-1} (bottom right).
\item {\it M31N~2007-11c.}
The location around the position of M31N~2007-11c was observed by {\it
  HST} with ACS/WFC using F814W and F475W filters on 2010 July
21. There is one resolvable source just within $3\sigma$ of the
calculated position. The source is 2.317 ACS/WFC pixels,
$0^{\prime\prime}\!\!.115$ or $2.80\sigma$ away from the defined
position. The local population density, which is resolved down to an
F814W magnitude of 24.7, suggests there is a 42.2\% chance of
coincidence at such a separation. The location of this nova is shown
in Figure~\ref{progenitor-grid-2} (top left).
\begin{figure*}
\begin{center}
\includegraphics[width=0.45\textwidth]{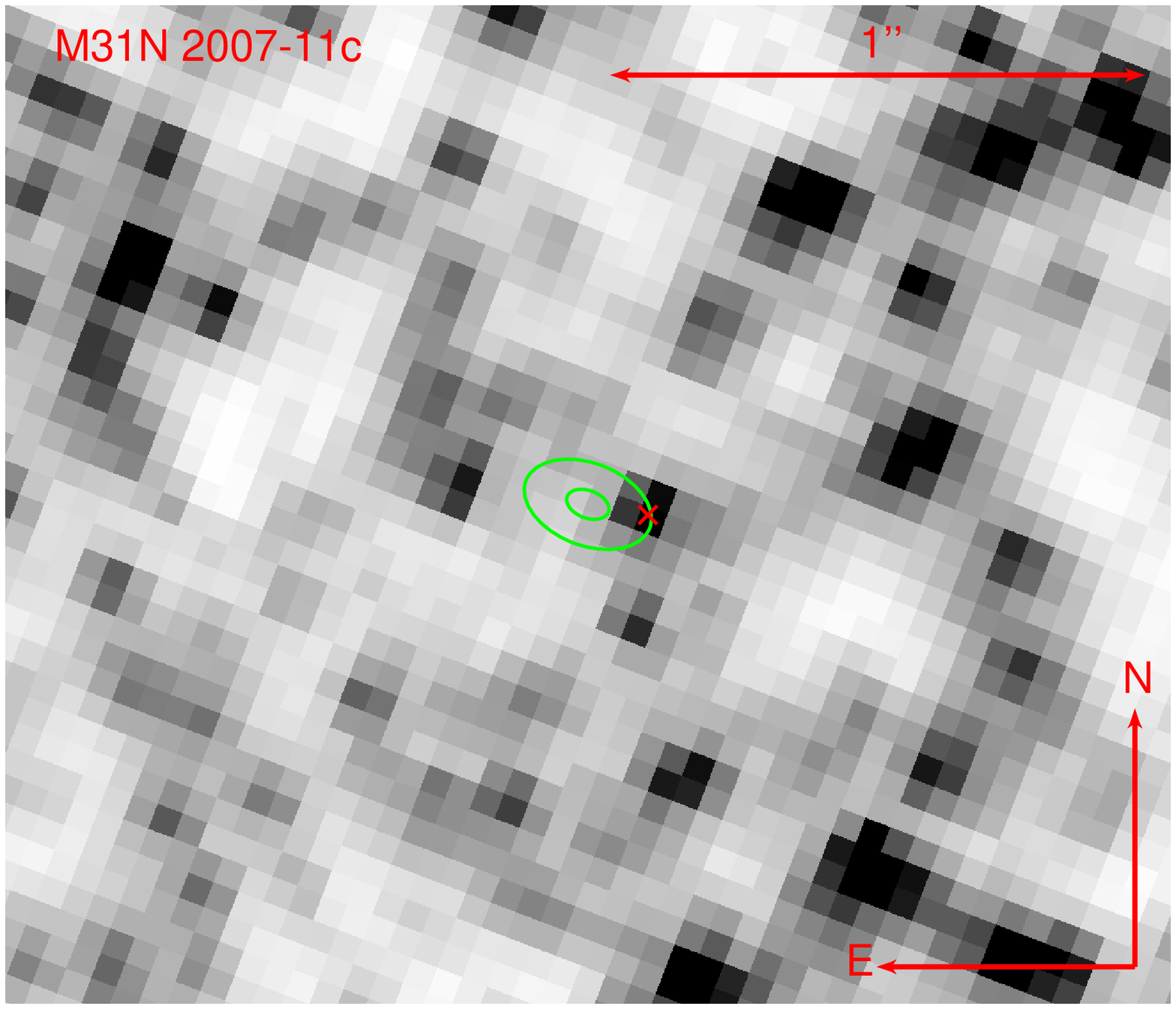}
\includegraphics[width=0.45\textwidth]{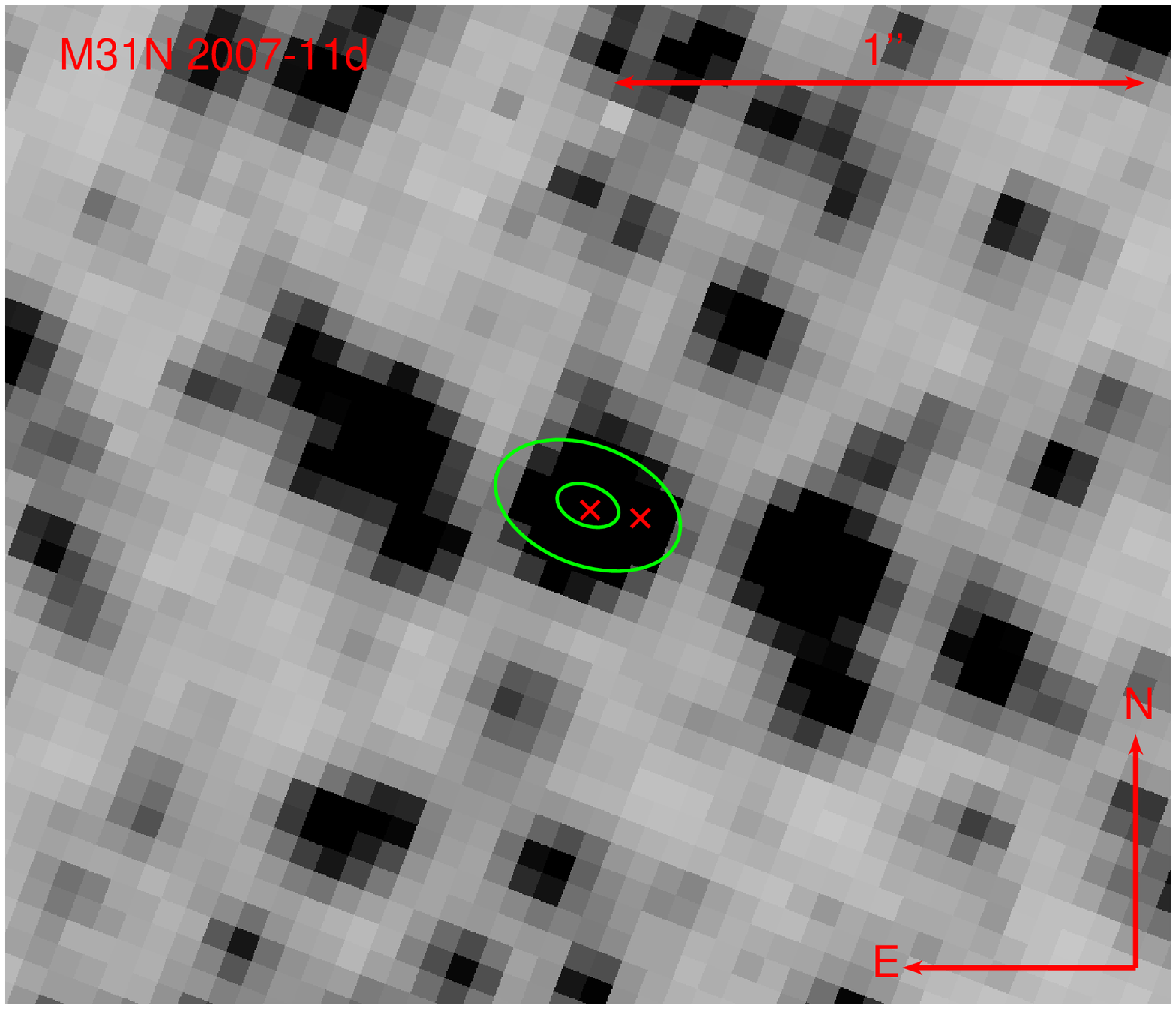}\\
\includegraphics[width=0.45\textwidth]{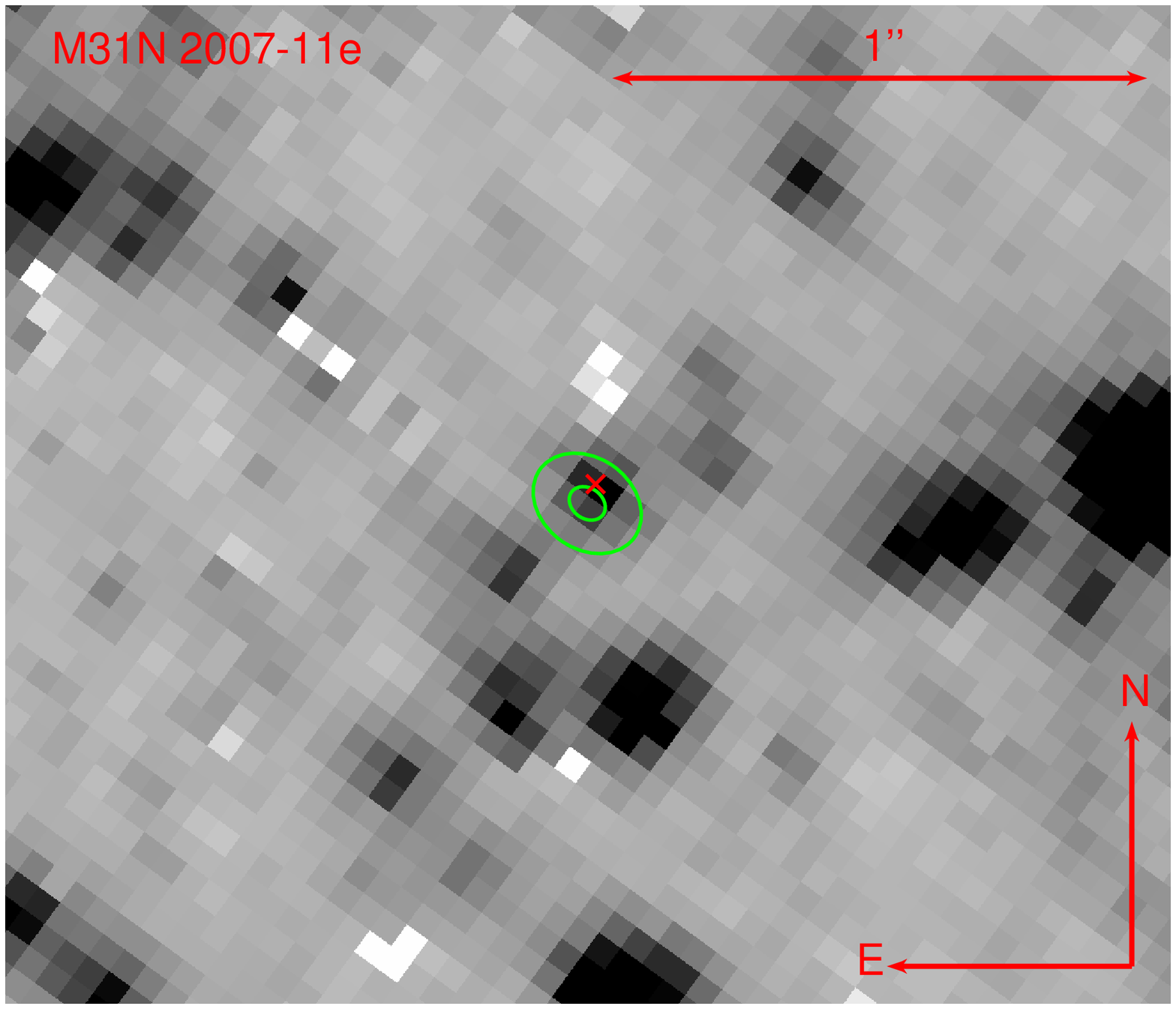}
\includegraphics[width=0.45\textwidth]{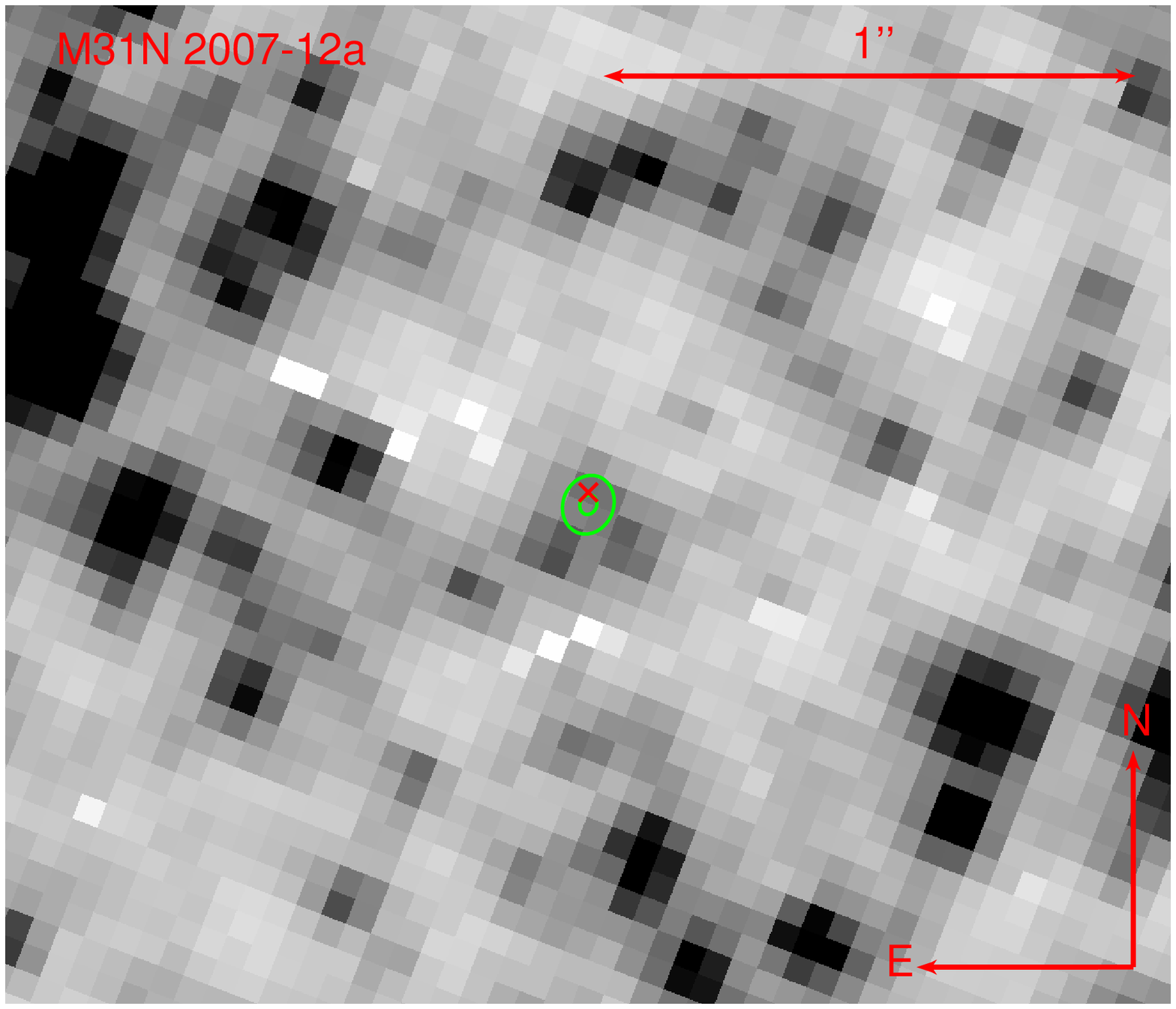}\\
\includegraphics[width=0.45\textwidth]{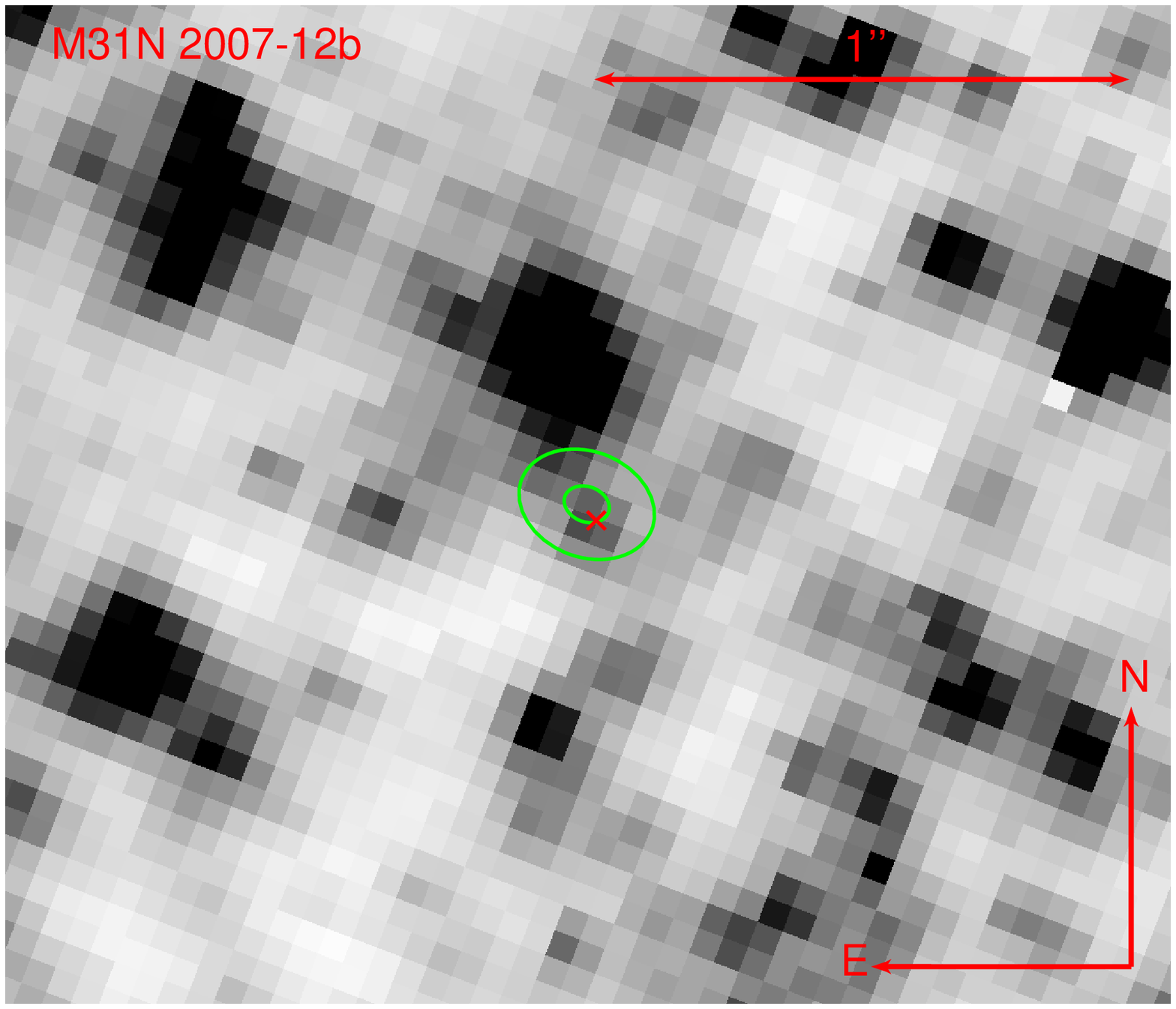}
\includegraphics[width=0.45\textwidth]{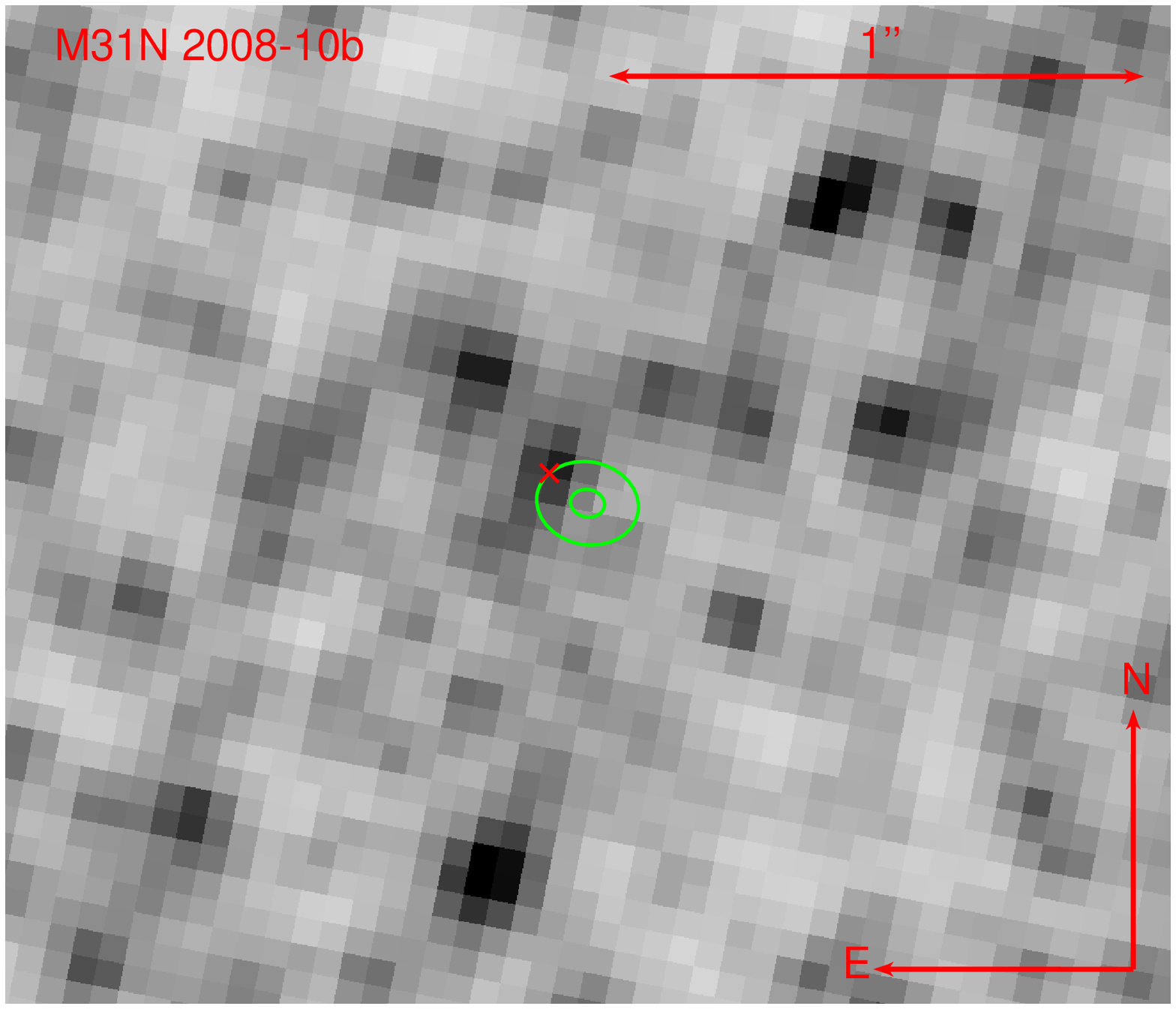}
\caption{As Figure~\ref{progenitor-grid-1}.  Top left: ACS/WFC F814W
  image, M31N~2007-11c eruption position determined from LT $i'$-band
  data.  Top right: ACS/WFC F814W image, M31N~2007-11d eruption
  position determined from LT $i'$-band data.  Middle left: ACS/WFC
  F814W image, M31N~2007-11e eruption position determined from LT
  $i'$-band data.  Middle right: ACS/WFC F814W image, M31N~2007-12a
  eruption position determined from LT $i'$-band data.  Bottom left:
  ACS/WFC F814W image, M31N~2007-12b eruption position determined from
  LT $i'$-band data.  Bottom right: ACS/WFC F435W image, M31N~2008-10b
  eruption position determined from LT $B$-band
  data. (A color version of this figure is available in the online journal.)\label{progenitor-grid-2}}
\end{center}
\end{figure*}
\item {\it M31N~2007-11d.}
{\it HST} images were taken of the position where M31N~2007-11d lies
with ACS/WFC using F814W and F475W filters on 2010 July 13. There is a
resolvable source within $1\sigma$ of the calculated position and
another within $2\sigma$. The closest candidate is 0.181 ACS/WFC
pixels, $0^{\prime\prime}\!\!.009$ or $0.25\sigma$ away from the
defined position. The local resolved stellar density, which is
resolved down to an F814W magnitude of 26.3, suggests there is only a
0.3\% chance of coincidence at such a separation. This candidate had
an F475W magnitude of $24.46\pm0.04$ and F814W magnitude of
$21.387\pm0.005$. This gives {\it I}-band magnitude of $21.44\pm0.05$,
{\it B}-band magnitude of $25.43\pm0.07$ and a $(B-I)$ color of
$3.99\pm0.09$. The other candidate is 2.018 ACS/WFC pixels,
$0^{\prime\prime}\!\!.101$ or $1.75\sigma$ away from the defined
position, with a 34.4\% chance of coincidence at this separation. We
can see from \citet{2005PhDT.........2D} that a nova at the apparent
position of M31N~2007-11d would be subject to $r'$-band extinction of
$A_{r'}=0.58$ if it were at the far side of the galaxy. Taking this
into account gives a {\it B}-band magnitude of $25.0\pm0.5$, {\it
  I}-band magnitude of $21.3\pm0.2$ and ($B-I$) color of $3.8\pm0.3$
for the closer of the two progenitor candidates. The area around
M31N~2007-11d is shown in Figure~\ref{progenitor-grid-2} (top right).
\item {\it M31N~2007-11e.}
{\it HST} images were taken with ACS/WFC using F814W and F475W filters
on 2012 February 5 that encompass the position of M31N~2007-11e. There
is a resolvable source within $2\sigma$ of the calculated {\it HST}
position and no other resolvable sources within $3\sigma$. The
candidate is 0.786 ACS/WFC pixels, $0^{\prime\prime}\!\!.039$ or
$1.40\sigma$ away from the defined position of the nova and has an
F814W magnitude of $24.19\pm0.03$ and F475W magnitude of
$25.5\pm0.1$. This gives an {\it I}-band magnitude of $24.19\pm0.04$,
{\it B}-band magnitude of $26.27\pm0.12$, with a $(B-I)$ color of
$2.07\pm0.13$. We estimate from \citet{2005PhDT.........2D} that a
nova at the apparent position of M31N~2007-11e may be subject to a
maximum $r'$-band extinction of $A_{r'}=0.42$. Taking this into
account gives a {\it B}-band magnitude of $26.0\pm0.3$, {\it I}-band
magnitude of $24.1\pm0.1$ and ($B-I$) color of $1.9\pm0.2$ for the
progenitor candidate. At the separation of $0.039^{\prime\prime}$, the
local population density suggests a coincidence probability of
4.2\%. These images have an F814W limiting magnitude of
26.4. M31N~2007-11e is shown in Figure~\ref{progenitor-grid-2} (middle
left).
\item {\it M31N~2007-12a.}
The {\it HST} images that were used to locate M31N~2007-12a were taken
with ACS/WFC using F814W and F475W filters on 2010 December 31. There
is only one resolvable source within $3\sigma$ of the calculated
position. The candidate is 0.446 ACS/WFC pixels,
$0^{\prime\prime}\!\!.022$ or $1.30\sigma$ away from the defined
position, with an F475W magnitude of $25.98\pm0.08$ and F814W
magnitude of $25.3\pm0.1$. This gives an $I$-band magnitude of
$25.3\pm0.1$ and $B$-band magnitude of $26.14\pm0.08$, with a $(B-I)$
color of $0.9\pm0.1$. We can see from \citet{2005PhDT.........2D} that
a nova at the apparent position of M31N~2007-12a would be subject to
$r'$-band extinction of  $A_{r'}=0.61$ if it were at the far side of
the galaxy. Therefore we derived a {\it B}-band magnitude of
$25.7\pm0.5$, {\it I}-band magnitude of $25.1\pm0.2$ and ($B-I$) color
of $0.6\pm0.3$. The local population density, which has an F814W
limiting magnitude of 26.1, suggests there is only a 1.7\% chance of
coincidence at such a separation. This nova is shown in
Figure~\ref{progenitor-grid-2} (middle right).
\item {\it M31N~2007-12b.}
The location of M31N~2007-12b was observed by {\it HST} with ACS/WFC
using F814W and F475W filters on 2010 July 21 and 22. There is a
resolvable source within $1\sigma$ of the calculated position and no
other resolvable sources within $3\sigma$. The candidate is 0.684
ACS/WFC pixels, $0^{\prime\prime}\!\!.034$ or $0.95\sigma$ away from
the defined position, with the local resolved stellar density
suggesting there is a 4.3\% chance of coincidence at such a
separation. The progenitor candidate had an F475W magnitude of
$25.36\pm0.04$ and F814W magnitude of $23.79\pm0.04$ at this time,
which gives an $I$-band magnitude of $23.80\pm0.04$, $B$-band
magnitude of $26.14\pm0.06$ and $(B-I)$ color of $2.34\pm0.07$. This
is the same candidate that was identified by
\citet{2009ApJ...705.1056B}, who used a different set of {\it HST}
images and found that the probability of a star being as close by
change was 3.4\% (the 2010 post-eruption observations were not
available at the time of that publication) They found the candidate to
have $I$-band apparent magnitude of $22.33\pm0.04$ and $(V-I)$ color
of $2.3\pm0.1$ in August 2004 (it is worth noting that
\citet{2009ApJ...705.1056B} indicated that their F814W observations
may have been affected by a cosmic ray). The location of the quiescent
M31N~2007-12b is shown in Figure~\ref{progenitor-grid-2} (bottom
left). \citet{2005PhDT.........2D} found that a nova at the apparent
position of M31N~2007-12b would be subject to $r'$-band extinction of
$A_{r'}=0.65$ if it were at the far side of the galaxy. We therefore
calculate an unextinguished {\it B}-band magnitude of $25.7\pm0.5$ and
{\it I}-band magnitude of $23.6\pm0.2$, and a de-reddened ($B-I$)
color of $2.1\pm0.3$.
\item {\it M31N~2008-10b.}
{\it HST} images, coincident with the position of M31N~2008-10b, were
taken with ACS/WFC using an F435W filter on 2004 October 2. There is
one resolvable source within $3\sigma$ of the calculated position. The
candidate is 1.815 ACS/WFC pixels, $0^{\prime\prime}\!\!.091$ or
$2.95\sigma$ away from the defined position. The local population
density, which is resolved down to an F435W magnitude of 26.4,
suggests there is a 30.9\% chance of coincidence at such a
separation. The {\it HST} images were only taken in one filter, so
were unable to measure the color of the source. The location of
M31N~2008-10b is shown in Figure~\ref{progenitor-grid-2} (bottom
right).  Additional post-eruption ACS/WFC F435W data taken on 2010
January 1 allow us to determine that the source listed above is not
related to the
nova, this is illustrated in Figure~\ref{2008-10bhst} (see also Section~\ref{sec:lc} for
further discussion).
\begin{figure}
\begin{center}
\includegraphics[width=0.45\textwidth]{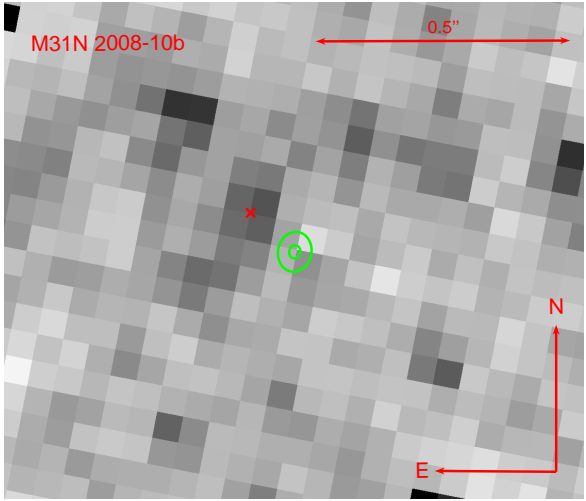}
\caption{As Figure~\ref{progenitor-grid-1}. {\it HST} ACS/WFC F435W FLT image
  with the M31N~2008-10b eruption position determined from additional post-eruption
  ACS/FWC F435W data. The marked source is the same as that marked in
  Figure~\ref{progenitor-grid-2} (bottom right). (A color version of this figure is available in the online journal.)\label{2008-10bhst}}
\end{center}
\end{figure}
\item {\it M31N~2008-12b.}
The location of nova M31N~2008-12b was observed by {\it HST} with
ACS/WFC using F814W and F475W filters on 2012 July 9. The closest
resolvable source, and the only one within $3\sigma$, is 1.431 ACS/WFC
pixels, $0^{\prime\prime}\!\!.072$ or $1.75\sigma$ away from the
defined position. It has an F475W magnitude of $24.75\pm0.05$, but is
not detected in the F814W data. However, it should be noted that due
to problems with the photometry and a short-exposure F814W image
having to be used, the limiting F814W magnitude was only 22.7, whereas
the F475W magnitude limit was 26.0. The local population density
suggests that there is an 18.0\% probability of such an alignment
occurring by chance. The closest source resolvable in the F814W data
was 4.558 ACS/WFC pixels, $0^{\prime\prime}\!\!.228$ or $5.80\sigma$
away from the defined position. The location of this nova eruption,
including the candidate only visible in the F475W image, is shown in
Figure~\ref{progenitor-grid-3} (top left).
\begin{figure*}
\begin{center}
\includegraphics[width=0.45\textwidth]{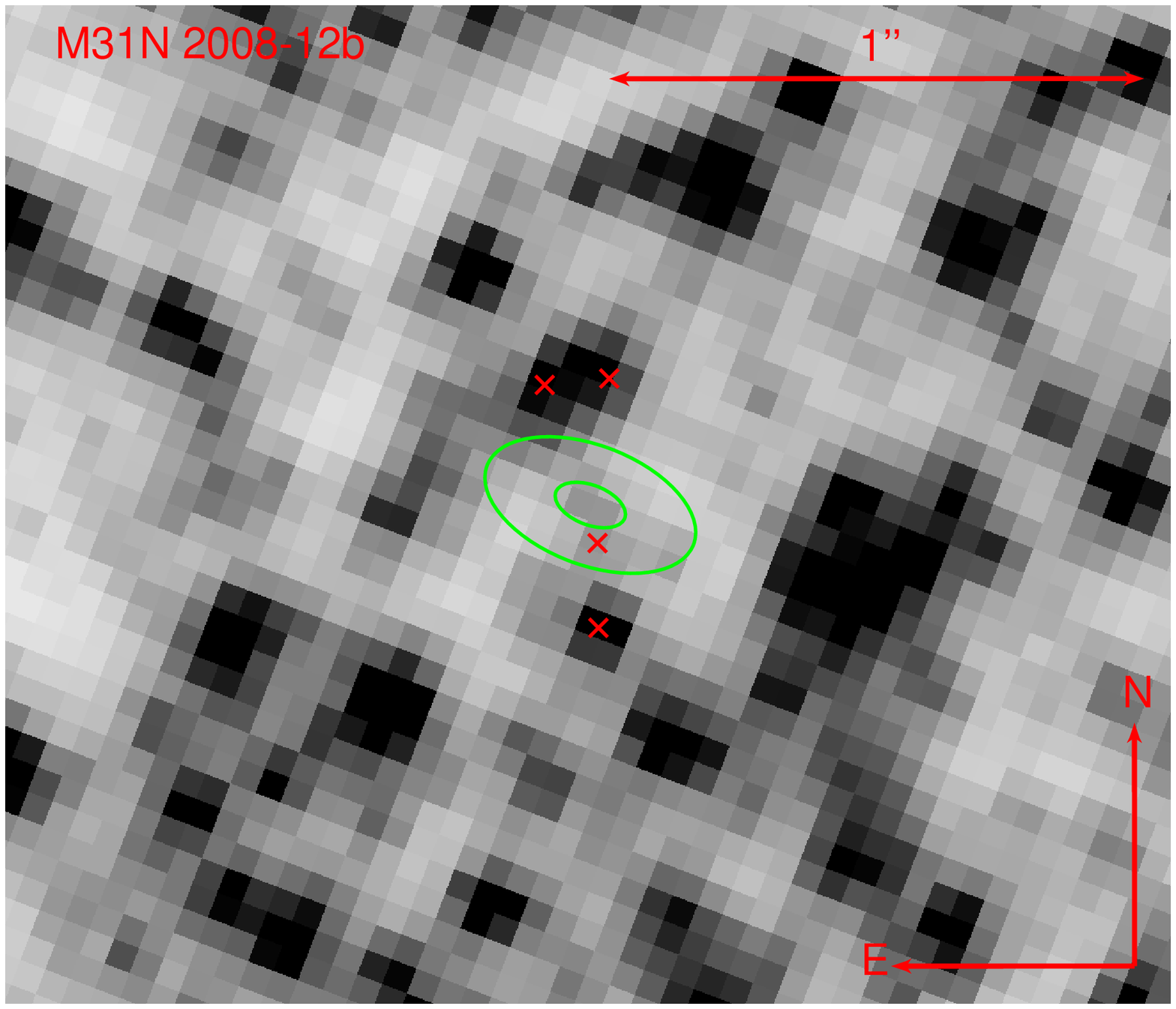}
\includegraphics[width=0.45\textwidth]{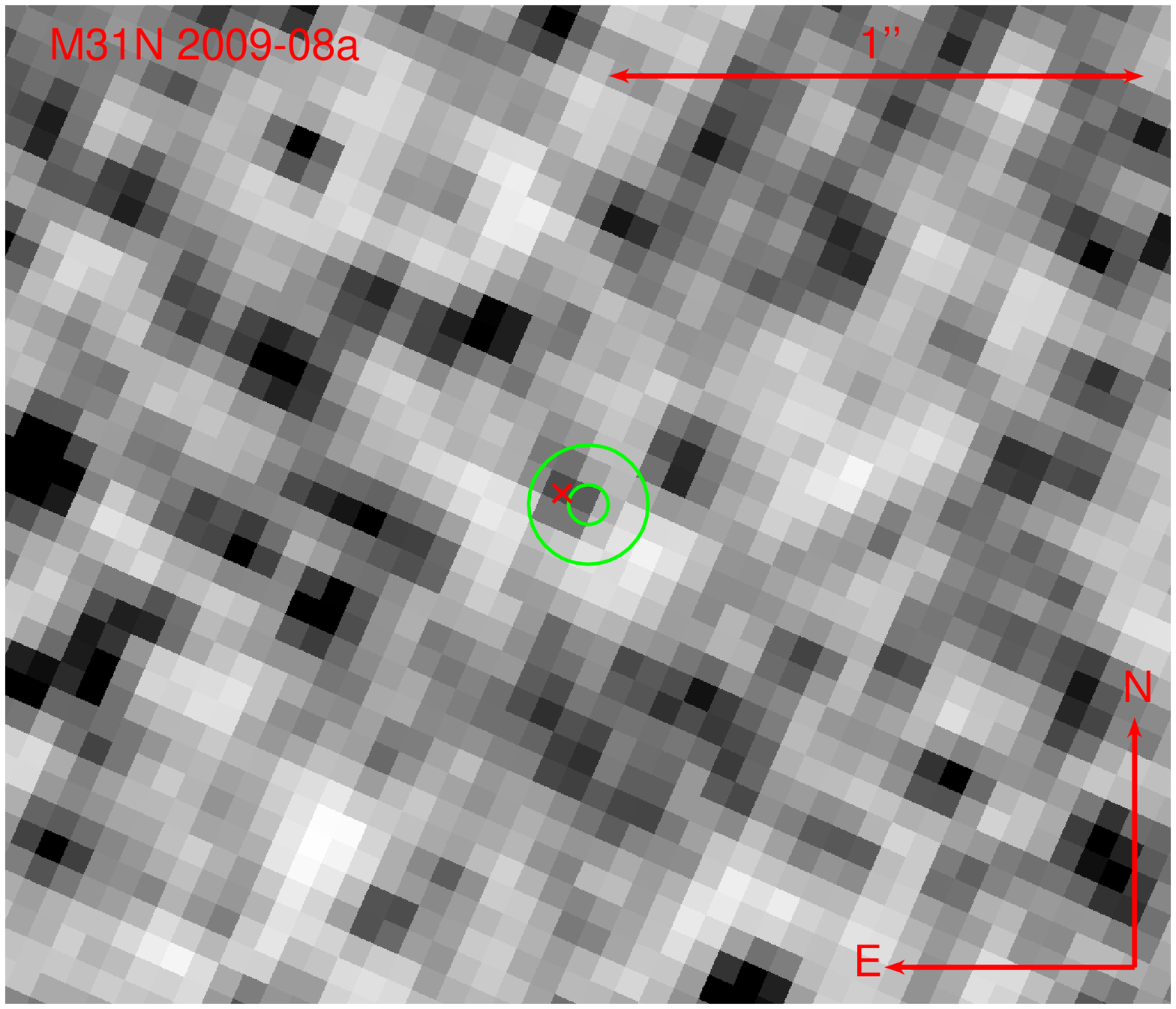}\\
\includegraphics[width=0.45\textwidth]{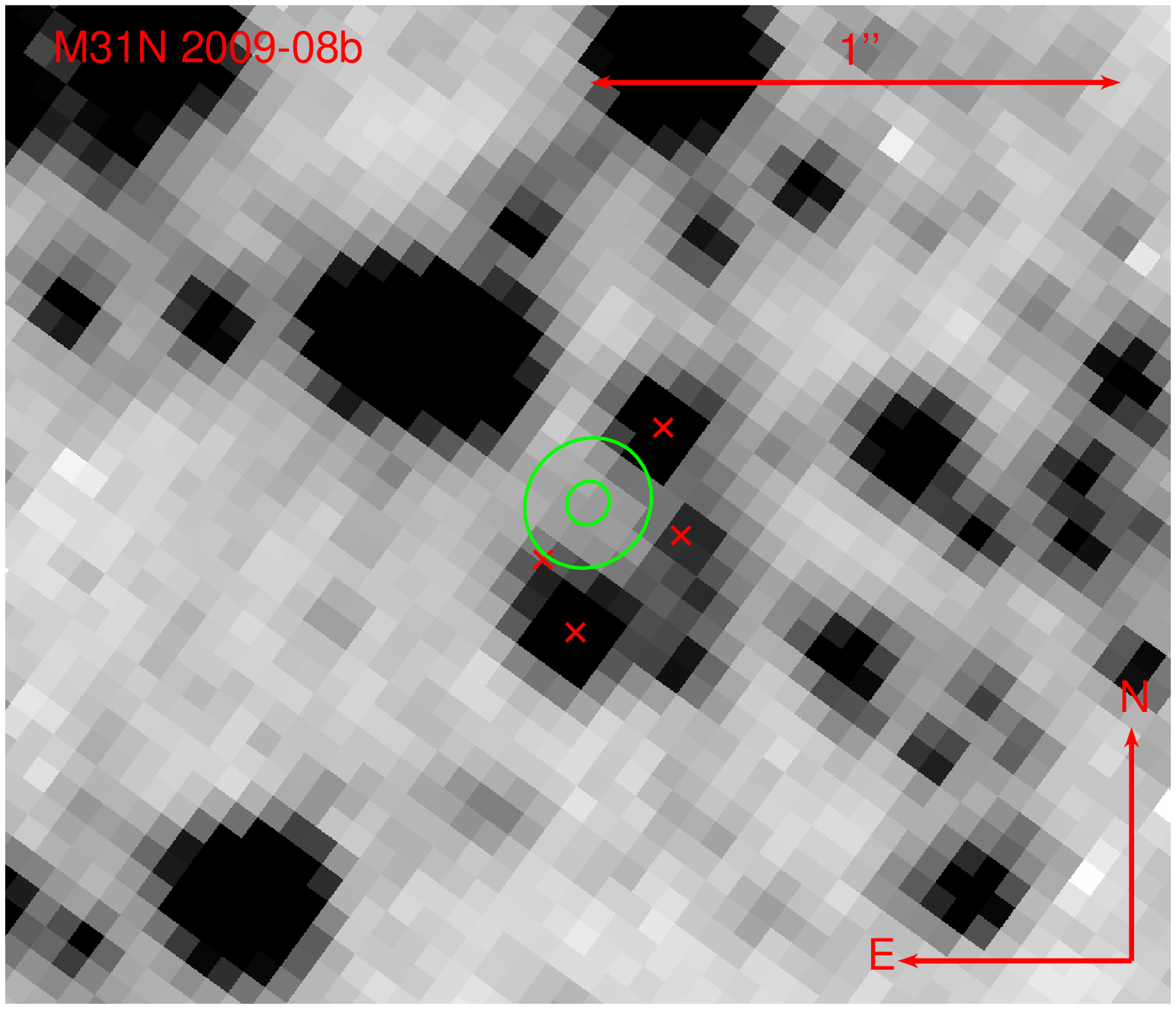}
\includegraphics[width=0.45\textwidth]{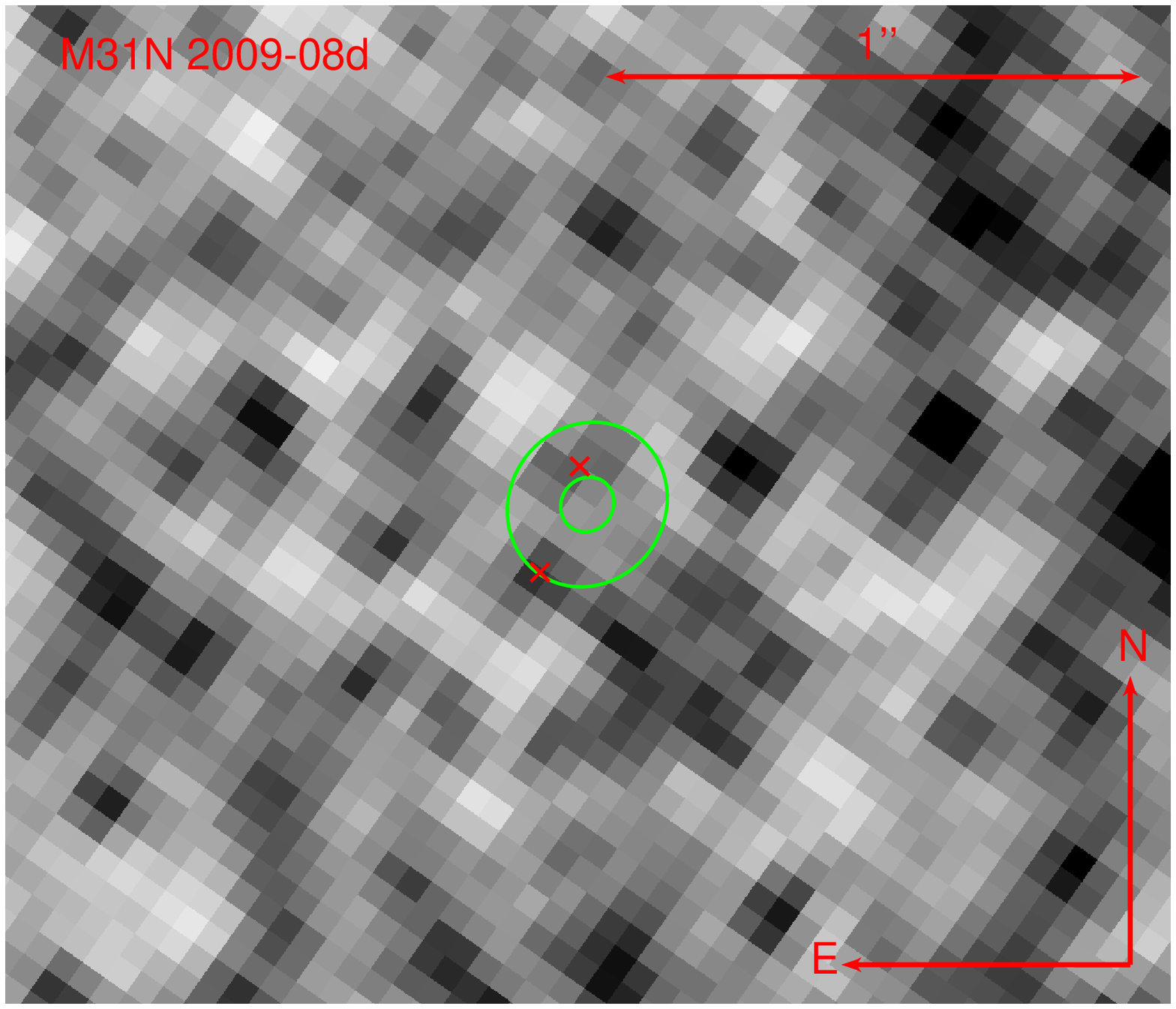}\\
\includegraphics[width=0.45\textwidth]{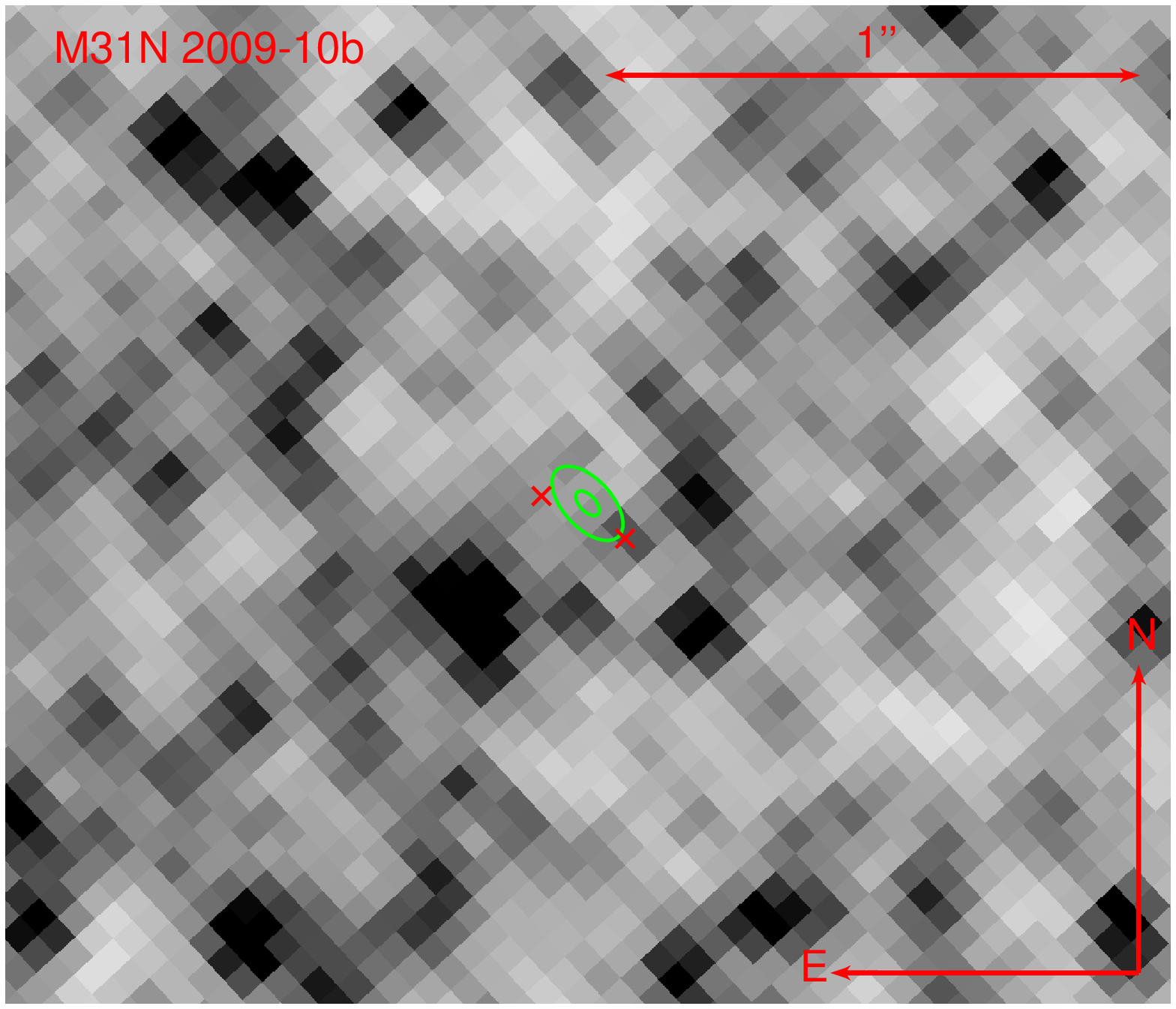}
\includegraphics[width=0.45\textwidth]{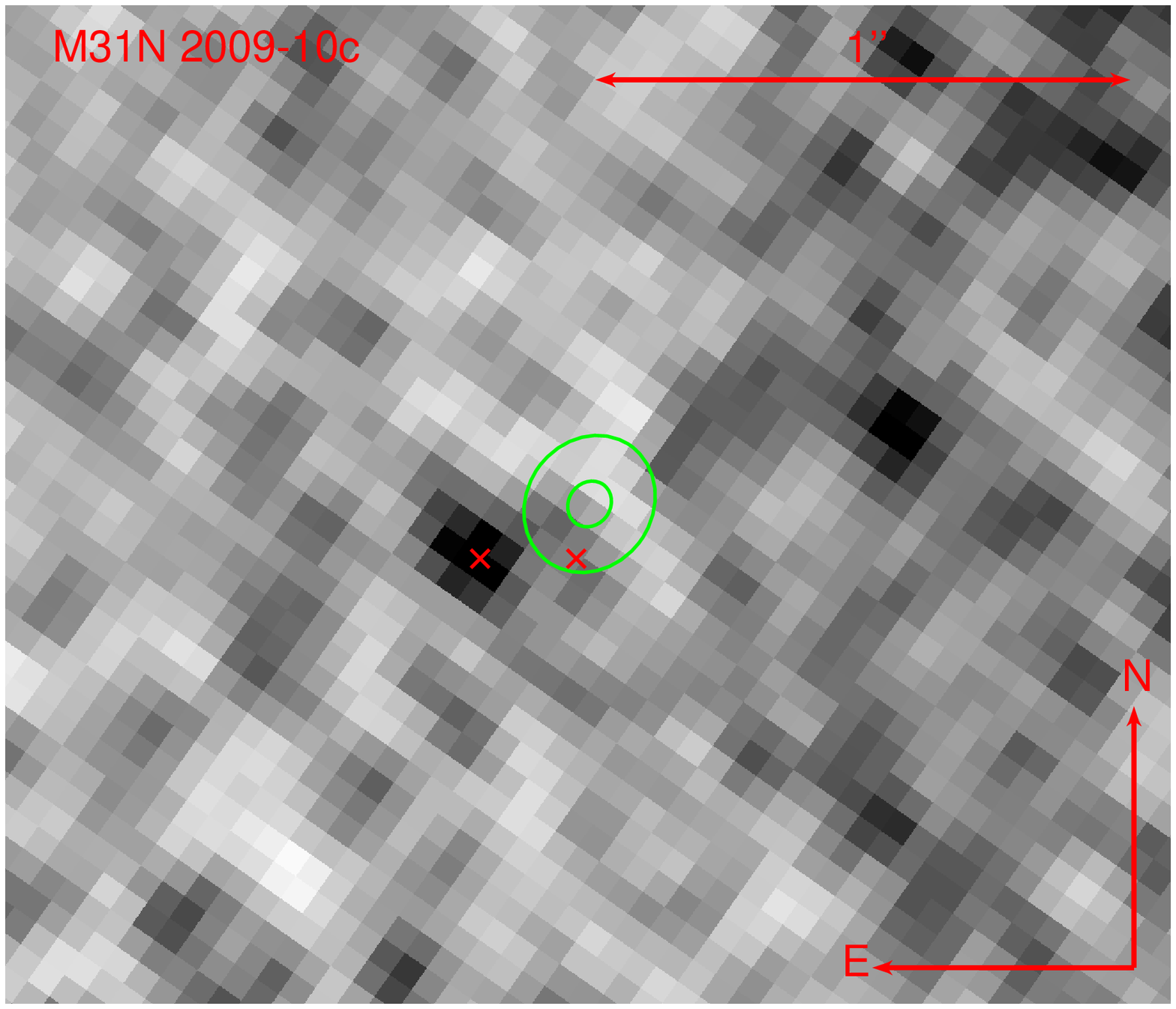}
\caption{As Figure~\ref{progenitor-grid-1}.  Top left: ACS/WFC F814W
  image, M31N~2008-12b eruption position determined from LT $i'$-band
  data.  Top right: ACS/WFC F435W image, M31N~2009-08a eruption
  position determined from LT $B$-band data.  Middle left: ACS/WFC
  F814W image, M31N~2009-08b eruption position determined from LT
  $i'$-band data. Middle right: ACS/WFC F435W image, M31N~2009-08d
  eruption position determined from LT $B$-band data.  Bottom left:
  ACS/WFC F435W image, M31N~2009-10b eruption position determined from
  LT $B$-band data.  Bottom right: ACS/WFC F435W image, M31N~2009-10c
  eruption position determined from LT $B$-band
  data. (A color version of this figure is available in the online journal.)\label{progenitor-grid-3}}
\end{center}
\end{figure*}
\item {\it M31N~2009-08a.} 
The {\it HST} images used to locate the quiescent M31N~2009-08a were
taken with ACS/WFC using an F435W filter on 2007 January 10. There is
a resolvable source within $2\sigma$ of the calculated position and no
other resolvable sources within $3\sigma$. The source is 1.118 ACS/WFC
pixels, $0^{\prime\prime}\!\!.056$ or $1.40\sigma$ away from the
defined position, with the local population density suggesting there
is a 12.1\% chance of coincidence at such a separation. The source has
an F435W magnitude of $25.50\pm0.05$ and the images have a limiting
magnitude of 26.0. The {\it HST} images were only taken in one filter,
so we were unable to calculate the color of the source, although we
estimate it to have a {\it B}-band magnitude of $\sim25.5$ with
internal extinction in M31 taking in to $\sim25.2$. The position of
the system, as determined from the LT images, is shown in
Figure~\ref{progenitor-grid-3} (top right). M31N~2009-08a was also
observed in eruption with ACS/WFC on 2010 July 21 and December 14,
which is described in Section~\ref{sec:lc}. Using this post-eruption
{\it HST} data, we were able to determine the position of the
progenitor without using LT data. Here we used the same method as
previously described, but used the photometrically determined
positions for the transformations. The closest resolvable source is
0.516 ACS/WFC pixels, $0^{\prime\prime}\!\!.026$ or $2.85\sigma$ away
from the defined position of the nova, with a 2.4\% probability of a
chance alignment. Due to the small errors associated with the
transformation and the source being relatively faint, the errors on
the position of the pre-eruption source may be more significant than
in the regular transformations using LT data. The position of the
nova, as determined from the post-eruption {\it HST} data, is shown in
Figure~\ref{2009-08ahst}.
\begin{figure}
\begin{center}
\includegraphics[width=0.45\textwidth]{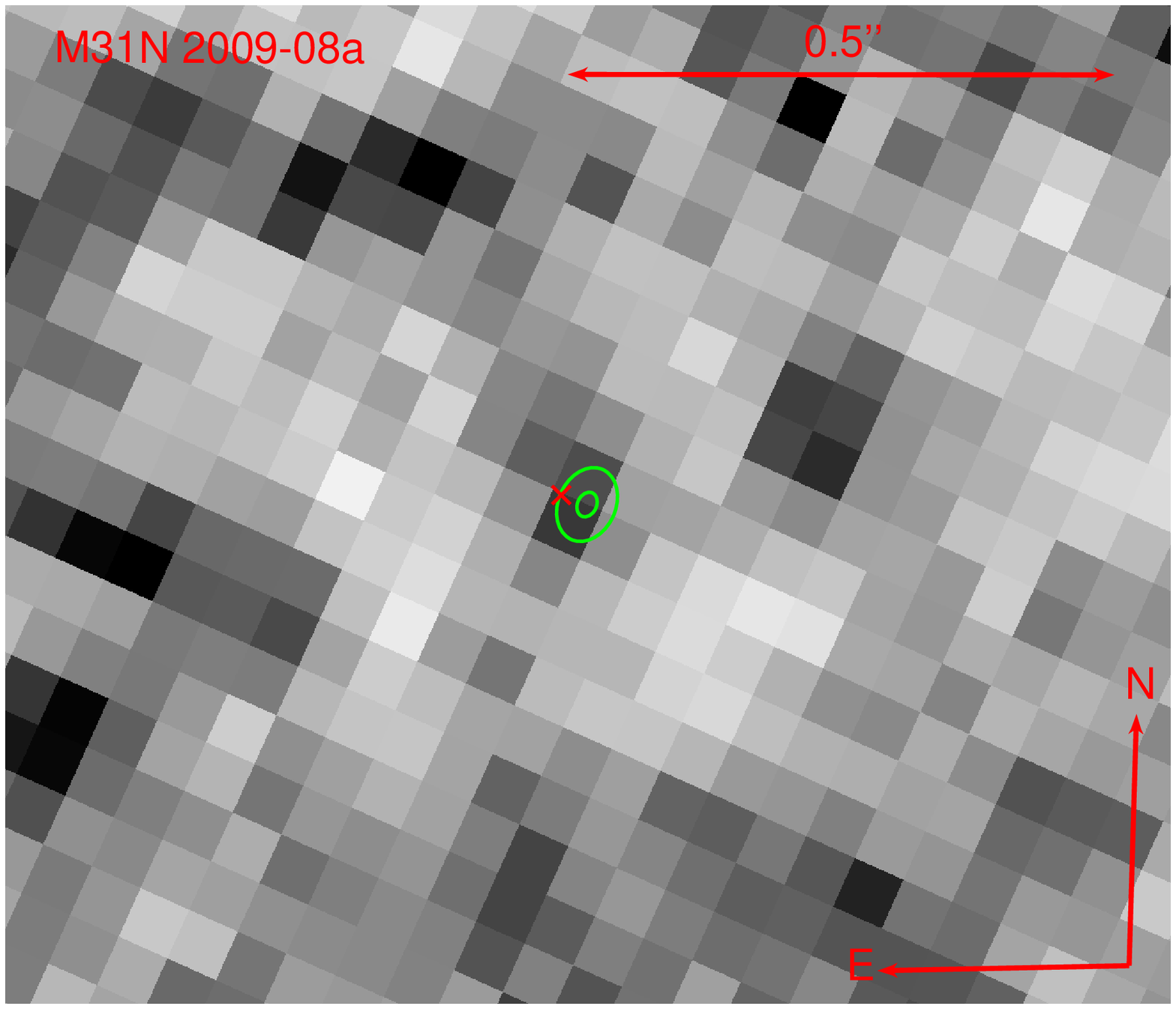}
\caption{As Figure~\ref{progenitor-grid-1}. ACS/WFC F435W FLT image
  with M31N~2009-08a eruption position determined from post-eruption
  F475W data. (A color version of this figure is available in the online journal.)\label{2009-08ahst}}
\end{center}
\end{figure}
\item {\it M31N~2009-08b.}
{\it HST} images, coincident with the position of M31N~2009-08b, were
taken with ACS/WFC using F814W and F475W filters on 2013 January
6. The closest resolvable source is 2.830 ACS/WFC pixels,
$0^{\prime\prime}\!\!.142$ or $3.25\sigma$ away from the defined
position. The local population density, which is resolved down to an
F435W magnitude of 26.6, suggests there is a 57.1\% chance of
coincidence at such a separation. Additionally, we note there is a
source visible around the $1\sigma$ area in the F475W image, but it is
too faint to determine the PSF and hence perform photometry or
accurate astrometry. The location of M31N~2009-08b is shown in
Figure~\ref{progenitor-grid-3} (middle left).
\item {\it M31N~2009-08d.}
The {\it HST} images used to locate position of the quiescent
M31N~2009-08d were taken with ACS/WFC using an F435W filter on 2004
January 23. The closest resolvable source is 3.236 ACS/WFC pixels,
$0^{\prime\prime}\!\!.163$ or $2.95\sigma$ away from the defined {\it
  HST} position. The local population density, which is resolved down
to an F435W magnitude of 25.9, suggests there is a 62.7\% chance of
coincidence at such a separation. There is also a very faint source
about $1.5\sigma$ away from the defined position, but it is too faint
to determine a PSF. The location of M31N~2009-08d is shown in
Figure~\ref{progenitor-grid-3} (middle right).
\item {\it M31N~2009-10b.} 
Nova M31N~2009-08d had coincident {\it HST} images taken with ACS/WFC
using an F435W filter on 2009 August 25. There is no resolvable source
within $3\sigma$ of the calculated position, with the closest source
being 1.803 ACS/WFC pixels, $0^{\prime\prime}\!\!.091$ or $4.30\sigma$
away from the defined position. The local population density suggests
there is a 29.1\% chance of coincidence at such a separation. The data
had an F435W limiting magnitude of 26.7. The location of M31N~2009-10b
in quiescence is shown in Figure~\ref{progenitor-grid-3} (bottom
left). It is clear from the post-eruption F475W {\it HST} images taken
on 2010 December 26 (see Section~\ref{sec:lc}) that neither of the
sources shown in the figure are the quiescent nova system.
\item {\it M31N~2009-10c.}
The {\it HST} images used to locate the position of the quiescent
M31N~2009-10c were taken with ACS/WFC using an F435W filter on 2004
January 23. There is one resolvable source within $3\sigma$ of the
calculated position. The source is 2.128 ACS/WFC pixels,
$0^{\prime\prime}\!\!.107$ or $2.30\sigma$ away from the defined
position and the local population density, which is resolved down to
an F435W magnitude of 25.9, suggests there is a 28.7\% chance of
coincidence at such a separation. The location of this system is shown
in Figure~\ref{progenitor-grid-3} (bottom right). It is clear from the
post-eruption F475W {\it HST} images taken on 2010 December 22 (see
Section~\ref{sec:lc}) that the source listed above is not the
quiescent nova.
\item {\it M31N~2009-11a.}
{\it HST} images, coincident with the position of M31N~2009-11a, were
taken with ACS/WFC using F814W and F555W filters on 2004 August
15. There are no resolvable sources within $3\sigma$ of the calculated
position, with the closest resolvable source being 3.999 ACS/WFC
pixels, $0^{\prime\prime}\!\!.201$ or $3.45\sigma$ away from the
defined position. The local population density, which is resolved to
an F814W magnitude of 26.5, suggests there is a 68.3\% chance of
coincidence at such a separation. The location of this system is shown
in Figure~\ref{progenitor-grid-4} (top left).
\begin{figure*}
\begin{center}
\includegraphics[width=0.45\textwidth]{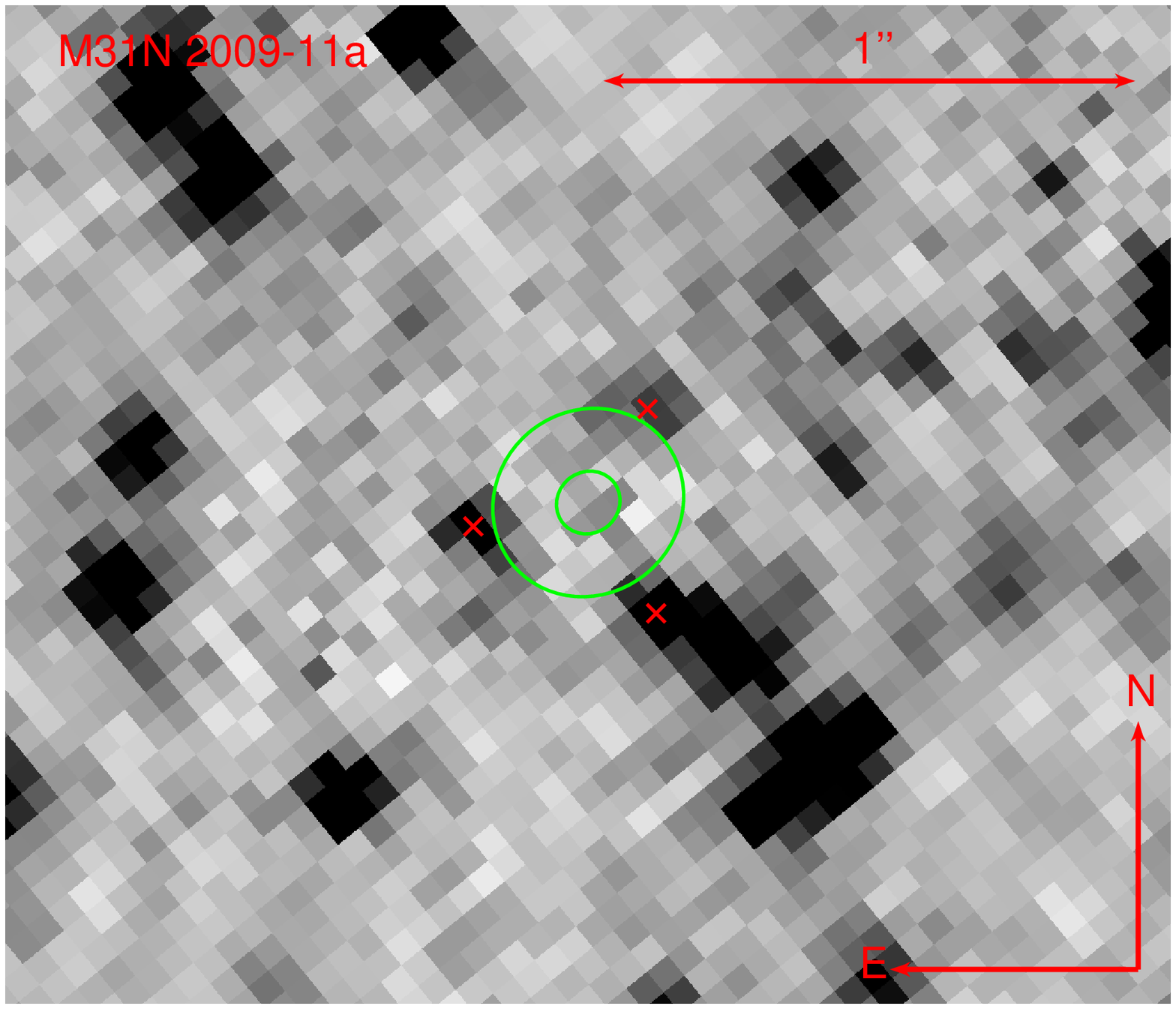}
\includegraphics[width=0.45\textwidth]{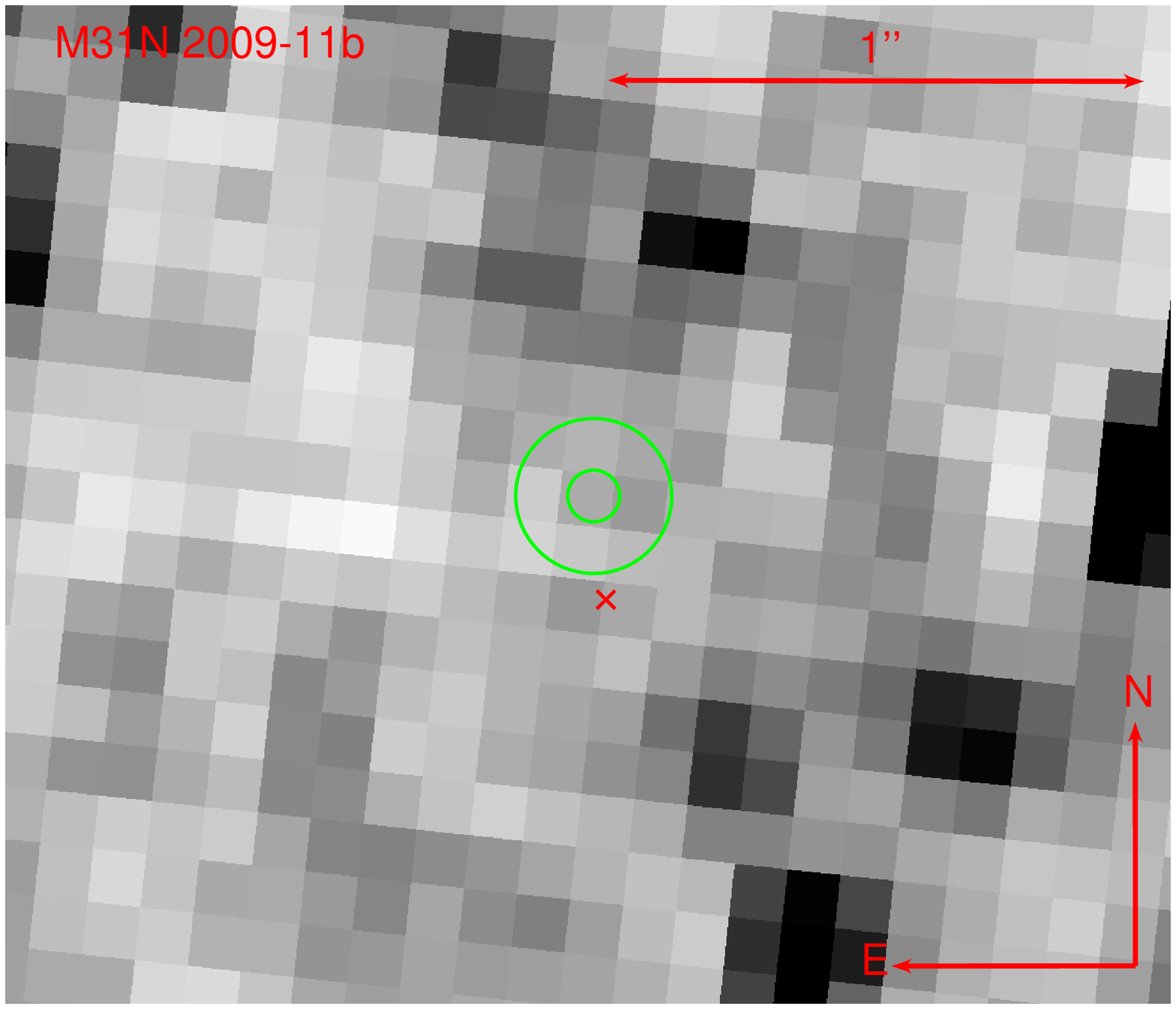}\\
\includegraphics[width=0.45\textwidth]{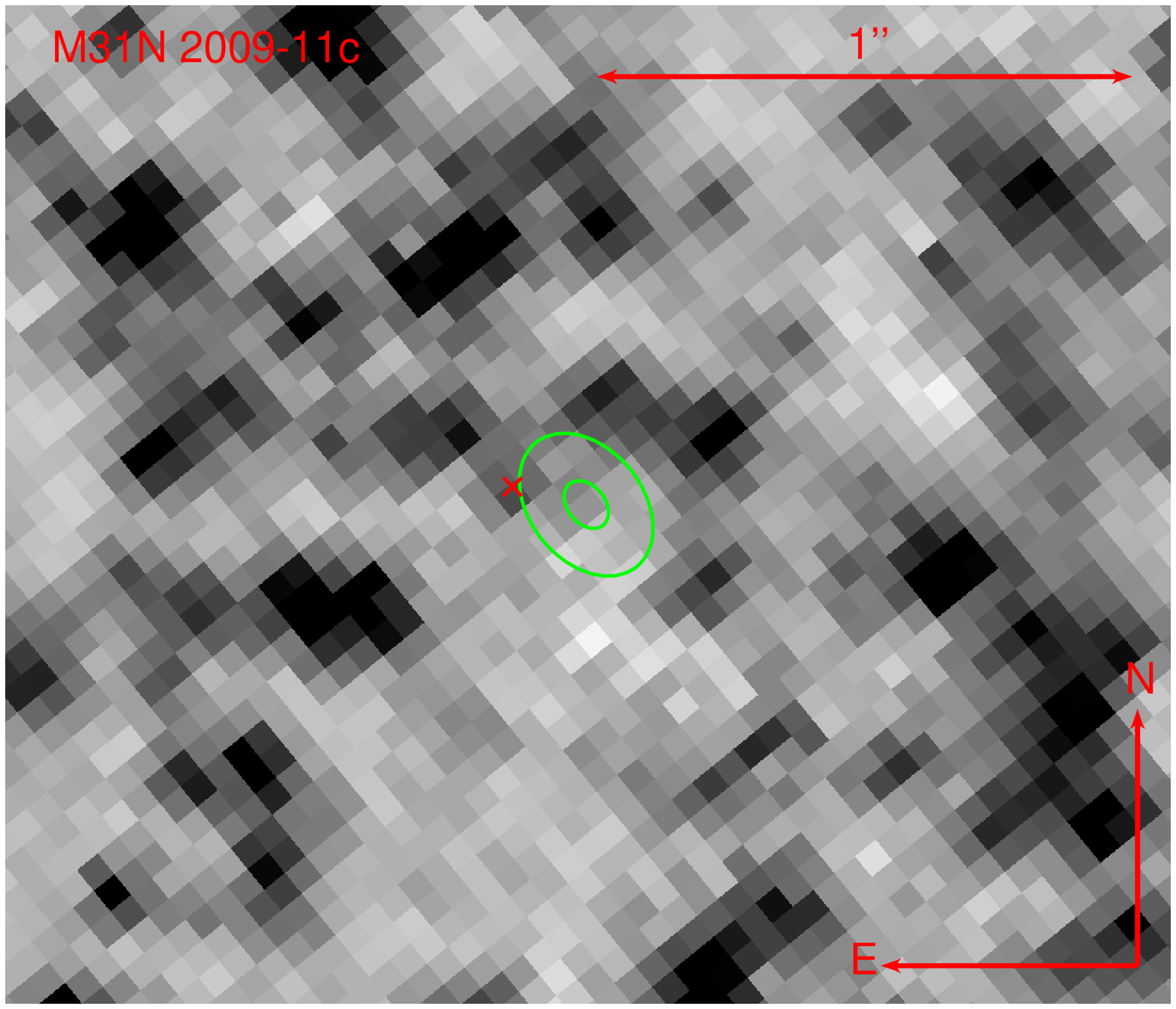}
\includegraphics[width=0.45\textwidth]{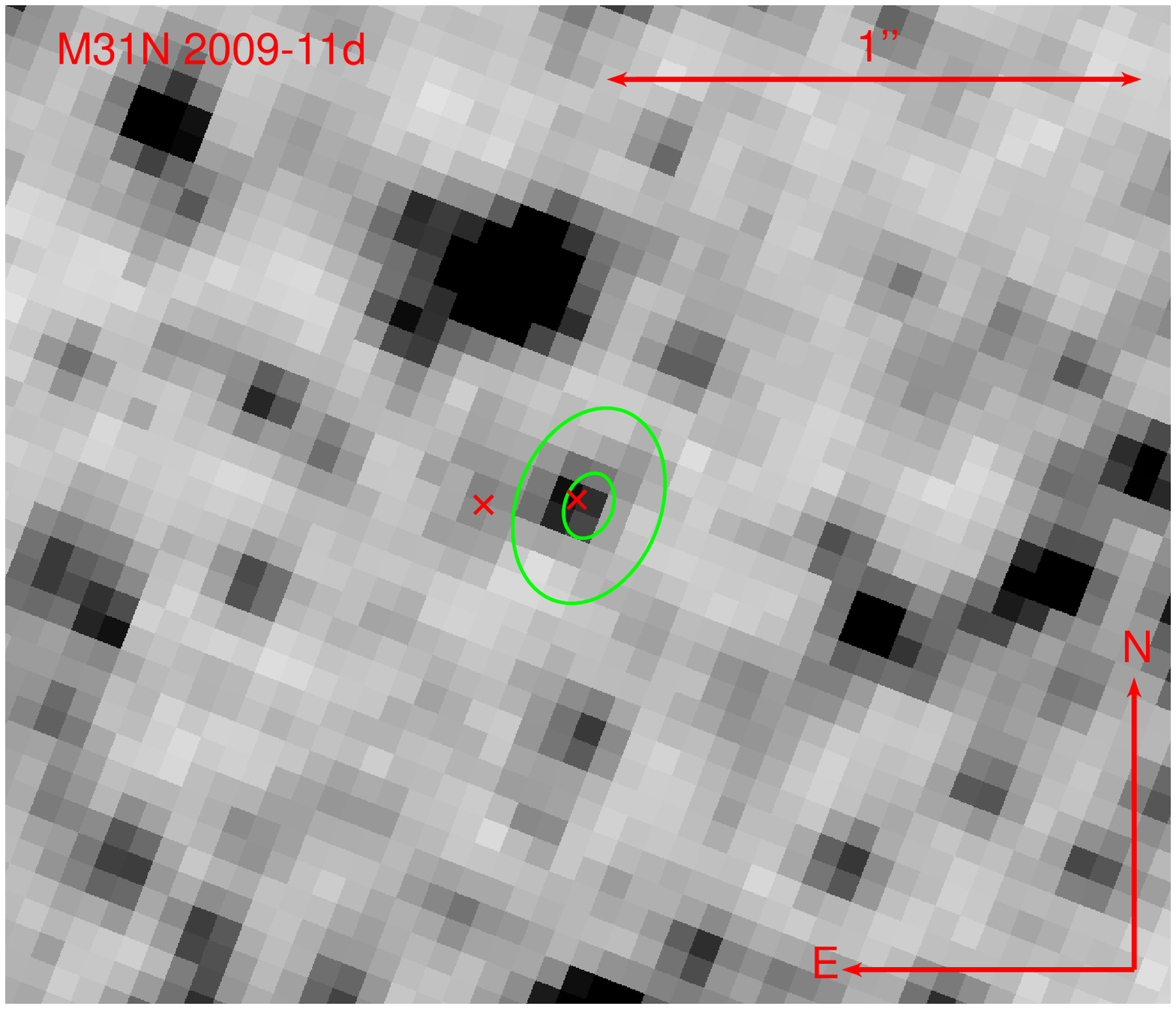}\\
\includegraphics[width=0.45\textwidth]{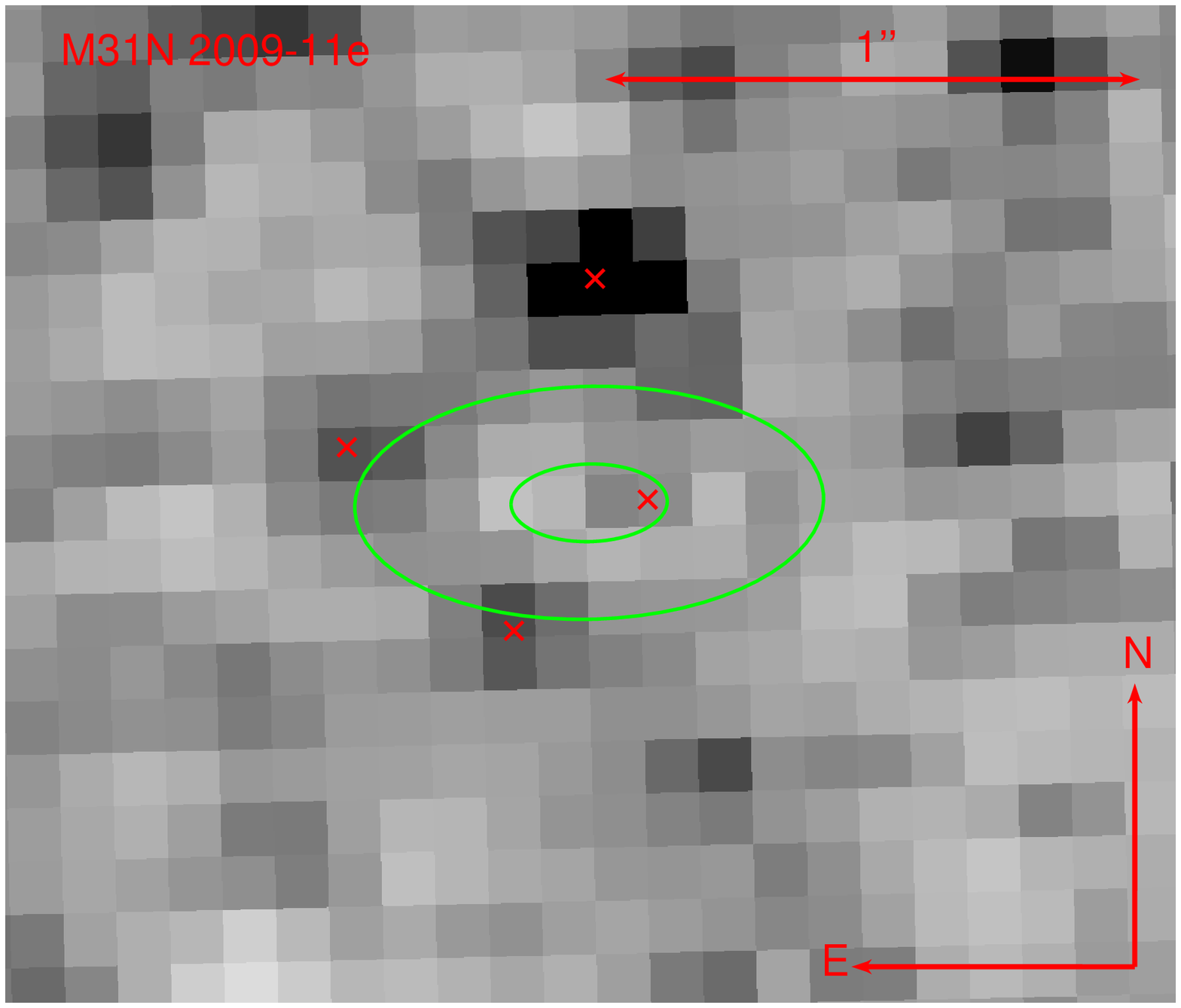}
\includegraphics[width=0.45\textwidth]{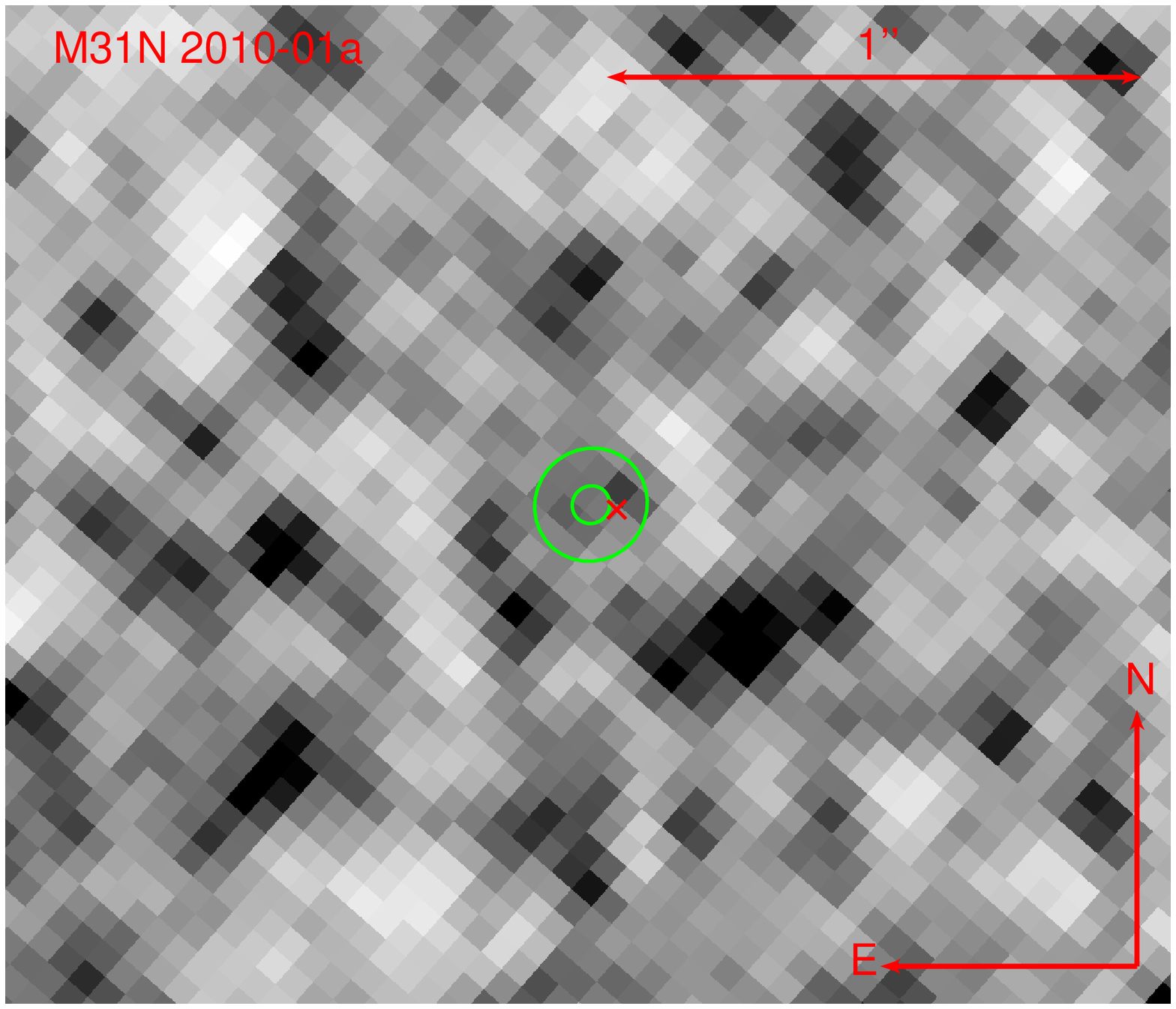}
\caption{As Figure~\ref{progenitor-grid-1}.  Top left: ACS/WFC F555W
  image, M31N~2009-11a eruption position determined from LT $V$-band
  data. Top right: WFPC2 F555W image, M31N~2009-11b eruption position
  determined from LT $V$-band data.  Middle left: ACS/WFC F555W image,
  M31N~2009-11c eruption position determined from LT $V$-band data.
  Middle right: ACS/WFC F475W image, M31N~2009-11d eruption position
  determined from LT $V$-band data.  Bottom left: WFPC2 F555W image,
  M31N~2009-11e eruption position determined from LT $V$-band data.
  Bottom right: ACS/WFC F435W image, M31N~2010-01a eruption position
  determined from LT $B$-band data. (A color version of this figure is available in the online journal.)\label{progenitor-grid-4}}
\end{center}
\end{figure*}
\item {\it M31N~2009-11b.}
Nova M31N~2009-11b is thought to be recurrent, with previous eruptions
having occurred in 1997 and 2001 (Shafter et~al.\ in preparation). The
location of M31N~2009-11b was observed by {\it HST} with WFPC2 using
F814W and F555W filters on 2004 August 22. There are no resolvable
sources within $3\sigma$ of the calculated position, with the closest
resolvable source being 1.971 WFPC2 pixels, $0^{\prime\prime}\!\!.197$
or $4.15\sigma$ away from the defined position. The local population
density suggests there is a 38.6\% probability of an object alignment
this close occurring by chance, although the limiting F814W magnitude
is only 23.8. If we were looking at a typical ACS/WFC image of M31, we
would expect an object at a distance of $0^{\prime\prime}\!\!.197$ to
have approximately an 81\% chance of being aligned by chance. The
location of M31N~2009-11b in the WFPC2 image is shown in
Figure~\ref{progenitor-grid-4} (top right).
\item {\it M31N~2009-11c.} 
The {\it HST} images used to locate M31N~2009-11c in quiescence were
taken with ACS/WFC using F814W and F555W filters on 2004 August
22. There are no resolvable sources within $3\sigma$ of the calculated
position. The closest resolvable source is 2.975 ACS/WFC pixels,
$0^{\prime\prime}\!\!.150$ or $3.35\sigma$ away from the defined
position, with the local population density, which is resolved to an
F814W magnitude of 25.8, suggesting there is a 55.0\% chance of
coincidence at such a separation. The location of this system is shown
in Figure~\ref{progenitor-grid-4} (middle left). It is clear from the
post-eruption F475W {\it HST} images taken on 2010 July 23 (see
Section~\ref{sec:lc}) that the source listed above is not the nova.
\item {\it M31N~2009-11d.}
Nova M31N~2009-11d had coincident {\it HST} images taken with ACS/WFC
using F475W and F814W filters on 2011 December 22. There is a
resolvable source within $1\sigma$ of the calculated position and no
other source within $3\sigma$. The source is 0.505 ACS/WFC pixels,
$0^{\prime\prime}\!\!.025$ or $0.60\sigma$ away from the defined
position. The local population density suggests there is only a 2.2\%
chance of an object alignment this close occurring by chance. The data
has an F814W limiting magnitude of 26.4. On 2011 December 22, the
source had an F814W magnitude of $25.1\pm0.2$ and F475W magnitude of
$25.67\pm0.04$. This gives a $B$-band magnitude of $25.81\pm0.06$,
$I$-band magnitude of $25.08\pm0.18$ and {\it (B-I)} color of
$0.7\pm0.2$. From \citet{2005PhDT.........2D} we can see that a nova
at the apparent position of M31N~2009-11d would be subject to
$r'$-band extinction of $A_{r'}=0.58$ if it were at the far side of
the galaxy. This gives a {\it B}-band magnitude of $25.4\pm0.4$, {\it
  I}-band magnitude of $24.9\pm0.3$ and ($B-I$) color of $0.5\pm0.3$
when the effects of internal extinction are included. M31N~2009-11d is
shown in quiescence in Figure~\ref{progenitor-grid-4} (middle right).
\item {\it M31N~2009-11e.} 
The location of M31N~2009-11e was observed by {\it HST} with WFPC2
using F814W and F555W filters on 1996 February 14. There is a
resolvable source within $1\sigma$ of the calculated position and no
other source within $3\sigma$. The source is 1.104 WFPC2 pixels,
$0^{\prime\prime}\!\!.110$ or $0.80\sigma$ away from the defined
position. The local population density, which is only resolvable to an
F814W magnitude of 22.8, suggests there is a 9.9\% chance of an object
alignment this close occurring by chance.  If a standard ACS/WFC image
of the M31 field had been available, we would predict an object to be
at least $0^{\prime\prime}\!\!.110$ from a random point in the image
approximately 33\% of the time. The location of this system is shown
in Figure~\ref{progenitor-grid-4} (bottom left). It is clear from the
post-eruption F475W {\it HST} images taken on 2010 December 25 (see
Section~\ref{sec:lc}) that the source listed above is not the nova.
\item {\it M31N~2010-01a.}
The {\it HST} images used to locate the quiescent nova M31N~2010-01a
were taken with ACS/WFC using an F435W filter on 2006 February
10. There is a resolvable source within $2\sigma$ of the calculated
position and no other source within $3\sigma$. The source is 1.058
ACS/WFC pixels, $0^{\prime\prime}\!\!.053$ or $1.45\sigma$ away from
the defined position. The local stellar density, which is resolved
down to an F435W magnitude of 26.1, suggests there is a 9.4\% chance
of an object alignment this close occurring by chance. The source had
an F435W magnitude of $24.79\pm0.04$, but the {\it HST} images were
only taken in one filter, so we were unable to calculate the color of
the source, although we estimate it to have had a {\it B}-band
magnitude of $\sim24.8$ with internal extinction in M31 taking in to
$\sim24.5$. The location of this nova is shown in
Figure~\ref{progenitor-grid-4} (bottom right). M31N~2010-01a was also
observed in eruption with ACS/WFC on 2010 July 21, which is described
in Section~\ref{sec:lc}. Using this post-eruption {\it HST} data, we
were able to determine the position of the progenitor without using LT
data. Here we used the same method as previously described, but used
the photometrically determined positions for the transformations. The
closest resolvable source is 0.641 ACS/WFC pixels,
$0^{\prime\prime}\!\!.032$ or 2.65$\sigma$ away from the defined
position of the nova, with a 3.8\% probability of a chance alignment
at this separation. As previously noted, due to the small errors
associated with the transformation and the source being relatively
faint, the errors on the position of the pre-eruption source may be
more significant than in the regular transformations using LT
data. The position of the quiescent nova, as determined from the
post-eruption {\it HST} data, is shown in
Figure~\ref{2010-01ahst}. Later in 2010 there was a nova eruption
(M31N~2010-12c) very close to the position of M31N~2010-01a, however
the precise positions reveal that they are not the same system
(\citealp{2010CBET.2610....1H}; Shafter et~al.\ in preparation).
\begin{figure}
\begin{center}
\includegraphics[width=0.45\textwidth]{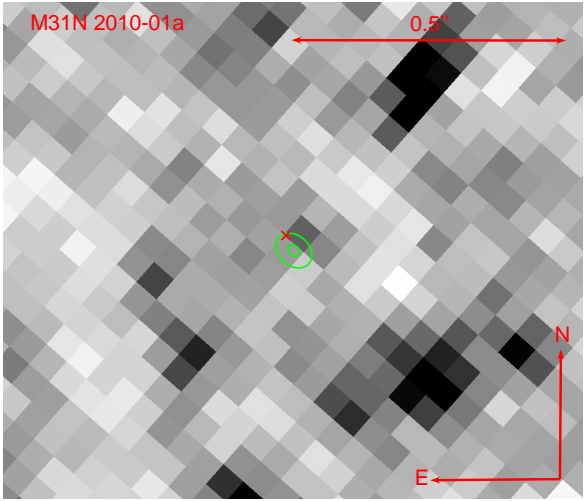}
\caption{As Figure~\ref{progenitor-grid-1}. ACS/WFC F435W FLT image
  with M31N~2010-01a eruption position determined from post-eruption
  F475W data. (A color version of this figure is available in the online journal.)\label{2010-01ahst}}
\end{center}
\end{figure}
\item {\it M31N~2010-05a.}
The location of M31N~2010-05a was observed by {\it HST} with ACS/WFC
using an F435W filter on 2004 January 23. There is one resolvable
source within $3\sigma$ of the calculated position. This source is
1.939 ACS/WFC pixels, $0^{\prime\prime}\!\!.098$ or $2.10\sigma$ away
from the defined position. The local population density suggests there
is a 36.9\% probability of an object alignment this close occurring by
chance. The data have a limiting F435W magnitude of 25.8. The location
of this nova is shown in Figure~\ref{progenitor-grid-5} (top left).
\begin{figure*}
\begin{center}
\includegraphics[width=0.45\textwidth]{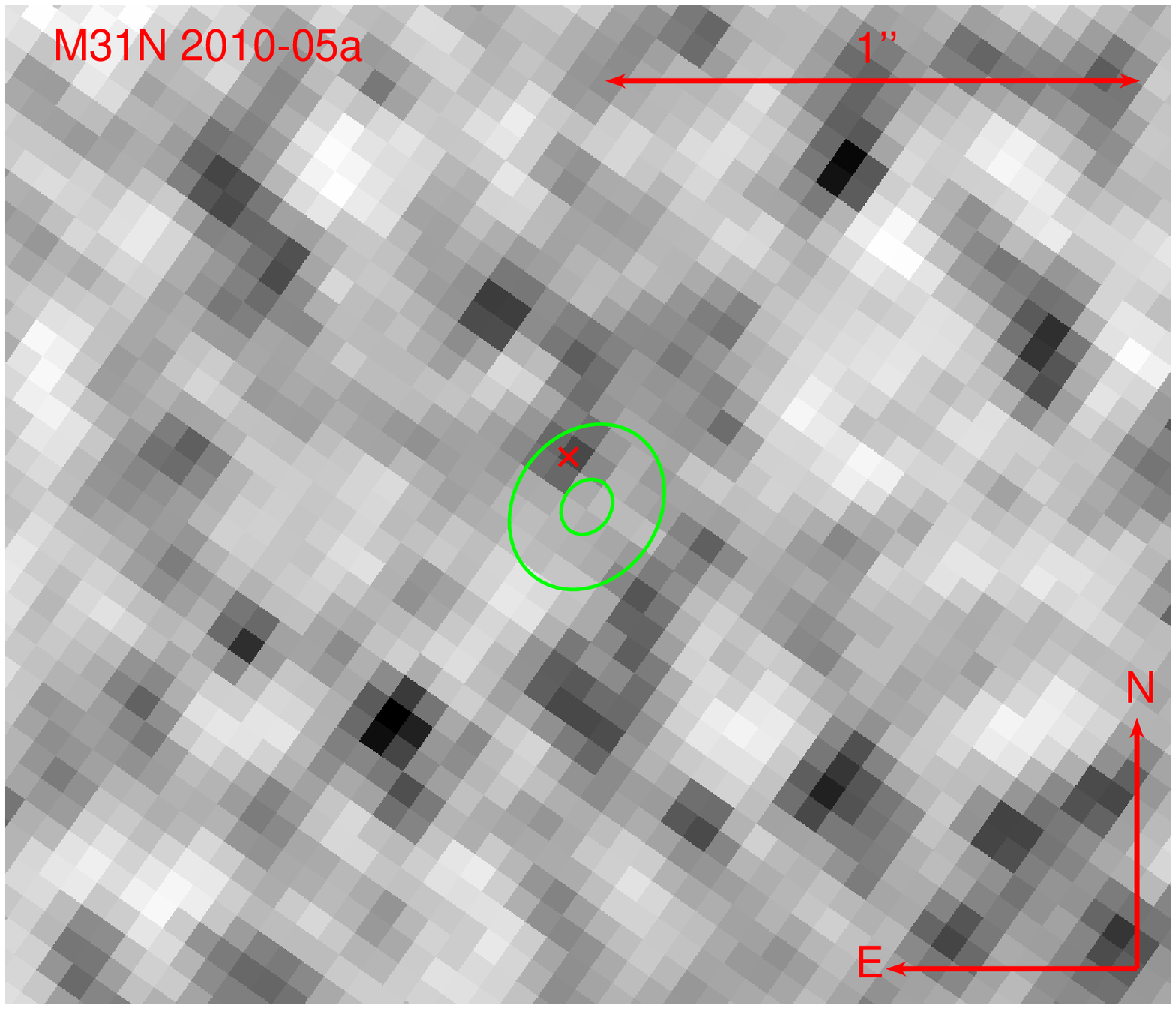}
\includegraphics[width=0.45\textwidth]{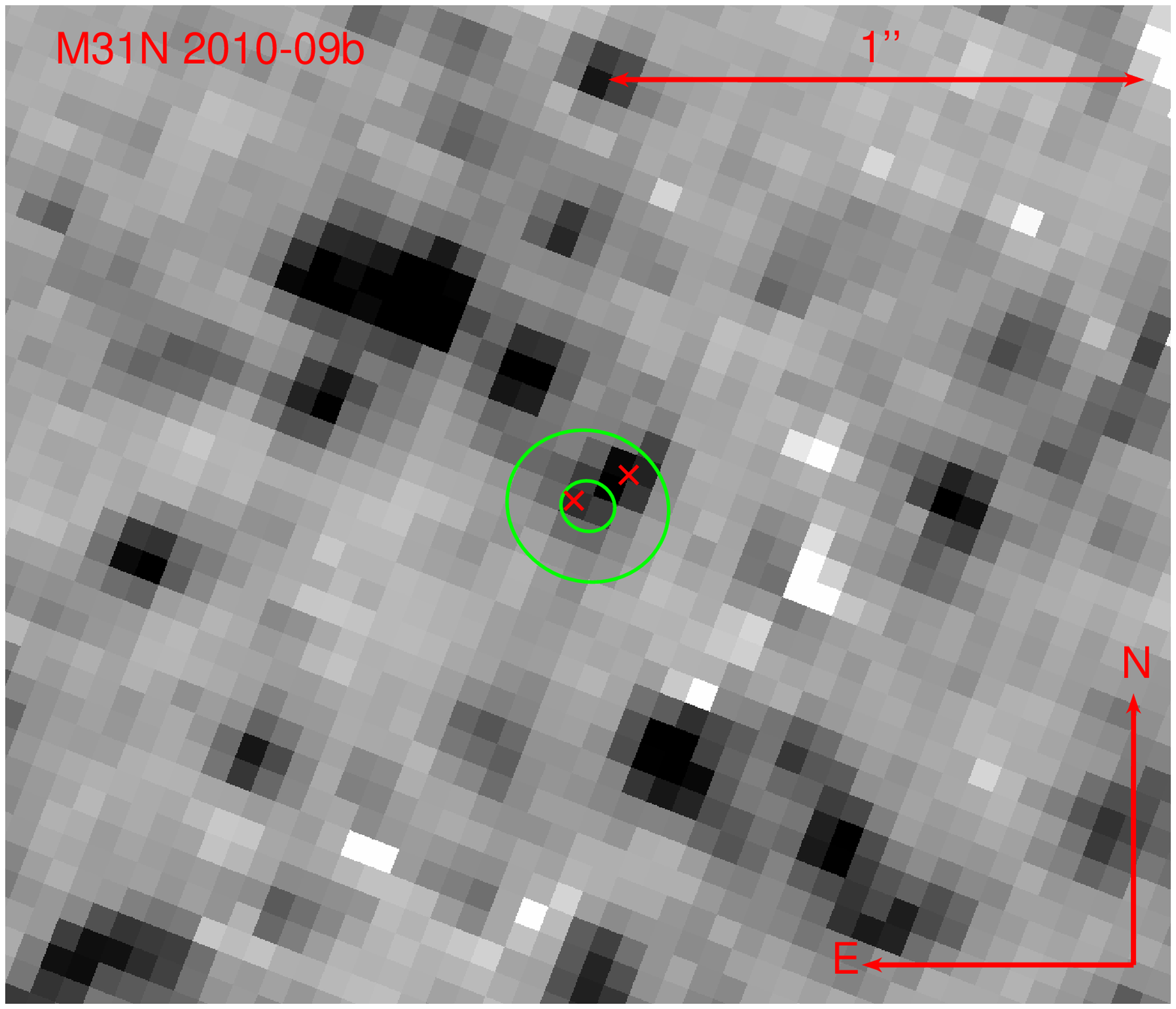}\\
\includegraphics[width=0.45\textwidth]{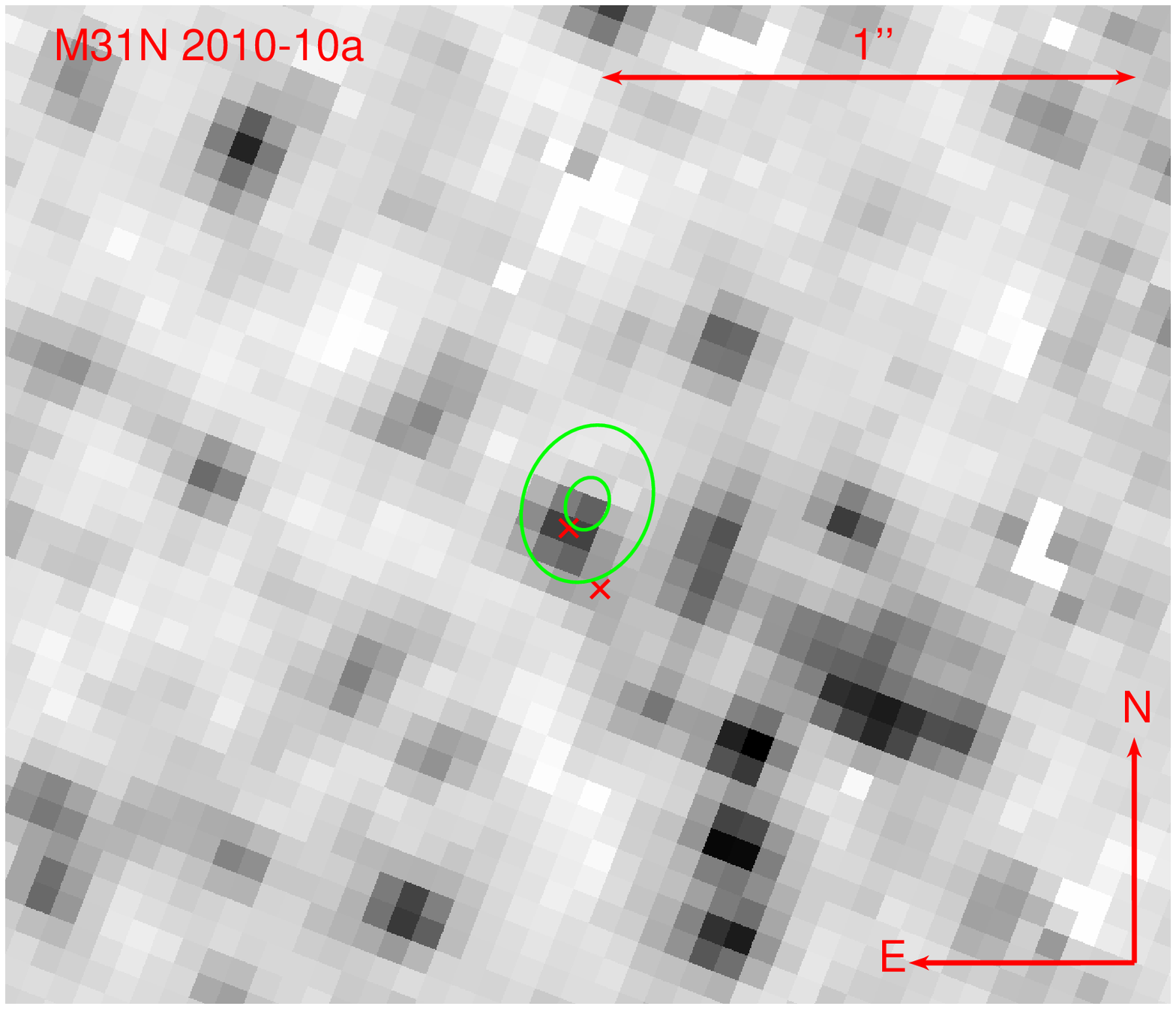}
\includegraphics[width=0.45\textwidth]{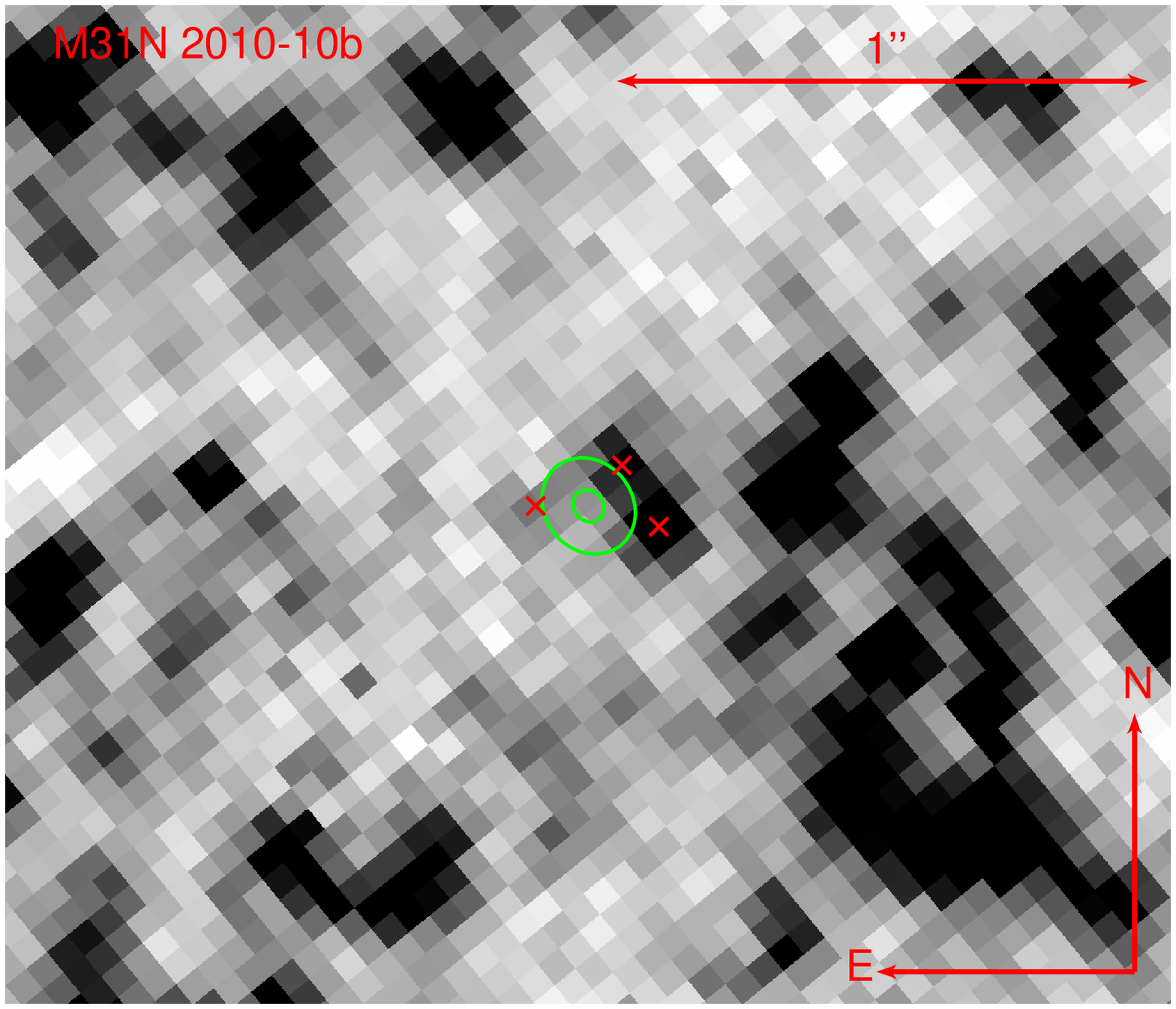}\\
\includegraphics[width=0.45\textwidth]{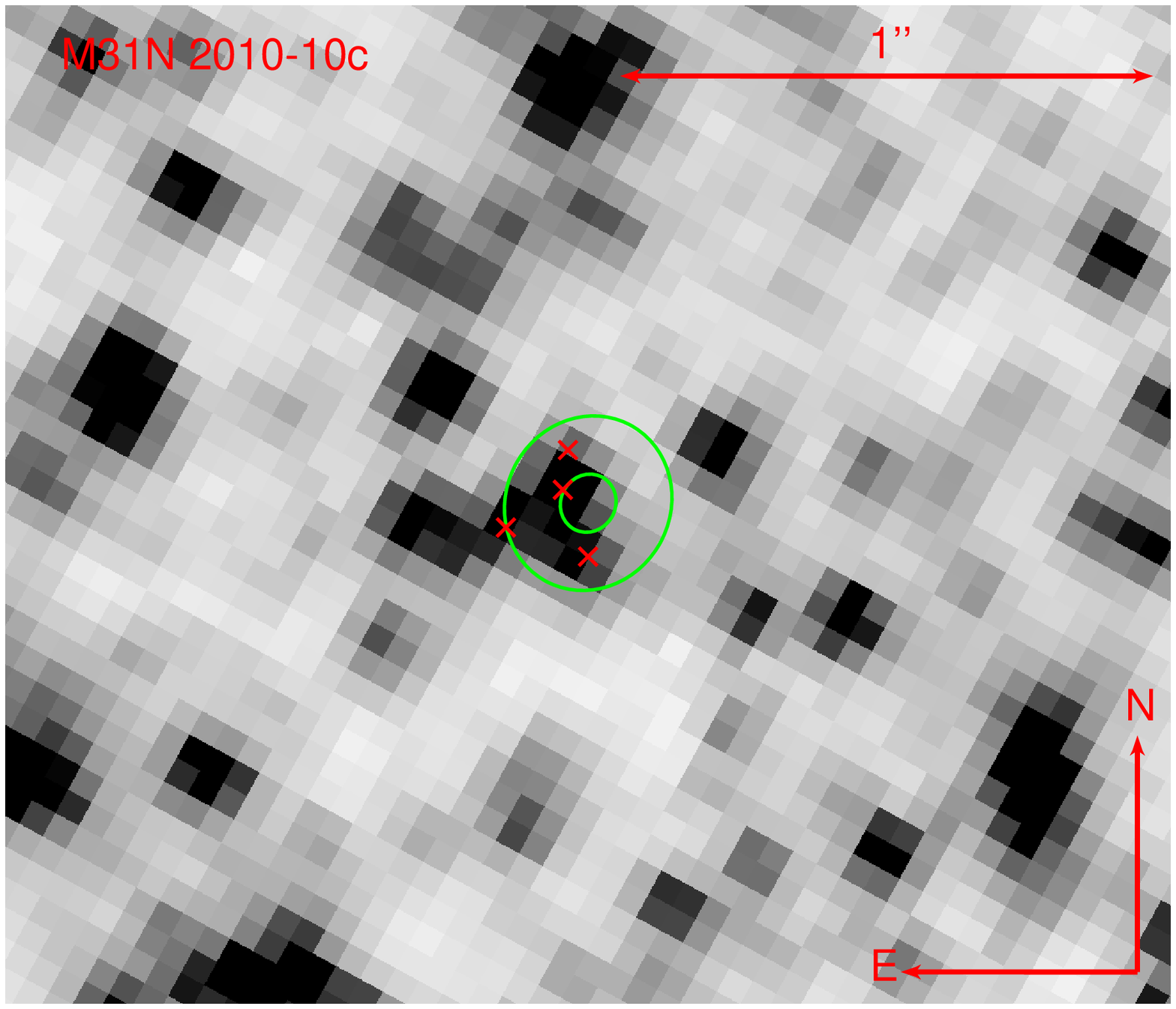}
\includegraphics[width=0.45\textwidth]{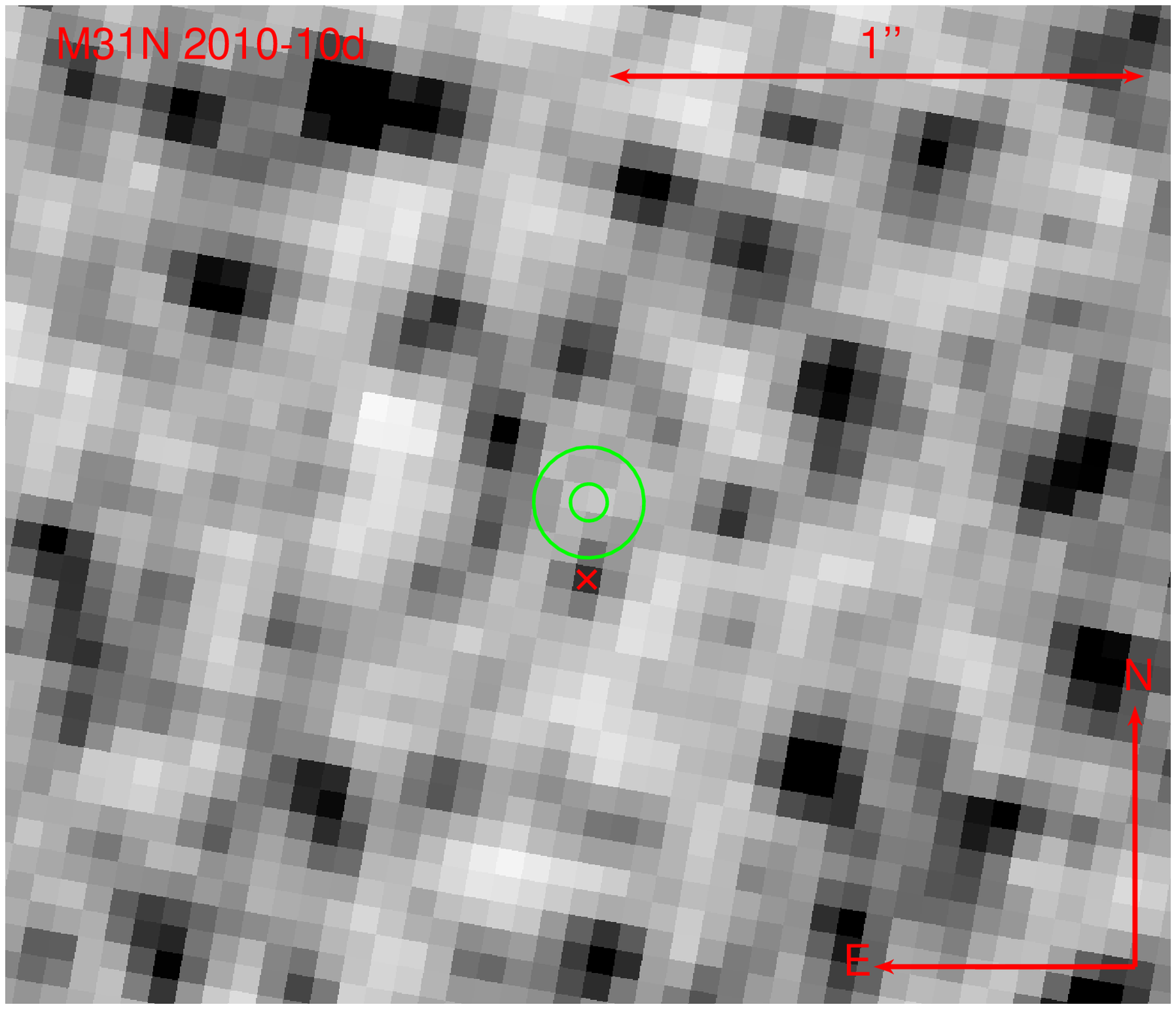}
\caption{As Figure~\ref{progenitor-grid-1}.  Top left: ACS/WFC F435W
  image, M31N~2010-05a eruption position determined from LT $B$-band
  data. Top right:  ACS/WFC F475W image, M31N~2010-09b eruption
  position determined from LT $B$-band data. Middle left: ACS/WFC
  F475W image, M31N~2010-10a eruption position determined from LT
  $B$-band data. Middle right: ACS/WFC F555W image, M31N~2010-10b
  eruption position determined from LT $V$-band data.  Bottom left:
  ACS/WFC F606W image, M31N~2010-10c eruption position determined from
  LT $V$-band data.  Bottom right: ACS/WFC F435W image, M31N~2010-10d
  eruption position determined from LT $B$-band
  data. (A color version of this figure is available in the online journal.)\label{progenitor-grid-5}}
\end{center}
\end{figure*}
\item {\it M31N~2010-09b.}
The {\it HST} images used to locate M31N~2010-09b in quiescence were
taken with ACS/WFC using F814W and F475W filters on 2011 December
9. There is a resolvable source within $1\sigma$ of the calculated
position and another resolvable source within $2\sigma$. The closest
source is 0.579 ACS/WFC pixels, $0^{\prime\prime}\!\!.029$ or
$0.55\sigma$ away from the defined position. The local stellar
density, which is resolvable down to an F814W magnitude of 26.3,
suggests there is a $2.6\%$ chance of coincidence at such a
separation. This source had an F814W magnitude of $24.7\pm0.1$ and
F475W magnitude of $26.30\pm0.07$. This gives a $B$-band magnitude of
$27.1\pm0.1$, $I$-band magnitude of $24.7\pm0.1$ and ($B-I$) color of
$2.4\pm0.1$. We can see from \citet{2005PhDT.........2D} that a nova
at the apparent position of M31N~2010-09b would be subject to
$r'$-band extinction of $A_{r'}=0.63$ if it were at the far side of
the galaxy. Therefore we calculated an unextinguished {\it B}-band
magnitude of $26.6\pm0.5$ and {\it I}-band magnitude of $24.5\pm0.2$,
and a de-reddened ($B-I$) color of $2.2\pm0.3$. The other source is
2.264 ACS/WFC pixels, $0^{\prime\prime}\!\!.114$ or $2.05\sigma$ away from the defined
position. The local population density suggests there is a $36.3\%$
chance of a resolvable source being within this distance of a random
object in the image. The location of M31N~2010-09b is shown in
Figure~\ref{progenitor-grid-5} (top right).
\item {\it M31N~2010-10a.}
{\it HST} observed the location of M31N~2010-10a with ACS/WFC using
F475W and F814W filters on 2012 December 15. Although the {\it HST}
data were taken more than two years post-eruption, the nova has a
maximum $B$-band $t_{2}$ of only $16\pm2$~days (see
Section~\ref{sec:lc}), so will have faded back to quiescence. There
is a resolvable source just outside 1$\sigma$ of the calculated
position and no other resolvable source within 3$\sigma$. The source
is 1.181 ACS/WFC pixels, $0^{\prime\prime}\!\!.059$ or 1.20$\sigma$
away from the defined position. The local population density suggests
there is a $11.2\%$ chance of coincidence at such a separation. The
location around this nova system is shown in
Figure~\ref{progenitor-grid-5} (middle left).
\item {\it M31N~2010-10b.}
The {\it HST} images used to locate the position of the quiescent
M31N~2010-10b were taken with ACS/WFC using F555W and F814W filters on
2004 August 14. There are no resolvable sources within 3$\sigma$ of
the calculated position, the closest resolvable source being 1.968
ACS/WFC pixels, $0^{\prime\prime}\!\!.099$ or $3.60\sigma$ away from
the defined position. The local population density suggests there is a
33.4\% probability of such an alignment occurring by chance. The
location of M31N~2010-10b is shown in Figure~\ref{progenitor-grid-5}
(middle right).
\item {\it M31N~2010-10c.}
The location of M31N~2010-10c was observed by {\it HST} with ACS/WFC
using F606W and F435W filters on 2005 July 22. There are three
resolvable sources within $2\sigma$ of the calculated position and
another resolvable source within $3\sigma$. The closest resolvable
source is 1.083 ACS/WFC pixels, $0^{\prime\prime}\!\!.055$ or
$1.05\sigma$ away from the defined position. The local population
density, which is resolvable down to an F606W magnitude of 27.1,
suggests there is a $12.4\%$ probability of such an alignment
occurring by chance. The second closest resolvable source is 1.988
ACS/WFC pixels, $0^{\prime\prime}\!\!.100$ or $1.90\sigma$ away from
the defined position. The local population density suggests there is a
$38.5\%$ probability of chance alignment and the third closest source
is 2.084 ACS/WFC pixels, $0^{\prime\prime}\!\!.105$ or $2.00\sigma$
away from the defined position, with a $41.6\%$ probability of chance
alignment. The fourth closest resolvable source is 3.405 ACS/WFC
pixels, $0^{\prime\prime}\!\!.171$ or $3.00\sigma$ away from the
defined position, with a $78.6\%$ probability of chance alignment. The
location of M31N~2010-10c, with the four closest progenitor
candidates, is shown in Figure~\ref{progenitor-grid-5} (bottom left).
\item {\it M31N~2010-10d.}
The {\it HST} images used to locate M31N~2010-10d in quiescence were
taken with ACS/WFC using an F435W filter on 2004 June 14. There are no
resolvable sources within $3\sigma$ of the calculated position, with
the closest resolvable source being 2.992 ACS/WFC pixels,
$0^{\prime\prime}\!\!.151$ or $4.25\sigma$ away from the defined
position. The local population density, which is resolved down to an
F435W magnitude of 26.3, suggests that the coincidence probability at
this separation is $65.6\%$. The location of this system is shown in
Figure~\ref{progenitor-grid-5} (bottom right). It can clearly be seen
from the {\it HST} data taken during eruption (see
Section~\ref{sec:lc}) that the above source is not the progenitor of
M31N~2010-10d.
\item {\it M31N~2010-10e.}
Nova M31N~2010-10e has been suggested as another eruption of
M31N~1963-09c (\citealp{2010ATel.3001....1P,2010ATel.3006....1S};
Shafter et~al.\ in preparation). The {\it HST} images were taken with
ACS/WFC using F814W and F555W filters on 2004 August 17. There is a
resolvable source within $1\sigma$ of the calculated position, one
within $2\sigma$ and another within $3\sigma$. The closest resolvable
source is 1.216 ACS/WFC pixels, $0^{\prime\prime}\!\!.061$ or
$0.80\sigma$ away from the defined position. The local population
density suggests there is a 15.6\% probability of such an alignment
occurring by chance. The next closest source is 1.859 ACS/WFC pixels,
$0^{\prime\prime}\!\!.094$ or $1.45\sigma$ away from the defined
position, with a 34.4\% probability of chance alignment.  The third
closest source is 4.310 ACS/WFC pixels, $0^{\prime\prime}\!\!.217$ or
$2.75\sigma$ away from the defined position, with a 92.3\% probability
of chance alignment. The {\it HST} data are resolvable down to an
F814W magnitude of 25.9. The position of M31N~2010-10e is shown in
Figure~\ref{progenitor-grid-6} (top left). The relatively large errors
on the position of the quiescent nova were caused by the nova being
very faint in the LT images. This is simply because M31N~2010-10e was
a fast nova and the first LT images were taken several days after
discovery.
\begin{figure*}
\begin{center}
\includegraphics[width=0.45\textwidth]{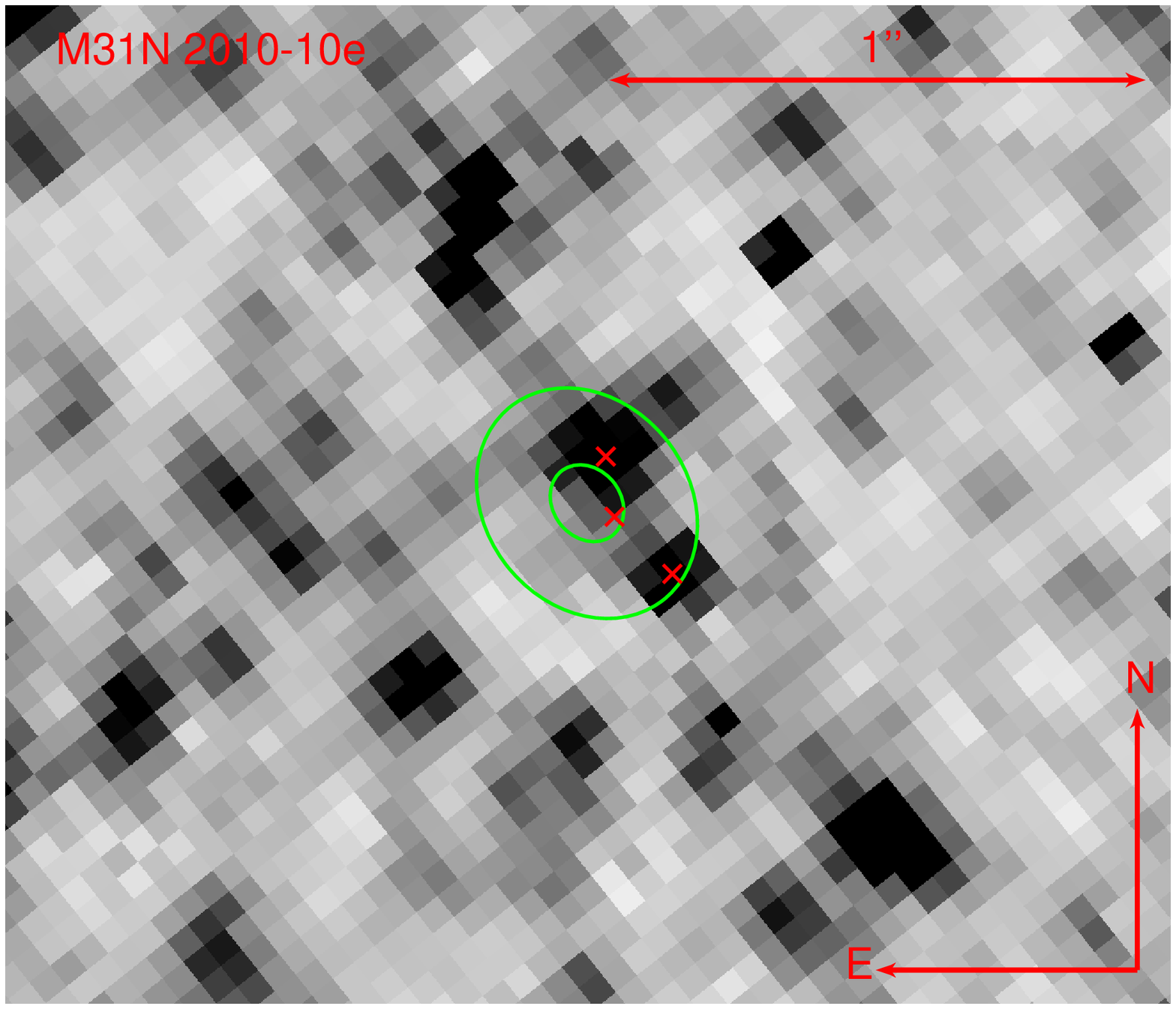}
\includegraphics[width=0.45\textwidth]{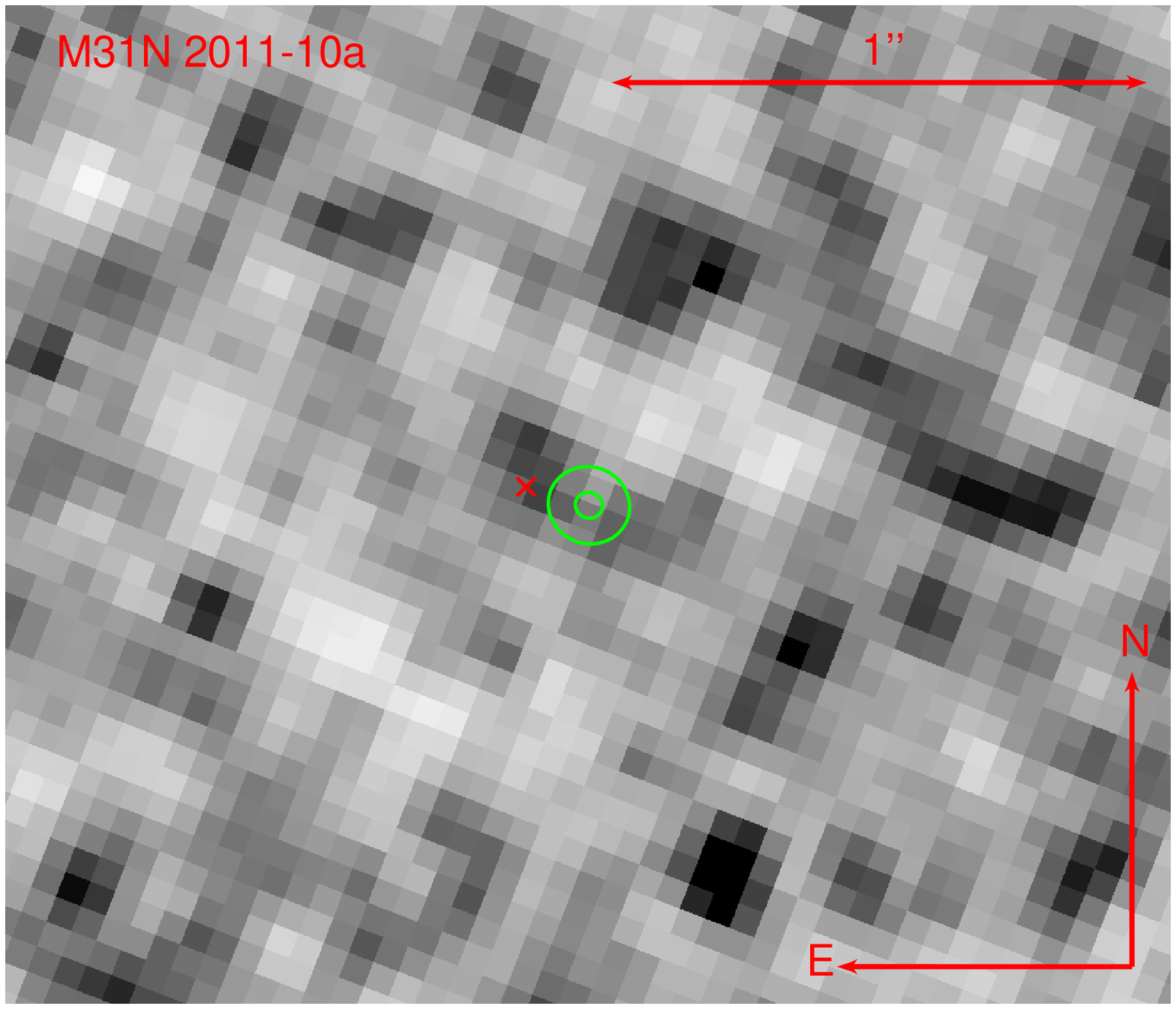}\\
\includegraphics[width=0.45\textwidth]{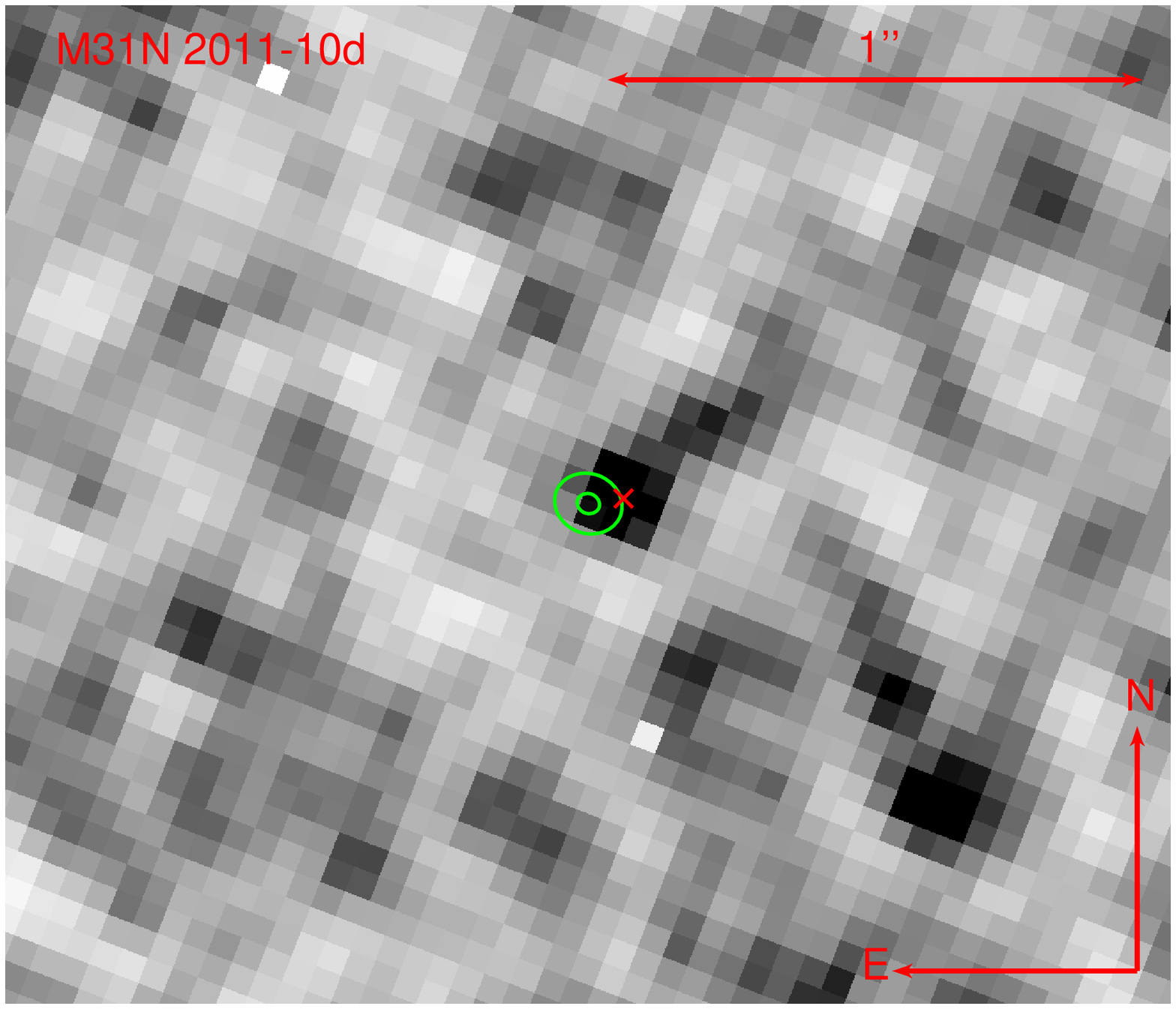}
\includegraphics[width=0.45\textwidth]{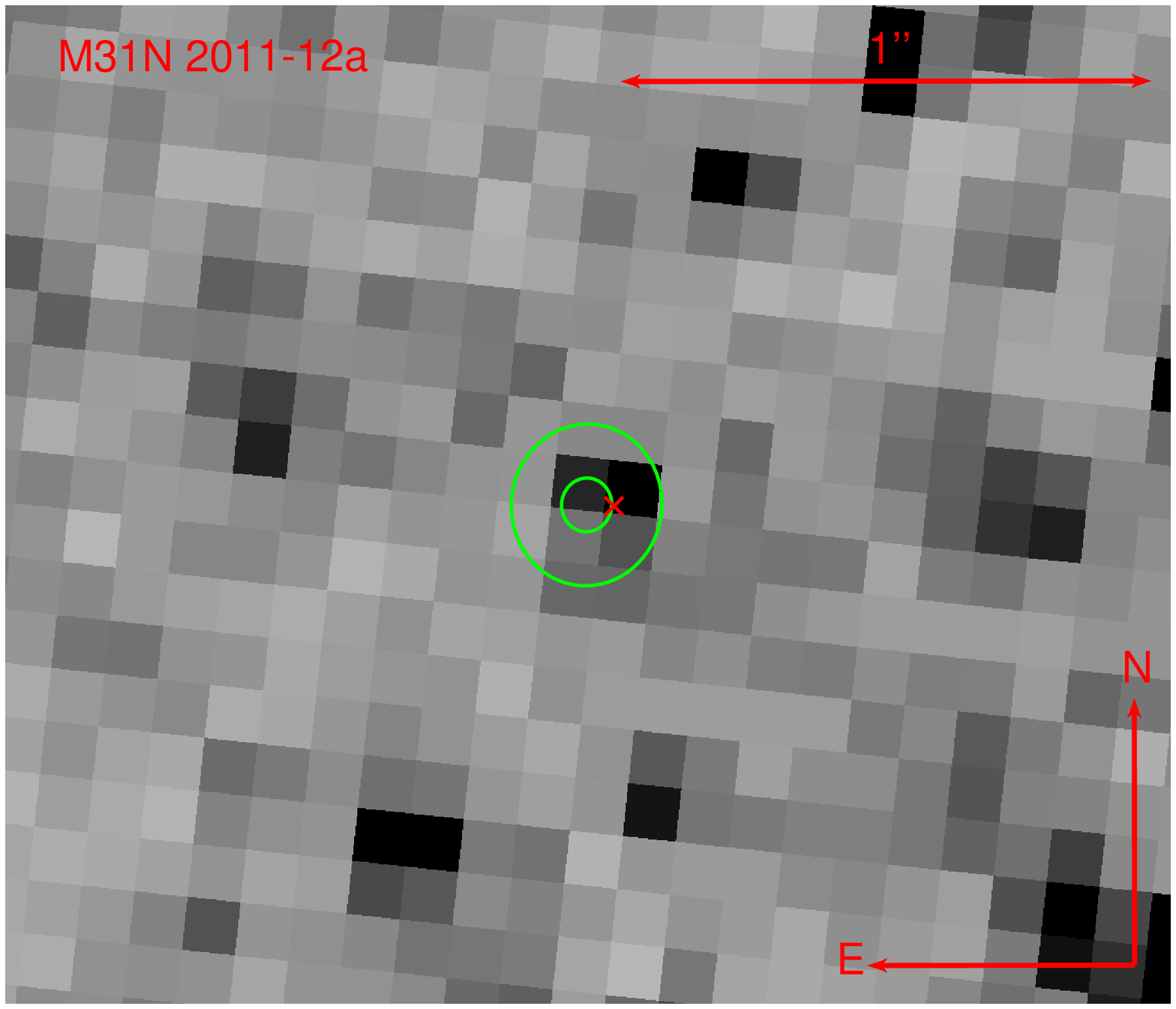}
\caption{As Figure~\ref{progenitor-grid-1}.  Top left: ACS/WFC F555W
  image, M31N~2010-10e eruption position determined from LT $V$-band
  data.  Top right: ACS/WFC F475W image, M31N~2011-10a eruption
  position determined from LT $B$-band data. Bottom left: ACS/WFC
  F475W image, M31N~2011-10d eruption position determined from LT
  $B$-band data. Bottom right: WFPC2 F555W image, M31N~2011-12a
  eruption position determined from LT $V$-band
  data. (A color version of this figure is available in the online journal.)\label{progenitor-grid-6}}
\end{center}
\end{figure*}
\item {\it M31N~2011-10a.}
The {\it HST} images used to locate the quiescent nova M31N~2011-10a
were taken with ACS/WFC using F814W and F475W filters on 2010 December
14. There are no resolvable sources within $3\sigma$ of the calculated
position and the closest resolvable source in the F475W image is 2.540
ACS/WFC pixels, $0^{\prime\prime}\!\!.128$ or $4.80\sigma$ away from
the defined position. The source was not detected in the F814W filter
due to problems with the photometry, leading to a magnitude limit of
22.6, whereas the F475W magnitude limit was 26.1. The local population
density suggests that the coincidence probability at this separation
is 46.8\%. The location around this nova system is shown in
Figure~\ref{progenitor-grid-6} (top right).
\item {\it M31N~2011-10d.}
The location of nova M31N~2011-10d was observed by {\it HST} with
ACS/WFC using F814W and F475W filters on 2010 December 14. There are
no resolvable sources within $3\sigma$ of the calculated position,
with the closest resolvable source bring 1.400 ACS/WFC pixels,
$0^{\prime\prime}\!\!.070$ or $3.25\sigma$ away from the defined
position. The local population density, which has an F435W limiting
magnitude of 26.1, suggests there is a 16.4\% probability of such an
alignment occurring by chance. The images have an F475W limiting
magnitude of 26.1, although as with M31N~2011-10a, the F814W limiting
magnitude is only 22.6. The location of the quiescent system is shown
in Figure~\ref{progenitor-grid-6} (bottom left).
\item {\it M31N~2011-12a.}
We located the position of the M31N~2011-12a system using {\it HST}
images taken with WFPC2 using F814W and F555W filters on 2004 August
14. There is a resolvable source within $2\sigma$ of the calculated
position and no other resolvable source within $3\sigma$. The source
is 1.129 WFPC2 pixels, $0^{\prime\prime}\!\!.113$ or $1.15\sigma$ away
from the defined position. The local population density, which is
resolved down to an F814W magnitude of 24.2, suggests there is a 3.6\%
probability of such an alignment occurring by chance. The source had
an F814W magnitude of $23.67\pm0.07$ and an F555W magnitude of
$21.58\pm0.03$. This gives a $V$-band magnitude of $23.67\pm0.07$ and
a $I$-band magnitude of $21.44\pm0.04$, with $(V-I)$ color
$2.16\pm0.08$.  However, from the overall distribution of stars in the
M31 field in a typical ACS/WFC image, we would expect a source to be
at least as close as $0^{\prime\prime}\!\!.113$ in a typical ACS/WFC
image about 35\% of the time. Therefore such a detection in ACS/WFC
would be relatively insignificant. For this reason we do not include
it in the list of systems with a high likelihood of a recovered
progenitor. The location of M31N~2011-12a is shown in
Figure~\ref{progenitor-grid-6} (bottom right).
\item {\it M31N~2012-01a.}
The {\it HST} images used to locate M31N~2012-01a in quiescence were
taken with ACS/WFC using F814W and F555W filters on 2011 February 16
and there is one resolvable source within $3\sigma$ of the calculated
{\it HST} position. The source is 1.712 ACS/WFC pixels,
$0^{\prime\prime}\!\!.086$ or $2.50\sigma$ away from the defined
position. The local population density suggests there is a 22.0\%
probability of such an alignment occurring by chance. Figure
\ref{fig:grid} shows the nova in the combined LT image taken during
eruption and how the coincident {\it HST} field overlaps its
position. The data in the vicinity of this system is shown in
Figure~\ref{progenitor-grid-7} (top left).
\begin{figure*}[ht]
\begin{center}
\includegraphics[width=0.45\textwidth]{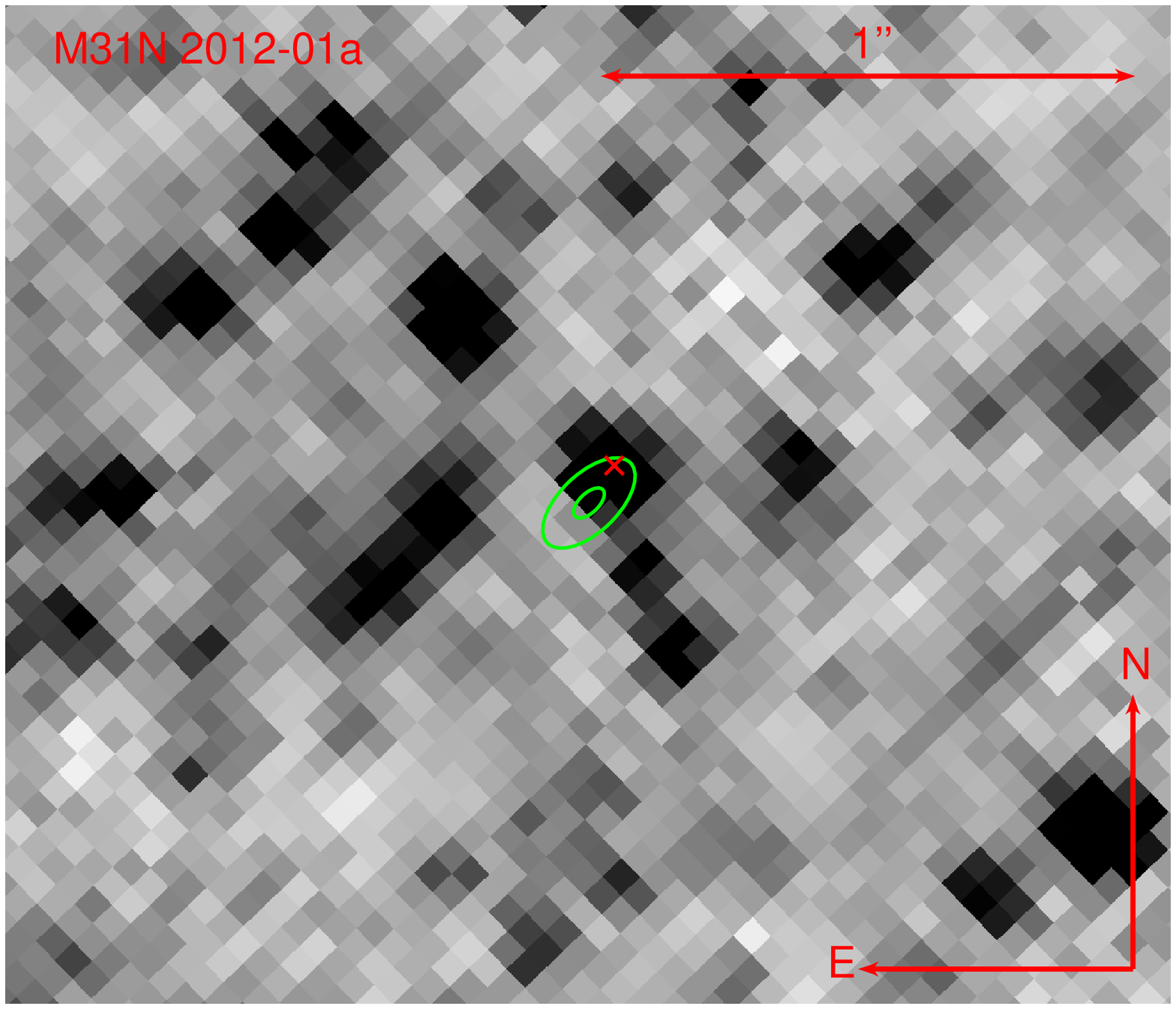}
\includegraphics[width=0.45\textwidth]{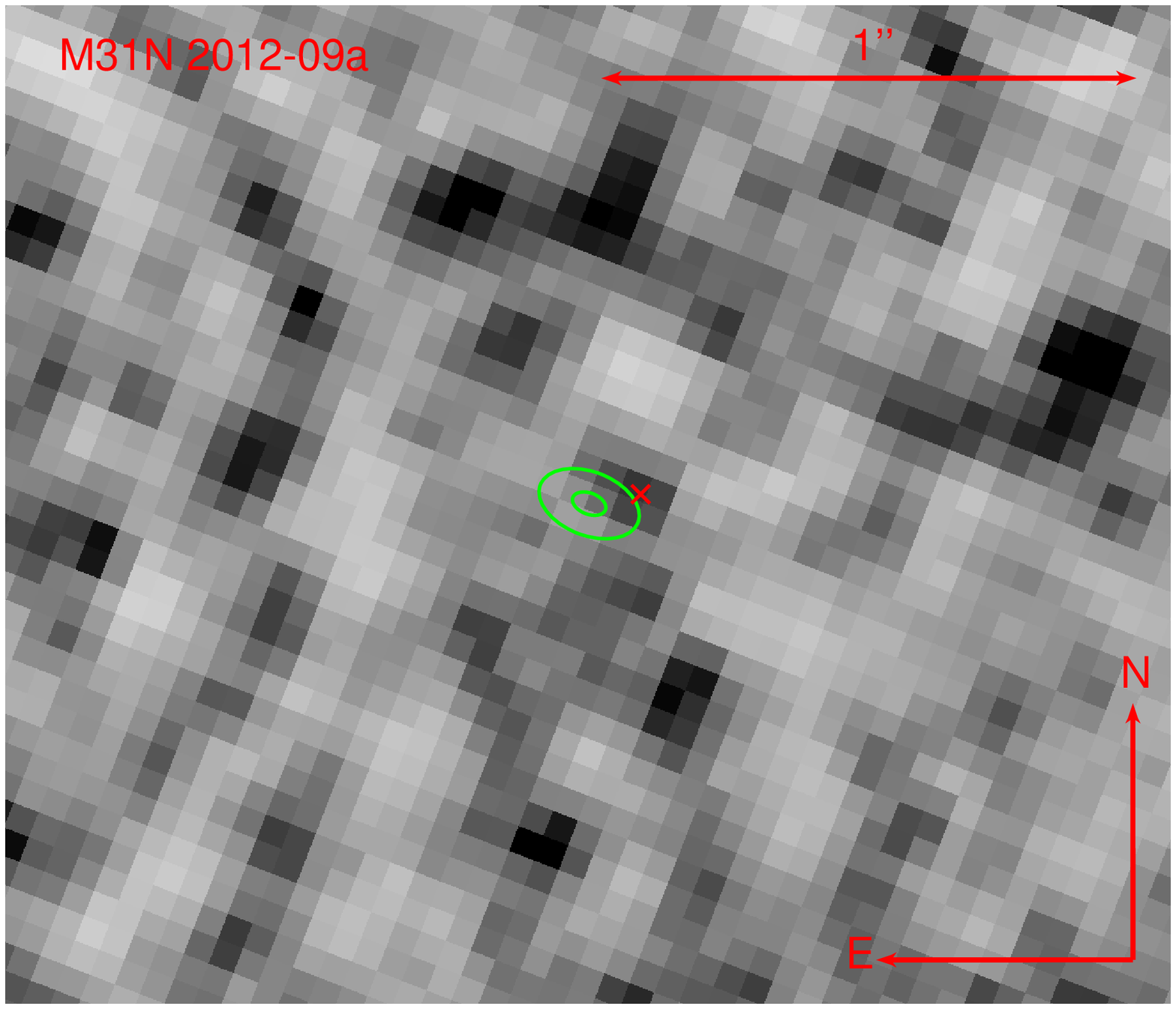}\\
\includegraphics[width=0.45\textwidth]{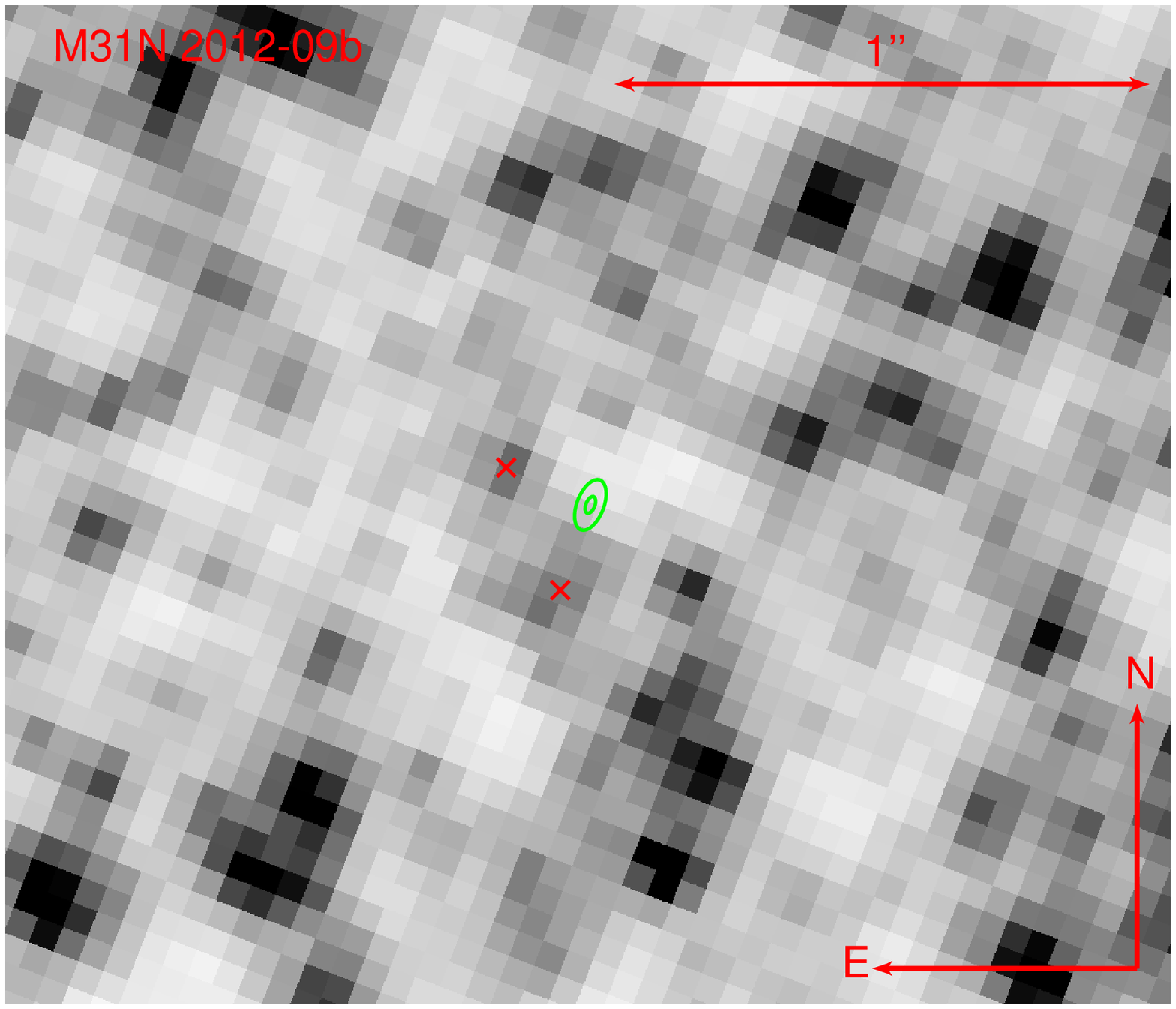}
\includegraphics[width=0.45\textwidth]{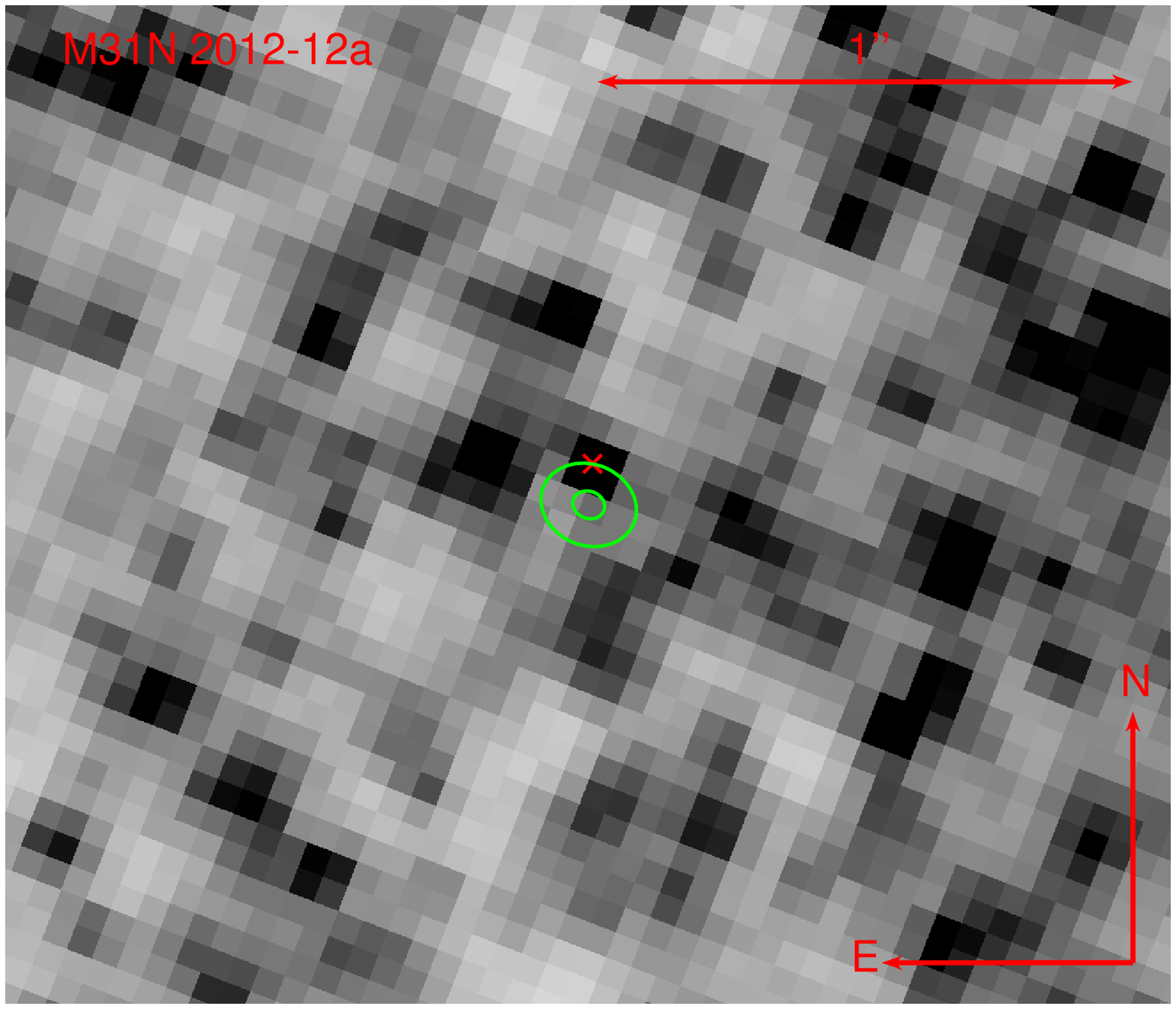}
\caption{As Figure~\ref{progenitor-grid-1}. Top left: ACS/WFC F555W
  image, M31N~2012-01a eruption position determined from LT $V$-band
  data. Top right: ACS/WFC F814W image, M31N~2012-09a eruption
  position determined from LT $r'$-band data. Bottom left: ACS/WFC
  F814W image, M31N~2012-09b eruption position determined from LT
  $r'$-band data. Bottom right: ACS/WFC F814W image, M31N~2012-12a
  eruption position determined from LT $r'$-band
  data. (A color version of this figure is available in the online journal.)\label{progenitor-grid-7}}
\end{center}
\end{figure*}
\item {\it M31N~2012-09a.}
Nova M31N~2012-09a is a recurrent candidate, with the first observed
eruption being M31N~1984-07a
\citep{2007A&A...465..375P,2012ATel.4368....1S}. The {\it HST} images
were taken with ACS/WFC using F814W and F475W filters on 2010 December
20 and 24. There are no resolvable sources within $3\sigma$ of the
calculated position, with the closest resolvable source being 2.152
ACS/WFC pixels, $0^{\prime\prime}\!\!.108$ or $3.65\sigma$ away from
the defined position. The local population density suggests that the
coincidence probability at this separation is 32.8\%. The F475
limiting magnitude of this data is 25.8, with that of F814W being
22.2. The location around this nova system is shown in
Figure~\ref{progenitor-grid-7} (top right).
\item {\it M31N~2012-09b.}
We located the expected position of the quiescent nova M31N~2012-09b
using {\it HST} images taken with ACS/WFC using F814W and F475W
filters on 2010 December 20. There are no resolvable sources within
$3\sigma$ of the calculated position. The closest resolvable source is
3.659 ACS/WFC pixels, $0^{\prime\prime}\!\!.184$ or $10.55\sigma$ away
from the defined position. The local population density, which is
resolvable down to an F814W magnitude of 24.6, suggests there is a
$78.4\%$ probability of such a distant alignment occurring by
chance. The F814W limiting magnitude was 24.6, although that of F475W
was 26.0 and there were no sources closer to the position of the nova
in this filter. The position of this nova is shown in
Figure~\ref{progenitor-grid-7} (bottom left). 
\item {\it M31N~2012-12a.}
The {\it HST} images used to find the position of M31N~2012-12a in
quiescence were taken with ACS/WFC using F814W and F475W filters on
2010 December 14. There is a resolvable source at $3.00\sigma$ of the
calculated position. The source is 1.532 ACS/WFC pixels or
$0^{\prime\prime}\!\!.077$ away from the defined position, with the
local population density suggesting that the coincidence probability
at this separation is 19.9\%. The F475W limiting magnitude is 25.6 and
the F814W limiting magnitude is 22.3. The location of M31N~2012-12a is
shown in Figure~\ref{progenitor-grid-7} (bottom right). 
\end{enumerate}

\section{Light curves} \label{sec:lc}

In this section we present additional, previously unpublished,
photometry following the eruptions of twelve M31 novae taken or
retrieved in the course of this survey.  This includes data from the
LT and serendipitous {\it HST} data. 

\begin{enumerate}
\item {\it M31N~2007-02b.}
Nova M31N~2007-02b had WFPC2 F555W data taken on 2454356~HJD. The nova had an F555W magnitude of $23.17\pm0.08$ at that
time. This was added to the light curve published by
\citet{2011ApJ...734...12S} and is shown in Figure~\ref{lc} (top
left). 
\begin{figure*}
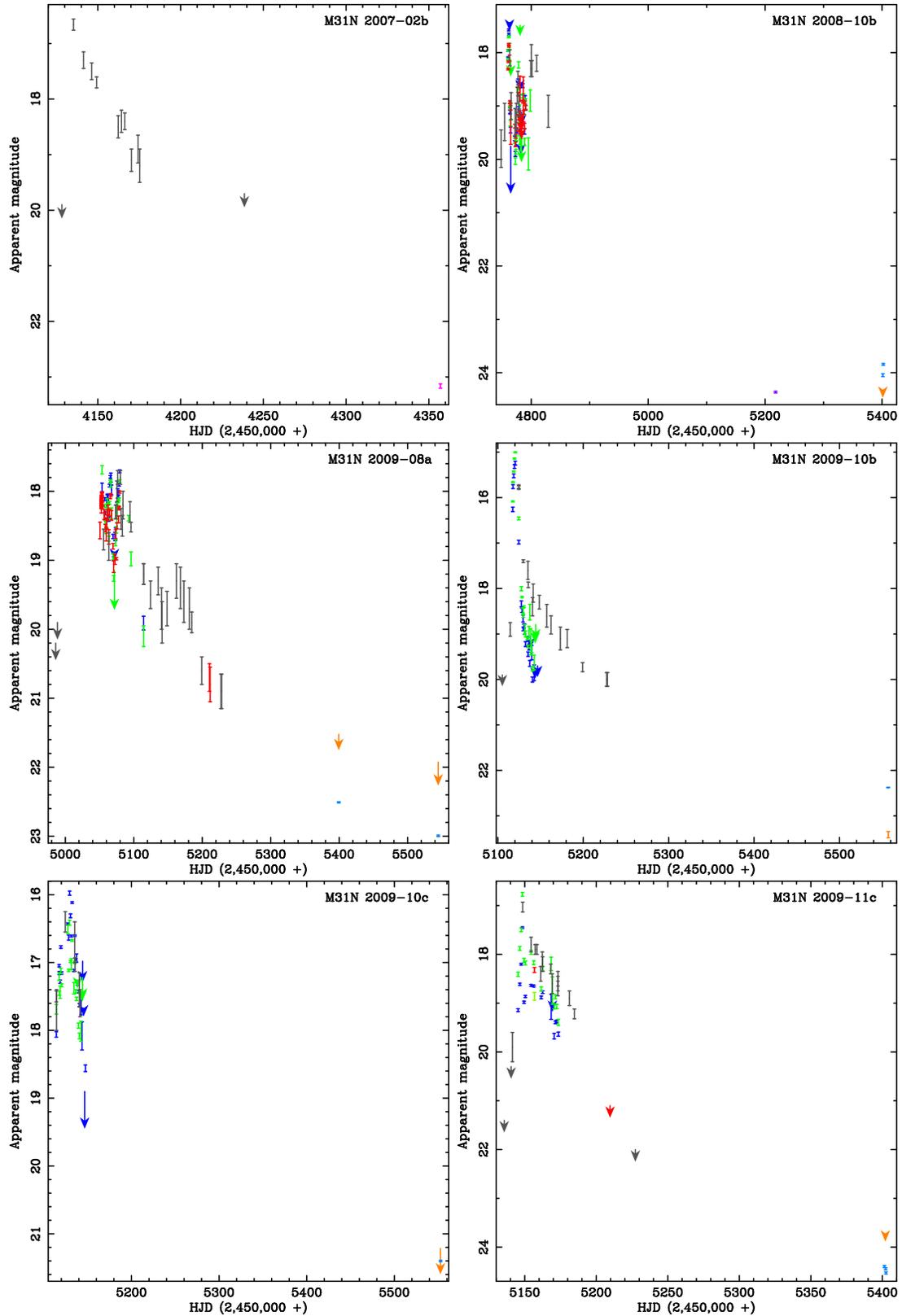

\begin{center}
\includegraphics[width=0.4\textwidth]{2007-02blc}
\includegraphics[width=0.4\textwidth]{2008-10blc}\\
\includegraphics[width=0.4\textwidth]{2009-08alc}
\includegraphics[width=0.4\textwidth]{2009-10blc}\\
\includegraphics[width=0.4\textwidth]{2009-10clc}
\includegraphics[width=0.4\textwidth]{2009-11clc}
\caption{Nova light curves (with 2009 and earlier eruptions being
  first published in \citealp{2011ApJ...734...12S}), with the
  following colors representing different bandpasses: $B$, royal
  blue; $V$, green; $R$, dark gray; $r'$ , red. We have
  extended these light curves using archival {\it HST} data where
  available.  Top left: M31N~2007-02b, magenta point is F555W filter.
  Top right: M31N~2008-10b, purple point is F435W filter, light blue
  point is F475W filter, orange arrow is F814W filter. Middle left:
  M31N~2009-08a, blue points are F475W filter.  Middle right:
  M31N~2009-10b.  Bottom left: M31N~2009-10c, blue point is F475W
  filter.  Bottom right: M31N~2009-11c, blue points are F475W filter,
  orange arrow is F814W filter. (A color version of this figure is available in the online journal.)\label{lc}}
\end{center}
\end{figure*}
\item {\it M31N~2008-10b.} 
Nova M31N~2008-10b had ACS/WFC F435W data taken on 2455217~HJD,
when the nova had a magnitude of $24.36\pm0.02$. The system also had
ACS/WFC data taken on 2455401 and 2455402 with F475W and F814W
filters. On 2455401 the nova had an F475W magnitude of
$23.85\pm0.02$ and on 2455402 it had an F475W magnitude of
$24.05\pm0.03$. The nova was not resolvable in the F814W images, but
for reference we added an upper limit for the magnitude, derived
simply by measuring the magnitude of a relatively faint nearby star,
clearly much brighter than the nova. This produced an F814W upper
limit of $24.41\pm0.07$ on 2455402. The points were added to the light
curve published by \citet{2011ApJ...734...12S} and are shown in
Figure~\ref{lc} (top right). As noted in Section~\ref{sec:prog}, it is
clear from the F435W eruption image taken on 2010 January 21 that the
candidate shown in Figure \ref{progenitor-grid-2} (bottom right) is
not the progenitor of this nova system. 
\item {\it M31N~2009-08a.} 
Nova M31N~2009-08a had ACS/WFC F475W and F814W data taken on 2455399 and 2455544~HJD. On 2455399 the system had an F475W magnitude
of $22.509\pm0.008$ and on 2455544 it had an F474W magnitude of
$23.00\pm0.01$. The nova was not resolvable in the F814W images on
either of these dates, in the first dataset we derived an upper limit
of $21.6\pm0.1$, with an upper limit of $22.1\pm0.2$ in the later set
of images. These data points were added to the light curve published
by \citet{2011ApJ...734...12S} and the extended light curve is shown
in Figure~\ref{lc} (middle left). These data were also used to help
locate the progenitor candidate (see Section~\ref{sec:prog}). 
\item {\it M31N~2009-10b.} 
Nova M31N~2009-10b had ACS/WFC F475W and F814W data taken on 2455557~HJD. At this time the nova had an F475W magnitude of
$22.378\pm0.007$ and F814W magnitude of $23.41\pm0.07$. These points
were added to the light curve published by \citet{2011ApJ...734...12S}
and are shown in Figure~\ref{lc} (middle right). These data were also
used to rule out some nearby sources as progenitor candidates (see
Section~\ref{sec:prog}). 
\item {\it M31N~2009-10c.} 
Nova M31N~2009-10c had one set of ACS/WFC F475W and F814W data taken on 2455552~HJD and another on 2455555. In
the latter set of images the system had an F475W magnitude of
$22.4\pm0.2$. In the images taken on 2455552, the nova had an
F475W magnitude of $21.40\pm0.01$. There is a source in the F814W
image that has a magnitude of $21.4\pm0.2$, but we cannot be certain
that this is the nova itself, so it is considered as an upper
limit. The points were added to the light curve published by
\citet{2011ApJ...734...12S} and are shown in Figure~\ref{lc} (bottom
left). Note that of the two F475W measurements, only the one taken on 2455552 is shown. This is due to the close proximity in the time
of the two measurements and the much lower error on the slightly
earlier measurement. These data were also used to rule out a nearby
source as progenitor candidate for this nova (see
Section~\ref{sec:prog}). 
\item {\it M31N~2009-11c.} 
Nova M31N~2009-11c had ACS/WFC F475W and F814W images taken on 2455401 and 2455402~HJD. On 2455401, the nova had an F475W magnitude of
$24.40\pm0.02$. It had an F475W magnitude of $24.45\pm0.02$ on 2455402.3
and $24.53\pm0.02$ on 2455402.6. There is a nearby candidate with an
F814W magnitude of $23.83\pm0.03$ on 2455401, but we cannot be certain
this is the nova, so it is considered as an upper limit. These points
were added to the light curve published by \citet{2011ApJ...734...12S}
and are shown in Figure~\ref{lc} (bottom right). These {\it HST} data
taken during eruption were also used to rule out a nearby source as
progenitor candidate (see Section~\ref{sec:prog}). 
\begin{figure*}
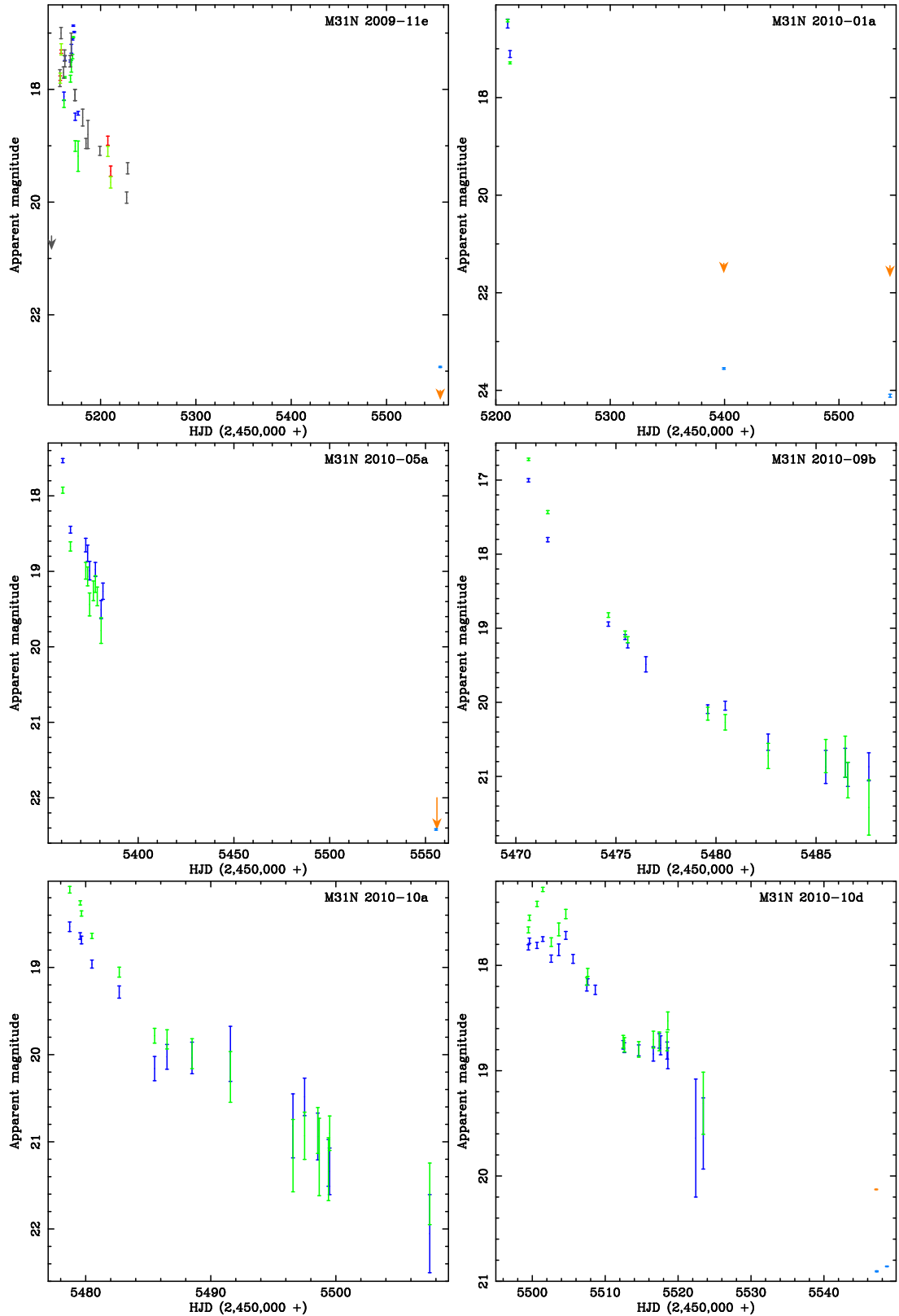

\begin{center}
\includegraphics[width=0.4\textwidth]{2009-11elc}
\includegraphics[width=0.4\textwidth]{2010-01alc}\\
\includegraphics[width=0.4\textwidth]{2010-05alc}
\includegraphics[width=0.4\textwidth]{2010-09blc}\\
\includegraphics[width=0.4\textwidth]{2010-10alc}
\includegraphics[width=0.4\textwidth]{2010-10dlc}
\caption{Nova light curves with archival {\it HST} data as in
  Figure~\ref{lc}.  Top left: M31N~2009-11e, light curve published by
  \citet{2011ApJ...734...12S} extended with {\it HST} points, blue
  point is F475W filter, orange arrow is F814W filter. Top right:
  M31N~2010-01a from LT data extended with {\it HST} points, light
  blue points are F475W filter, orange arrows are F814W filter. Middle
  left: M31N~2010-05a from LT data extended with {\it HST} points,
  light blue point is F475W filter, orange arrow is F814W
  filter. Middle right: M31N~2010-09b from LT data. Bottom left:
  M31N~2010-10a from LT data. Bottom right: M31N~2010-10d from LT data
  extended with {\it HST} points, light blue point is F475W filter,
  orange point in F814W filter. (A color version of this figure is available in the online journal.)\label{lc2}}
\end{center}
\end{figure*}
\item {\it M31N~2009-11e.} 
Nova M31N~2009-11e had ACS/WFC F475W and F814W data taken on 2455556~HJD. At this time the nova had an F475W magnitude
of $22.93\pm0.01$, but was not visible in the F814W image. For
reference we have added the magnitude of a nearby faint (but clearly
brighter than the nova) star as an upper limit. The points were added
to the light curve published by \citet{2011ApJ...734...12S} and shown
in Figure~\ref{lc2} (top left). These data were also used to rule out
a nearby source as progenitor candidate (see Section~\ref{sec:prog}). 
\item {\it M31N~2010-01a.} 
Nova M31N~2010-01a was observed by the LT on 2455210.36 and 2455212.39~HJD with {\it B} and {\it V} filters. The observations taken on
January 13.86 appear to be close to the optical maximum. The object
increased its $R$-band brightness between 2455208.59 and 2455209.59
\citep{2010CBET.2124....1B} and our observations show that the nova
had started to fade by 2455212.39. In Addition to the LT
photometry, the system was observed with ACS/WFC on 212455399 and
2455544 with F475W and F814W filters. On 2455399 the nova had
an F475W magnitude of $23.55\pm0.02$ and F814W upper limit of
$21.5\pm0.1$. On 2455544 it had F475W magnitude of $24.11\pm0.03$
and an F814W upper limit of $21.6\pm0.1$. The F814W photometry listed
above are likely to be the nova itself, however as we do not have any
quiescent F814W data and we cannot be certain we are seeing the nova,
we list them as upper limits. The light curve showing the LT and {\it
  HST} data points is shown in Figure~\ref{lc2} (top right). From the
above, we know that the nova must have reached maximum between 2455208.59 and 2455210.36. Therefore to calculate the upper $t_{2}$ limits we
linearly extrapolate the measurements taken on 2455208.59 and
2455209.59 to compute the expected brightness at 2455210.36. We then
linearly extrapolate between the two LT measurements. This gives a $B$
and $V$ lower $t_{2}$ limits of 7 and 5~days respectively. To
calculate the upper limit we extrapolate between the second LT points
and the first {\it HST} point, taking the first LT point as the
maximum. This gives a $B$ and $V$ $t_{2}$ upper limits of 49 and
40~days respectively. This approach yields mean $B$ and $V$ $t_{2}$
values of $30\pm20$~days and $20\pm20$~days respectively. The large
errors are unsurprising given the lack of data. The {\it HST} data
taken during eruption were also used to help locate the progenitor
candidate (see Section~\ref{sec:prog}). 
\item {\it M31N~2010-05a.} 
Nova M31N~2010-05a was observed by the LT in $B$ and $V$ filters
multiple times between 2455360.65 and 2455381.62~HJD. In our data, the
nova was at its brightest in both the $B$- and $V$-bands on 2455360.65, when it had a magnitude of $17.9\pm0.1$ in $V$-band and
$17.53\pm0.07$ in $B$-band. The nova brightened in the {\it R}-band
between 2455344.55 and 2455351.28 \citep{2010CBET.2305....1H,2010CBET.2305....2N} and then
appeared to remain near the maximum for several days (see
\citealp{2010CBET.2319....1H,2010CBET.2341....2H}). The LT
observations appear to start just as the nova began to fade. The
system was also observed with ACS/WFC in F475W and F814W filters
between 2455555.38 and 2455556.11. At this time the nova had an
F475W magnitude of $22.42\pm0.01$. Although the nova was not
resolvable in the F814W image, we have added an upper limit of
$22.1\pm0.2$ for reference. This was calculated from a nearby faint
star that was clearly brighter than the nova. A light curve of the LT
and {\it HST} points is shown in Figure~\ref{lc2} (middle left). As
the nova appears to stay near maximum for a few days, we cannot
calculate $t_{2}$ directly. We also note that there is a significant
gap between the end of the LT observations and the time  of the {\it
  HST} observations. To calculate the lower $t_{2}$ limit we assume
that the first LT observations were at maximum and then linearly
extrapolate the LT points (excluding the first two measurements) until
$t_{2}$ is reached. This gives a $B$ and $V$ lower limit of 22 and
23~days respectively. For the upper limit we assume that the nova was
at peak (and the same magnitude as the first LT measurements) when it
first appeared to plateau on 2455351.28. We then extrapolate between the
final LT point and the F475W {\it HST} point. This gives a $B$ and $V$
upper limit of 56 and 50~days respectively. Therefore producing a
$B$-band $t_{2}$ estimate of $40\pm20$~days and a $V$-band estimate of
$40\pm10$~days. 
\item {\it M31N~2010-09b.} 
Nova M31N~2010-09b was observed by the LT in {\it B} and {\it V}
filters between 2455470.63 and 2455487.64~HJD. In these data, the nova
was at its brightest in both $B$ and $V$ filters in the images taken
on 2455470.63, when it had a $V$-band magnitude of $16.72\pm0.04$
and $B$-band magnitude of $17.00\pm0.08$. The nova brightened
significantly in the $R$-band between 2455469.91 and 2455470.68 \citep{2010ATel.2896....1P}. Our observations also constrain that
the nova must have faded significantly by 2455471.60. Therefore it
appears that the LT started observing the nova when it was at, or very
close to, maximum brightness. A light curve of the LT points is shown
in Figure~\ref{lc2} (middle right). Assuming the first set of
observations were indeed taken at maximum, a $V$-band $t_{2}$ of
$3.8\pm0.2$~days and a $B$-band $t_{2}$ of $4.3\pm0.6$~days were
derived. 
\item {\it M31N~2010-10a.} 
Nova M31N~2010-09b was observed by the LT between 2455478.72
and 2455507.51~HJD with {\it B} and {\it V} filters. The light curve
for this nova is shown in Figure~\ref{lc2} (bottom left). The nova
brightened in the {\it R}-band between 2455475.05 and 2455476.04
\citep{2010CBET.2483....1N}. The LT images show that the nova fades in
both {\it B} and {\it V} bands between 2455478.72 and 2455479.56. Therefore it is likely that the nova reached peak between 2455475.05 and 2455478.72. As the peak may have been missed we cannot
calculate an accurate $t_{2}$. However by calculating how long it
takes the nova to fade by two magnitudes from the first LT observation
we can calculate an upper limit on $t_{2}$, which for $B$-band is
$16\pm2$~days and for $V$-band a maximum of $11\pm5$~days. From the
{\it R}-band observations \citet{2012ApJ...752..133C} constrained the
decline time to $t_{2}>9$~days. 
\item {\it M31N~2010-10d.} 
Nova M31N~2010-10d was observed by the LT in $B$ and $V$ filters
regularly between 2455499.47 and 2455523.45~MJD. It is unclear
from the LT data when maximum light was reached in the $B$-band and
there appear to be two clear peaks in the $V$-band. Similarly, it is
also unclear from the light curve presented in
\citet{2012ApJ...752..133C} when the peak brightness is reached. The
nova was observed through the {\it HST} ACS/WFC F475W and F814W
filters on 2455547 and 2455548. On 2455547 the nova had an
F475W magnitude of $20.906\pm0.005$ and an F814W magnitude of
$20.128\pm0.003$. On 2455548 it had an F475W magnitude of
$20.859\pm0.003$. A light curve of the LT and {\it HST} points is
shown in Figure~\ref{lc2} (bottom right). From the $V$-band maximum
(i.e.\ the first peak) we calculate a $t_{2}$ of $22.8\pm0.9$~days. As
the $B$-band LT data do not cover the nova fading by two magnitudes,
we need to extrapolate to the first F475W data point. If we take the
$B$-band point that corresponds to the $V$-band maximum we calculate a
$t_{2}$ of $25\pm5$~days. If we take the point of the highest $B$-band
flux as the maximum we calculate a $t_{2}$ of $21\pm5$~days. These are
the upper and lower limits respectively, which gives a $B$-band
$t_{2}$ of $23\pm7$~days. As noted in Section~\ref{sec:prog}, the {\it
  HST} data taken during eruption were also used to rule out a nearby
source as progenitor candidate. 
\end{enumerate}

\section{Discussion}
Eleven nova systems ($29\%$ of the original catalog of 38 novae) are
positionally aligned with a resolved point source in archival {\it
  HST} data, with the probability of such a close alignment occurring
by chance being $\leq5\%$ in each case. The novae with candidate
progenitor systems are M31N~2007-02b, -10a, -11b, -11d, -11e, -12a,
-12b, 2009-08a, -11d, 2010-01a and -09b \citep[the progenitor system
of M31N~2007-12b had already been identified
by][]{2009ApJ...705.1056B}.  The photometry of these eleven candidate
progenitor systems is summarised in Tables~\ref{tab:flight}
and~\ref{tab:cand}.  

\begin{deluxetable*}{lllllll}
\tablecaption{Progenitor raw photometry in native ACS/WFC or WFPC2 system.\label{tab:flight}}
\tablewidth{0pt}
\tablehead{
\colhead{Nova} & \multicolumn{6}{c}{Broadband Photometry}\\
\colhead{}    & \colhead{F435W}   & \colhead{F475W}  & \colhead{F555W}  & \colhead{F606W}  & \colhead{F625W}  & \colhead{F814W}
}
\startdata
M31N~2007-02b  & \nodata                & \nodata                & \nodata                & $25.95\pm0.05$     & \nodata                & $24.82\pm0.03$\\
M31N~2007-10a  & \nodata                & \nodata                & \nodata                & \nodata                  & $22.397\pm0.008$ &\nodata\\
M31N~2007-11b  & \nodata                & \nodata                & $22.6\pm0.1$     & \nodata                  & \nodata                & $20.44\pm0.06$\\
M31N~2007-11d  & \nodata                & $24.46\pm0.04$   & \nodata                & \nodata                  & \nodata                & $21.387\pm0.005$\\            
M31N~2007-11e  & \nodata                & $25.5\pm0.1$     & \nodata                & \nodata                  & \nodata                & $24.19\pm0.03$\\
M31N~2007-12a  & \nodata                & $25.98\pm0.08$   & \nodata                & \nodata                  & \nodata                & $25.3\pm0.1$\\
M31N~2007-12b  & \nodata                & $25.36\pm0.04$   & \nodata                & \nodata                  & \nodata                & $23.79\pm0.04$\\
M31N~2009-08a  & $25.50\pm0.05$   & \nodata                & \nodata                & \nodata                  & \nodata                & \nodata\\
M31N~2009-11d  & \nodata                & $25.67\pm0.04$   & \nodata                & \nodata                  & \nodata                & $25.1\pm0.2$\\
M31N~2010-01a  & $24.79\pm0.04$   & \nodata                & \nodata                & \nodata                  & \nodata                & \nodata\\
M31N~2010-09b  & \nodata                & $26.30\pm0.07$   & \nodata                & \nodata                  & \nodata                & $24.7\pm0.1$
\enddata
\end{deluxetable*}

\begin{deluxetable*}{lcccccc} 
\tablecaption{Progenitor photometry converted to {\it UBVRI} system,
  where available. The magnitudes were transformed using conversions
  from \citet{2005PASP..117.1049S}. These magnitude also include the
  extinction internal to M31.\label{tab:cand}}
\tablewidth{0pt}
\tablehead{
\colhead{Nova} & \multicolumn{4}{c}{Broadband Photometry}                          & \multicolumn{2}{c}{color}\\
\colhead{}     & \colhead{$B$}       & \colhead{$V$}  & \colhead{$R$}  & \colhead{$I$} & \colhead{$(B-I)$} & \colhead{$(V-I)$}
}
\startdata
M31N~2007-02b  &\nodata                & $26.0\pm0.4$      &\nodata          & $24.6\pm0.2$     &\nodata & $1.4\pm0.2$\\
M31N~2007-10a  &\nodata                &\nodata                  &$\sim22.0$ & \nodata                &\nodata &\nodata\\
M31N~2007-11b  &\nodata                & $22.3\pm0.3$      &\nodata          & $20.1\pm0.2$     &\nodata & $2.1\pm0.2$ \\
M31N~2007-11d  & $25.0\pm0.5$    &\nodata                  &\nodata          & $21.3\pm0.2$     & $3.8\pm0.3$  &\nodata\\
M31N~2007-11e  & $26.0\pm0.3$    &\nodata                  &\nodata          & $24.1\pm0.1$     & $1.9\pm0.2$  &\nodata\\
M31N~2007-12a  & $25.7\pm0.5$    &\nodata                  &\nodata          & $25.1\pm0.2$     & $0.6\pm0.3$  &\nodata\\
M31N~2007-12b  & $25.7\pm0.5$    &\nodata                  &\nodata          & $23.6\pm0.2$     & $2.1\pm0.3$  &\nodata\\
M31N~2009-08a  &$\sim25.5$       &\nodata                  &\nodata          & \nodata                &\nodata &\nodata\\
M31N~2009-11d  & $25.4\pm0.4$    &\nodata                  &\nodata          & $24.9\pm0.3$     & $0.5\pm0.3$  &\nodata\\
M31N~2010-01a  & $\sim24.5$      &\nodata                  &\nodata          & \nodata                &\nodata &\nodata\\
M31N~2010-09b  & $26.6\pm0.5$    &\nodata                  &\nodata          & $24.5\pm0.2$     & $2.2\pm0.3$  &\nodata
\enddata
\end{deluxetable*}

The only confirmed RG-novae in our Galaxy are RS~Oph
\citep{1999A&A...344..177A}, T~Coronae Borealis \citep{1987AJ.....93..938K},
V3890~Sagittarii and V745~Scorpii \citep{1993AJ....105..320H}, all of which are
RNe. Although this is a small percentage of the overall Galactic nova
population of almost 400 \citep{2010AN....331..160B}, in many cases
the systems have simply not been observed in quiescence. KT~Eri is
also believed to contain a red giant secondary
\citep{2012A&A...537A..34J}, as are EU Sct and V2487~Ophiuchi
\citep{2012ApJ...746...61D}. This suggests the RG-nova rate in the
Galaxy is probably higher than previously believed and there has been
no study systematic enough to produce a reliable population estimate.  

\begin{figure*}
\begin{center}
\includegraphics[width=0.49\textwidth]{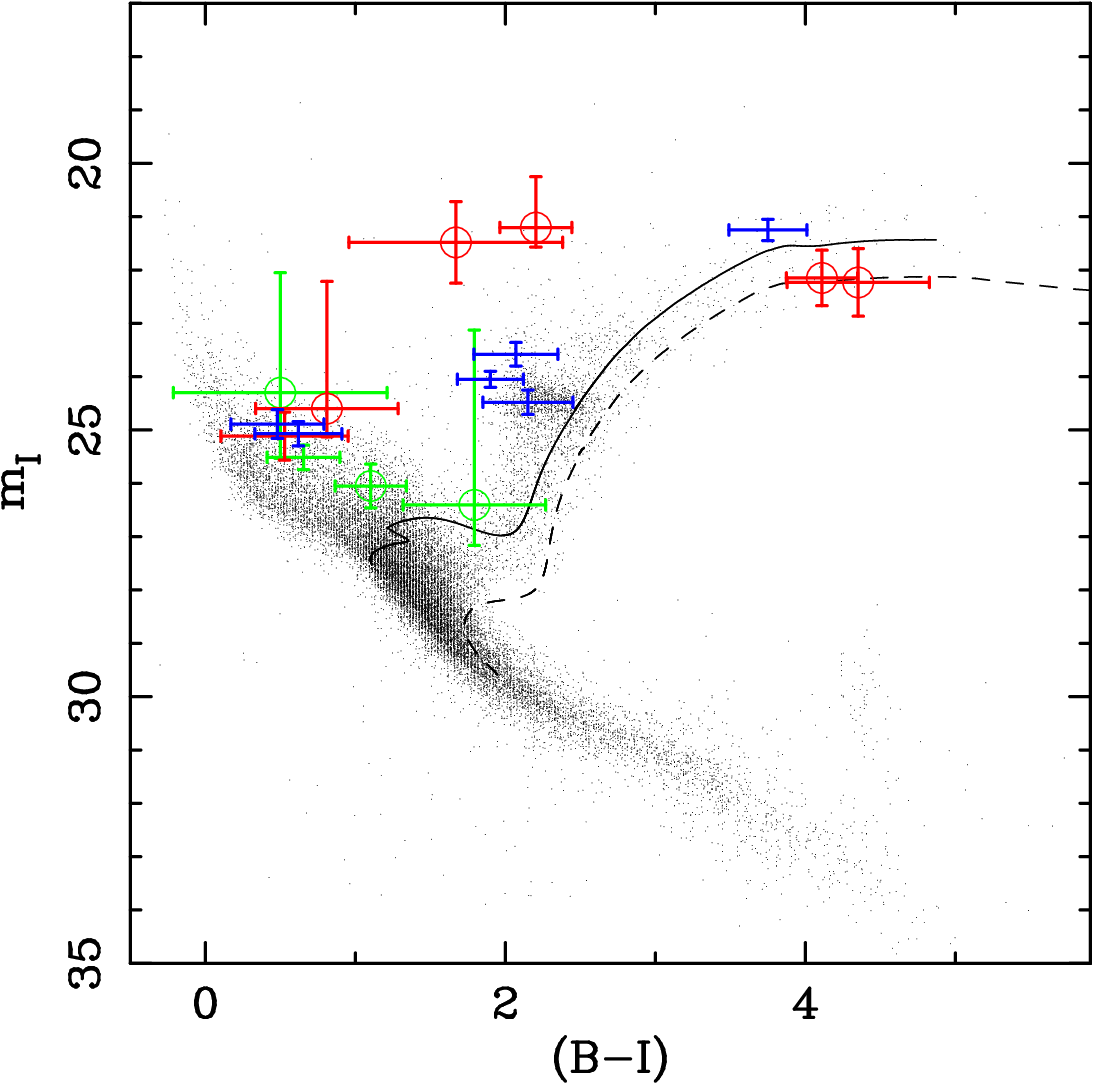}
\includegraphics[width=0.49\textwidth]{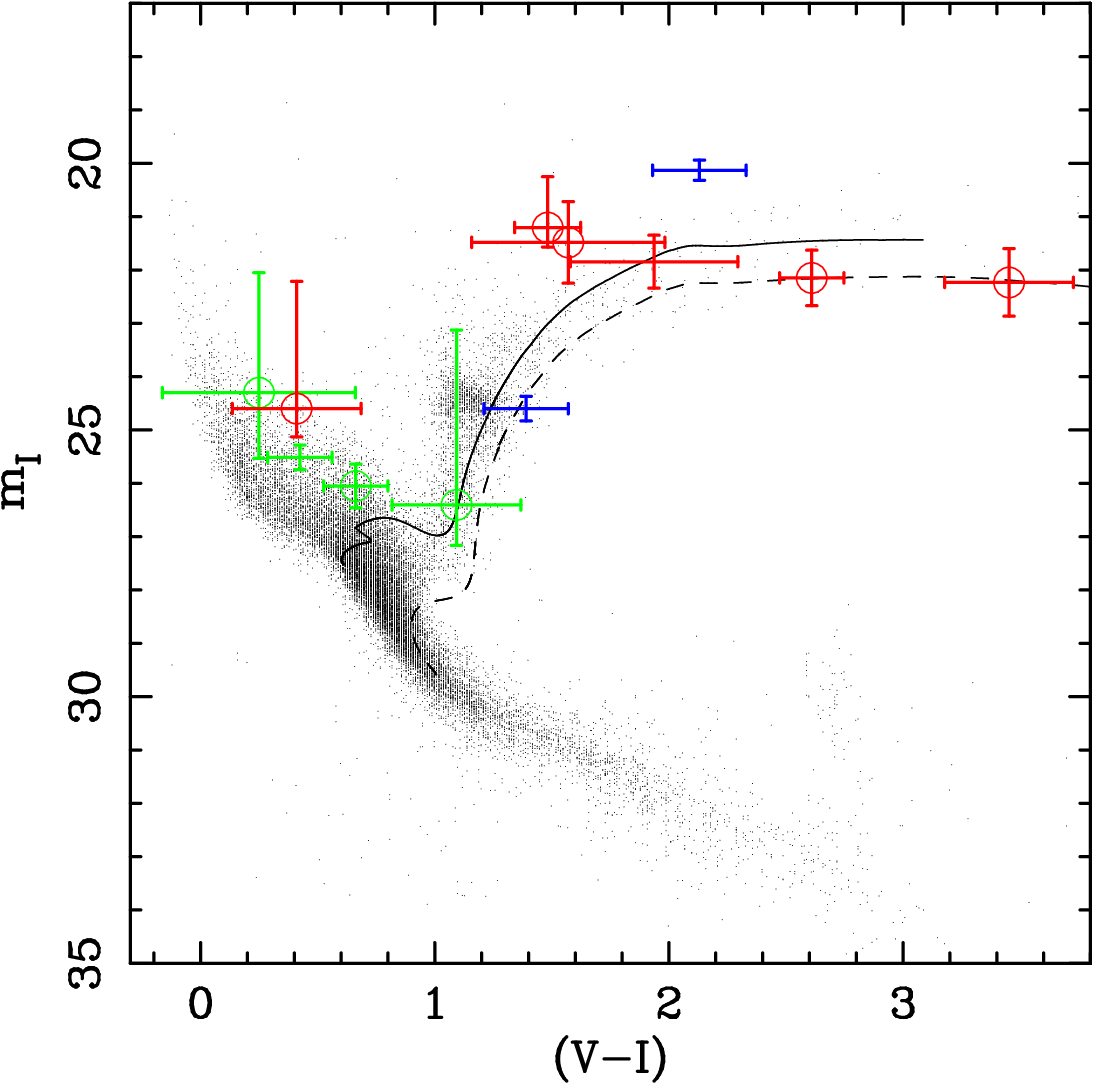}
\caption{Color$-$magnitude diagrams showing the {\it Hipparcos} data
  \citep{1997ESASP1200.....P} shifted to the distance of M31, assuming
  $(m-M)_{0}=24.43$ \citep{1990ApJ...365..186F} and extinction
  of $E_{B-V}=0.1$~mag towards M31
  \citep{1992ApJS...79...77S}. Possible extinction internal to M31 is
  calculated for each nova separately and is discussed in the
  text. The blue points show the M31 progenitor candidates found in
  this work. The red points represent Galactic RG-novae and the green
  points represent Galactic SG-novae (see
  \citealp{2010ApJS..187..275S}, \citealp{2012ApJ...746...61D} and
  references therein). Left plot: $(B-I)$ color against $I$-band
  magnitude. Right plot: $(V-I)$ color against $I$-band
  magnitude. (A color version of this figure is available in the online journal.)\label{colmag}}
\end{center}
\end{figure*}

Eight of the eleven candidate progenitor systems identified by this
survey have quiescent photometry in at least two bands.  As such, the
positions of these systems are plotted in the color--magnitude
diagrams shown in Figure~\ref{colmag}.  Inspection of this figure
shows that the majority of progenitor candidates (six of eight) lie on
or near the red giant branch.  Of course, it should be noted that if
we had simply chosen a random sample of resolved stars in {\it HST}
data of M31 then this would still be the case.   The other two
candidate progenitor systems (M31N~2007-12a and 2010-09b) both have
$(B-I)<1$.  The color--magnitude diagram position of these two
quiescent systems is consistent with high mass, luminous main-sequence
stars.  However, as is also indicated in Figure~\ref{colmag}, this
position is similar to a pair of Galactic RG-novae.  These Galactic
novae are the suspected recurrent KT~Eri, which is also likely to
harbor a red giant secondary
\citep{2012A&A...537A..34J,2012ApJ...746...61D} and the recurrent
V2487~Oph \citep{2012ApJ...746...61D}.  The particularly blue color of
KT~Eri and V2487~Oph is likely due to a bright accretion disk inclined
towards the observer.  This pair of M31 quiescent systems is also
coincident with the color--magnitude position of a number of SG-novae
or suspected SG-novae. So if we are indeed observing M31N~2007-12a and
2009-11d in quiescence they may be RG-novae with colors affected by a
strong accretion disk or they may be luminous quiescent SG-novae
similar to the recurrent SG-nova U~Scorpii. 

\begin{figure}
\includegraphics[width=\columnwidth]{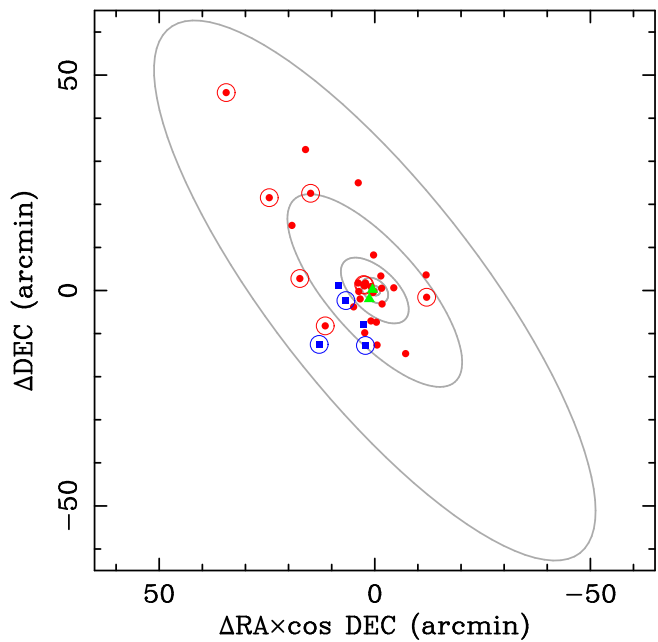}
\caption{The spatial distribution and spectral class of the 38 novae
  in our catalog as in Figure~\ref{spatialplot}, the eleven novae with
  candidate progenitor systems are further circled. (A color version of this figure is available in the online journal.)\label{spatialcand}
}
\end{figure}

In Figure~\ref{spatialcand}, we show the spatial distribution of the
novae in our survey as in Figure~\ref{spatialplot}, but also indicate
the eleven systems with candidate progenitor systems. This shows that
a much higher proportion of novae in the disk appear to have a
recovered progenitor when compared to the systems in the central
bulge. Although increased crowding near the center of M31 (rendering
any detection less significant) may have some influence on this, it
does not seem to be the main factor. Indeed, the relationship between
the distance of the nearest source from the nova and the probability
of a chance alignment is largely uniform for the regions of M31
studied in this work. This indicates that there may be a higher
proportion of RG-novae in younger stellar populations. Although of
course the background light is brighter near the center of M31 and
this may obscure some faint sources that would have been visible in
the outer regions of the galaxy. 

As is noted in Section~\ref{sec:prog}, there are problems with using
the WFPC2 data. For example, as can be seen in Section~\ref{sec:prog},
M31N~2011-12a had a progenitor candidate detected that only had 3.6\%
probability of such an alignment occurring by chance.  However, due to
the relatively bright limiting magnitude of the WFPC2 data, the same
detection in a typical ACS/WFC image would be relatively
insignificant. It could also be the case that, for a system where no
candidate was detected in WFPC2, if the same region was imaged with
ACS/WFC, it may reveal a coincident source. 

The most direct method to identify RNe is to find coincident eruptions. However, due to the large population of M31 nova candidates found over the last 100 years (over 900; see \citealp{2010AN....331..187P}, and their online catalog\footnote{\url{http://www.mpe.mpg.de/m31novae/opt/m31/index.php}}) many coincident sources will simply be
chance alignments (see \citealp{2013arXiv1307.2296S}; Shafter et~al.,
in preparation). Therefore the positions we publish here may help
eliminate some of the recurrent candidates that are different systems
aligned by chance. 

There are three RN candidates in our catalog: M31N~2009-11b, 2010-10e
and 2012-09a (see also Section~\ref{sec:prog}). It might be expected
that these novae would have a higher chance of harboring red giant
secondaries, as in our Galaxy half of the ten RNe are RG-novae (see
e.g.\ \citealp{2012ApJ...746...61D}). It can be seen in
Figure~\ref{progenitor-grid-4} (top right) that there is no resolvable
source within $3\sigma$ of the position of M31N~2009-11b, although the
limiting magnitude of the data is 23.8, meaning that we can exclude a
luminous red giant system (such as one similar to RS Oph) as the
progenitor, but it is possible that the system may contain a fainter
red giant or sub-giant secondary. Although M31N~2010-10e does not have
a progenitor that we can locate with a high confidence, it can be seen
from Figure~\ref{progenitor-grid-6} (top left) that a red giant
companion is not completely ruled out for this system. Indeed there is
a faint candidate within $1\sigma$ and a relatively bright source just
outside $1\sigma$. M31N~2012-09a does not have any sources within
$3\sigma$ of the calculated position, although the limiting magnitudes
indicate a lower-luminosity red giant or sub-giant companion may not
be resolvable in the data. 

We can, however, all but rule out M31N~2006-11a, 2008-10b, 2009-08b,
-11a, -11c, -11e and 2010-10b as RG-novae.  These seven novae either
had no resolved sources within $3\sigma$ of the eruption position or
had post-eruption {\it HST} data yielding extremely precise positions.
Additionally, the entirety of the red giant branch was resolvable in
the {\it HST} images associated with these novae. 

It is well worth looking in more detail at the possibility or effects
of coincidental alignment.  As we have already stated, the resolved
stellar population in M31 that is available to {\it HST} is dominated
by red giant stars, the same type of objects that we are looking for.
So there is a huge potential background of ``false positives''.
However, we believe that the Monte Carlo technique employed to compute
the probability of each alignment (between eruption position and
candidate progenitor) occurring through chance is robust (down to the
magnitude limit stated for each nova).  Indeed the criterion of
$\leq5\%$ probability of chance alignment that we employ is fairly
conservative.  The purpose of this paper is to present the entire
catalog of progenitor systems, whilst we intend to fully explore the
statistics of this survey and the uncovered population in a follow-up
paper (Williams et~al.\ in preparation).  However, it is worth
exploring the coincidence probabilities of the population of eleven
objects here in more detail.  Given our sample of eleven progenitor
systems we can simply compute the probability that all eleven are just
chance alignments.  The probability that all eleven systems align as
well as they do by pure chance is $10^{-20}$.  Extending this
approach, the probability that exactly ten of these systems are chance
alignments, but one is genuine, is $5\times10^{-14}$.  At the other
end of the spectrum, the probability that all eleven are genuine and
there are no coincidental alignments is 0.76.  If we are concerned
with $n\sigma$ results, then we can be confident at beyond the
$3\sigma$ level that, at most, two of the eleven systems are chance
alignments, with nine being genuine.  At the $5\sigma$ level, we can
still be confident that, at most, five are chance alignments, with the
majority, six, being genuine. 

There is a vast array of selection effects that influenced the
original input catalog, such as the luminosity of each nova, the speed
class, the position in M31, the time of year, the availability of
ground-based telescopes, the availability of archival {\it HST} data,
to name just a few.  The full effect of these selection effects on
this catalog and the significance of this result will be explored in a
follow-up paper (Williams et~al.\ in preparation).  However, should
these 38 nova turn out to be a fairly representative sample, then this
result is extremely significant.  Just $\sim1\%$ of Galactic novae are
known recurrent novae with red giant secondaries, another $\sim1\%$
are RNe with sub-giant secondaries.  If we include all Galactic novae
with known evolved (non-main-sequence) secondaries then they account
for only $\sim3\%$ of Galactic nova population.  In M31 we have found
that 29\% of our original catalog have evolved (most likely red giant)
secondaries, or at least 24\% and 16\% at the 3 and $5\sigma$ levels
respectively.  Such a large population of novae with evolved
secondaries would have considerable impact, particularly with regards
to their potential as a significant SN~Ia progenitor channel. 

\section{Conclusions}
In this paper we have presented the results of a survey for the
progenitor system of each of a catalog of 38 spectroscopically
confirmed novae in M31.  The eruption position of each nova was
determined using LT data and archival {\it HST}
data were used to search for the progenitors.  Here we summarise the
main conclusions. 

\begin{itemize}
\item We have recovered a resolved, quiescent, progenitor system for
  eleven of the original 38 novae.  For each of these systems the
  probability of a chance alignment is $<5\%$. 
\item These systems are M31N~2007-02b, -10a, -11b, -11d, -11e, -12a,
  -12b, 2009-08a, -11d, 2010-01a and -09b.  Photometry of each of
  these progenitors is consistent with quiescent Galactic RG-novae. 
\item The archival data also allow us to all but rule out red giant
  secondaries for seven novae, M31N~2006-11a, 2008-10b, 2009-08b,
  -11a, -11c, -11e and 2010-10b. 
\item Although misidentification due to chance alignments is possible,
  we can be confident at the $3\sigma$ level that at least nine of the
  candidate progenitor systems are genuine.  At the $5\sigma$ level
  this reduces to at least six. 
\item If the input catalog is representative of the M31 nova
  population then up to 29\%, 24\% or 16\% of M31 novae contain
  evolved (most likely red giant) secondaries at the 1, 3, and
  $5\sigma$ limits respectively. This is much higher than that found
  so far in the Milky Way.  
\end{itemize}

Additionally, we have produced several uniquely deep light curves for
M31 novae using archival {\it HST} data points. 

In a follow up to this paper, Williams et~al.\ (in preparation) will
examine the statistics of the catalog and of the underlying nova
population of M31 as a whole. We will produce an estimate of the
RG-nova rate within M31 and explore the effect it has on the case for
them being a major channel for producing SNe Ia. 

\acknowledgements The Liverpool Telescope is operated on the island of
La Palma by Liverpool John Moores University in the Spanish
Observatorio del Roque de los Muchachos of the Instituto de
Astrofisica de Canarias with financial support from the UK Science and
Technology Facilities Council.  Some of the data presented in this
paper were obtained from the Multimission Archive at the Space
Telescope Science Institute (MAST).  STScI is operated by the
Association of Universities for Research in Astronomy, Inc., under
NASA contract NAS5-26555.  Support for MAST for non-{\it HST} data is
provided by the NASA Office of Space Science via grant NNX09AF08G and
by other grants and contracts.  This publication makes use of data
products from the Two Micron All Sky Survey, which is a joint project
of the University of Massachusetts and the Infrared Processing and
Analysis Center/California Institute of Technology, funded by the
National Aeronautics and Space Administration and the National Science
Foundation. S.C.W. is currently supported by a STFC PhD studentship.
M.J.D. would like to thank Andrew Dolphin for advice in utilising the
HSTphot and DOLPHOT packages. A.W.S. acknowledges support from NSF
grant AST1009566.  We thank an anonymous referee for valuable comments on the initial submitted version of this paper.

Facilities: \facility{Hubble Space Telescope}, \facility{Hobby Eberley
  Telescope}, \facility{Liverpool Telescope}, \facility{Faulkes
  Telescope North}.

\bibliographystyle{apj}

\begin{thebibliography}{}
\expandafter\ifx\csname natexlab\endcsname\relax\def\natexlab#1{#1}\fi

\bibitem[{{Ansari} {et~al.}(2004){Ansari}, {Auri{\`e}re}, {Baillon}, {Bouquet},
  {Coupinot}, {Coutures}, {Ghesqui{\`e}re}, {Giraud-H{\'e}raud}, {Gillieron},
  {Gondolo}, {Hecquet}, {Kaplan}, {Kim}, {Le Du}, {Melchior}, {Moniez},
  {Picat}, \& {Soucail}}]{2004A&A...421..509A}
{Ansari}, R., {Auri{\`e}re}, M., {Baillon}, P., {et~al.} 2004, \aap, 421, 509

\bibitem[{{Anupama}(2008)}]{2008ASPC..401...31A}
{Anupama}, G.~C. 2008, in Astronomical Society of the Pacific Conference
  Series, Vol. 401, RS Ophiuchi (2006) and the Recurrent Nova Phenomenon, ed.
  A.~{Evans}, M.~F. {Bode}, T.~J. {O'Brien}, \& M.~J. {Darnley} (San Francisco, CA:ASP), 31

\bibitem[{{Anupama} \& {Miko{\l}ajewska}(1999)}]{1999A&A...344..177A}
{Anupama}, G.~C., \& {Miko{\l}ajewska}, J. 1999, \aap, 344, 177

\bibitem[{{Auri{\`e}re} {et~al.}(2001){Auri{\`e}re}, {Baillon}, {Bouquet},
  {Carr}, {Cr{\'e}z{\'e}}, {Evans}, {Giraud-H{\'e}raud}, {Gould}, {Hewett},
  {Kaplan}, {Kerins}, {Lastennet}, {Le Du}, {Melchior}, {Paulin-Henriksson},
  {Smartt}, \& {Valls-Gabaud}}]{2001ApJ...553L.137A}
{Auri{\`e}re}, M., {Baillon}, P., {Bouquet}, A., {et~al.} 2001, \apjl, 553,
  L137

\bibitem[{{Barsukova} {et~al.}(2009){Barsukova}, {Afanasiev}, {Fabrika},
  {Valeev}, {Hornoch}, \& {Pietsch}}]{2009ATel.2251....1B}
{Barsukova}, E., {Afanasiev}, V., {Fabrika}, S., {et~al.} 2009, The
  Astronomer's Telegram (ATel), 2251

\bibitem[{{Bode}(2010)}]{2010AN....331..160B}
{Bode}, M.~F. 2010, Astronomische Nachrichten, 331, 160

\bibitem[{{Bode} {et~al.}(2009){Bode}, {Darnley}, {Shafter}, {Page},
  {Smirnova}, {Anupama}, \& {Hilton}}]{2009ApJ...705.1056B}
{Bode}, M.~F., {Darnley}, M.~J., {Shafter}, A.~W., {et~al.} 2009, \apj, 705,
  1056

\bibitem[{{Burwitz} {et~al.}(2010){Burwitz}, {Pietsch}, {Henze}, {Updike},
  {Milne}, {Williams}, {Hartmann}, {Rodriguez}, {Holmes}, {Kolb}, \&
  {Lucas}}]{2010CBET.2124....1B}
{Burwitz}, V., {Pietsch}, W., {Henze}, M., {et~al.} 2010, Central Bureau
  Electronic Telegrams (CBET), 2124, 1

\bibitem[{{Cao}(2011)}]{2011ATel.3702....1C}
{Cao}, Y. 2011, ATel, 3702

\bibitem[{{Cao} {et~al.}(2011){Cao}, {Kasliwal}, \&
  {Yaron}}]{2011ATel.3701....1C}
{Cao}, Y., {Kasliwal}, M.~M., \& {Yaron}, O. 2011, ATel,
  3701

\bibitem[{{Cao} {et~al.}(2012){Cao}, {Kasliwal}, {Neill}, {Kulkarni}, {Lou},
  {Ben-Ami}, {Bloom}, {Cenko}, {Law}, {Nugent}, {Ofek}, {Poznanski}, \&
  {Quimby}}]{2012ApJ...752..133C}
{Cao}, Y., {Kasliwal}, M.~M., {Neill}, J.~D., {et~al.} 2012, \apj, 752, 133

\bibitem[{{Cardelli} {et~al.}(1989){Cardelli}, {Clayton}, \&
  {Mathis}}]{1989ApJ...345..245C}
{Cardelli}, J.~A., {Clayton}, G.~C., \& {Mathis}, J.~S. 1989, \apj, 345, 245

\bibitem[{{Ciroi} {et~al.}(2007){Ciroi}, {Di Mille}, {Rafanelli}, \&
  {Temporin}}]{2007ATel.1292....1C}
{Ciroi}, S., {Di Mille}, F., {Rafanelli}, P., \& {Temporin}, S. 2007, ATel, 1292

%\bibitem[{{Cutri} {et~al.}(2003){Cutri}, {Skrutskie}, {van Dyk}, {Beichman},
%  {Carpenter}, {Chester}, {Cambresy}, {Evans}, {Fowler}, {Gizis}, {Howard},
%  {Huchra}, {Jarrett}, {Kopan}, {Kirkpatrick}, {Light}, {Marsh}, {McCallon},
%  {Schneider}, {Stiening}, {Sykes}, {Weinberg}, {Wheaton}, {Wheelock}, \&
%  {Zacarias}}]{2003yCat.2246....0C}
%{Cutri}, R.~M., {Skrutskie}, M.~F., {van Dyk}, S., {et~al.} 2003, 2MASS All-Sky Catalog of Point Sources NASA/IPAC Infrared Science Archive (Pasadena, CA: CalTech)

\bibitem[{{Darnley}(2005)}]{2005PhDT.........2D}
{Darnley}, M.~J. 2005, PhD thesis, Liverpool
  John Moores University, UK

\bibitem[{{Darnley} {et~al.}(2012){Darnley}, {Ribeiro}, {Bode}, {Hounsell}, \&
  {Williams}}]{2012ApJ...746...61D}
{Darnley}, M.~J., {Ribeiro}, V.~A.~R.~M., {Bode}, M.~F., {Hounsell}, R.~A., \&
  {Williams}, R.~P. 2012, \apj, 746, 61

\bibitem[{{Darnley} {et~al.}(2011){Darnley}, {Ribeiro}, {Bode}, \&
  {Munari}}]{2011A&A...530A..70D}
{Darnley}, M.~J., {Ribeiro}, V.~A.~R.~M., {Bode}, M.~F., \& {Munari}, U. 2011,
  \aap, 530, A70

\bibitem[Darnley et 
al.(2014)]{2014A&A...563L...9D} Darnley, M.~J., Williams, S.~C., Bode, M.~F., et al.\ 2014, \aap, 563, L9 

\bibitem[{{Darnley} {et~al.}(2004){Darnley}, {Bode}, {Kerins}, {Newsam}, {An},
  {Baillon}, {Novati}, {Carr}, {Cr{\'e}z{\'e}}, {Evans}, {Giraud-H{\'e}raud},
  {Gould}, {Hewett}, {Jetzer}, {Kaplan}, {Paulin-Henriksson}, {Smartt},
  {Stalin}, \& {Tsapras}}]{2004MNRAS.353..571D}
{Darnley}, M.~J., {Bode}, M.~F., {Kerins}, E., {et~al.} 2004, \mnras, 353, 571

\bibitem[{{Darnley} {et~al.}(2006){Darnley}, {Bode}, {Kerins}, {Newsam}, {An},
  {Baillon}, {Belokurov}, {Calchi Novati}, {Carr}, {Cr{\'e}z{\'e}}, {Evans},
  {Giraud-H{\'e}raud}, {Gould}, {Hewett}, {Jetzer}, {Kaplan},
  {Paulin-Henriksson}, {Smartt}, {Tsapras}, \& {Weston}}]{2006MNRAS.369..257D}
---. 2006, \mnras, 369, 257

\bibitem[Darnley et al.(2008)]{2008ASPC..401..203D} Darnley, M.~J., 
Hounsell, R.~A., 
\& Bode, M.~F.\ 2008, in Astronomical Society of the Pacific Conference
  Series, Vol. 401, RS Ophiuchi (2006) and the Recurrent Nova Phenomenon, ed.
  A.~{Evans}, M.~F. {Bode}, T.~J. {O'Brien}, \& M.~J. {Darnley} (San Francisco, CA:ASP), 203 

\bibitem[{{Darnley} {et~al.}(2013){Darnley}, {Bode}, {Harman}, {Hounsell},
  {Munari}, {Ribeiro}, {Surina}, {Williams}, \&
  {Williams}}]{2013arXiv1303.2711D}
{Darnley}, M.~J., {Bode}, M.~F., {Harman}, D.~J., {et~al.} 2013, to
appear in Stella Novae: Future and Past Decades, P. A. Woudt \&
V. A. R. M. Ribeiro (eds), ASPCS, ArXiv
  e-prints, arXiv:1303.2711

\bibitem[{{della Valle} \& {Livio}(1996)}]{1996ApJ...473..240D}
{della Valle}, M., \& {Livio}, M. 1996, \apj, 473, 240

\bibitem[{{Di Mille} {et~al.}(2009){Di Mille}, {Ciroi}, {Navasardyan}, {Orio},
  {Rafanelli}, \& {Bianchini}}]{2009ATel.2171....1D}
{Di Mille}, F., {Ciroi}, S., {Navasardyan}, H., {et~al.} 2009, ATel, 2171

\bibitem[{{Di Mille} {et~al.}(2007){Di Mille}, {Ciroi}, {Rafanelli},
  {Navasardyan}, \& {Bufano}}]{2007ATel.1325....1D}
{Di Mille}, F., {Ciroi}, S., {Rafanelli}, P., {Navasardyan}, H., \& {Bufano},
  F. 2007, ATel, 1325

\bibitem[{{Di Mille} {et~al.}(2010){Di Mille}, {Orio}, {Ciroi}, {Bianchini},
  {Rafanelli}, \& {Nelson}}]{2010AN....331..197D}
{Di Mille}, F., {Orio}, M., {Ciroi}, S., {et~al.} 2010, Astronomische
  Nachrichten, 331, 197

\bibitem[{{Dilday} {et~al.}(2012){Dilday}, {Howell}, {Cenko}, {Silverman},
  {Nugent}, {Sullivan}, {Ben-Ami}, {Bildsten}, {Bolte}, {Endl}, {Filippenko},
  {Gnat}, {Horesh}, {Hsiao}, {Kasliwal}, {Kirkman}, {Maguire}, {Marcy},
  {Moore}, {Pan}, {Parrent}, {Podsiadlowski}, {Quimby}, {Sternberg}, {Suzuki},
  {Tytler}, {Xu}, {Bloom}, {Gal-Yam}, {Hook}, {Kulkarni}, {Law}, {Ofek},
  {Polishook}, \& {Poznanski}}]{2012Sci...337..942D}
{Dilday}, B., {Howell}, D.~A., {Cenko}, S.~B., {et~al.} 2012, Science, 337, 942

\bibitem[{{Dolphin}(2000)}]{2000PASP..112.1383D}
{Dolphin}, A.~E. 2000, \pasp, 112, 1383

\bibitem[{{Dolphin}(2009)}]{2009PASP..121..655D}
---. 2009, \pasp, 121, 655

\bibitem[{{Fabrika} {et~al.}(2009){Fabrika}, {Sholukhova}, {Valeev}, {Hornoch},
  \& {Pietsch}}]{2009ATel.2240....1F}
{Fabrika}, S., {Sholukhova}, O., {Valeev}, A., {Hornoch}, K., \& {Pietsch}, W.
  2009, ATel, 2240

\bibitem[{{Freedman} \& {Madore}(1990)}]{1990ApJ...365..186F}
{Freedman}, W.~L., \& {Madore}, B.~F. 1990, \apj, 365, 186

\bibitem[{{Gal-Yam} \& {Quimby}(2007)}]{2007ATel.1236....1G}
{Gal-Yam}, A., \& {Quimby}, R. 2007, ATel, 1236

\bibitem[{{Gaposchkin}(1957)}]{1957gano.book.....G}
{Gaposchkin}, C.~H.~P. 1957, {The Galactic Novae}, ed. C.~H.~P. Gaposchkin (Amsterdam: North-Holland)

\bibitem[Hachisu et al.(2000)]{2000ApJ...528L..97H} Hachisu, I., Kato, M., 
Kato, T., \& Matsumoto, K.\ 2000, \apjl, 528, L97 

\bibitem[Hachisu et al.(2007)]{2007ApJ...659L.153H} Hachisu, I., Kato, M., 
\& Luna, G.~J.~M.\ 2007, \apjl, 659, L153 

\bibitem[{{Haiman} {et~al.}(1994){Haiman}, {Magnier}, {Lewin}, {Lester}, {van
  Paradijs}, {Hasinger}, {Pietsch}, {Supper}, \&
  {Truemper}}]{1994A&A...286..725H}
{Haiman}, Z., {Magnier}, E., {Lewin}, W.~H.~G., {et~al.} 1994, \aap, 286, 725

\bibitem[{{Harrison} {et~al.}(1993){Harrison}, {Johnson}, \&
  {Spyromilio}}]{1993AJ....105..320H}
{Harrison}, T.~E., {Johnson}, J.~J., \& {Spyromilio}, J. 1993, \aj, 105, 320

\bibitem[{{Henze} {et~al.}(2008){Henze}, {Meusinger}, \&
  {Pietsch}}]{2008A&A...477...67H}
{Henze}, M., {Meusinger}, H., \& {Pietsch}, W. 2008, \aap, 477, 67

\bibitem[Henze et 
al.(2014)]{2014A&A...563L...8H} Henze, M., Ness, J.-U., Darnley, M.~J., et al.\ 2014, \aap, 563, L8 

\bibitem[{{Hornoch} {et~al.}(2010{\natexlab{a}}){Hornoch}, {Khan}, {Bird},
  {Pejcha}, {Garnavich}, {Littlefield}, {Paul}, \&
  {Bouzid}}]{2010CBET.2319....1H}
{Hornoch}, K., {Khan}, R., {Bird}, J., {et~al.} 2010{\natexlab{a}}, CBET, 2319, 1

\bibitem[{{Hornoch} \& {Pejcha}(2009)}]{2009CBET.2061....5H}
{Hornoch}, K., \& {Pejcha}, O. 2009, CBET, 2061,
  5

\bibitem[{{Hornoch} {et~al.}(2009{\natexlab{a}}){Hornoch}, {Pejcha}, \&
  {Kusnirak}}]{2009CBET.2058....3H}
{Hornoch}, K., {Pejcha}, O., \& {Kusnirak}, P. 2009{\natexlab{a}}, CBET, 2058, 3

\bibitem[{{Hornoch} {et~al.}(2009{\natexlab{b}}){Hornoch}, {Pejcha}, \&
  {Wolf}}]{2009CBET.2062....1H}
{Hornoch}, K., {Pejcha}, O., \& {Wolf}, M. 2009{\natexlab{b}}, CBET, 2062, 1

\bibitem[{{Hornoch} {et~al.}(2009{\natexlab{c}}){Hornoch}, {Pejcha}, {Zasche},
  \& {Kusnirak}}]{2009CBET.2057....3H}
{Hornoch}, K., {Pejcha}, O., {Zasche}, P., \& {Kusnirak}, P.
  2009{\natexlab{c}}, CBET, 2057, 3

\bibitem[{{Hornoch} {et~al.}(2010{\natexlab{b}}){Hornoch}, {Prieto}, {Khan}, \&
  {Hornochova}}]{2010CBET.2610....1H}
{Hornoch}, K., {Prieto}, J., {Khan}, R., \& {Hornochova}, P.
  2010{\natexlab{b}}, CBET, 2610, 1

\bibitem[{{Hornoch} {et~al.}(2010{\natexlab{c}}){Hornoch}, {Prieto}, {Khan},
  {Pejcha}, {Kubanek}, {Gorosabel}, {Martorell}, \&
  {Jelinek}}]{2010CBET.2127....1H}
{Hornoch}, K., {Prieto}, J., {Khan}, R., {et~al.} 2010{\natexlab{c}}, CBET, 2127, 1

\bibitem[Hornoch et al.(2010d)]{2010CBET.2305....1H} Hornoch, K., Wolf, M., 
Hornochova, P., et al.\ 2010d, CBET, 2305, 1 

\bibitem[{{Hornoch} {et~al.}(2010{\natexlab{e}}){Hornoch}, {Zasche}, {Hornoch},
  {Wolf}, {Hornochova}, {Kubanek}, {Gorosabel}, {Lara-Gil}, \&
  {Jelinek}}]{2010CBET.2341....2H}
{Hornoch}, K., {Zasche}, P., {Hornoch}, K., {et~al.} 2010{\natexlab{e}},
  CBET, 2341, 2

\bibitem[{{Jurdana-{\v S}epi{\'c}} {et~al.}(2012){Jurdana-{\v S}epi{\'c}},
  {Ribeiro}, {Darnley}, {Munari}, \& {Bode}}]{2012A&A...537A..34J}
{Jurdana-{\v S}epi{\'c}}, R., {Ribeiro}, V.~A.~R.~M., {Darnley}, M.~J.,
  {Munari}, U., \& {Bode}, M.~F. 2012, \aap, 537, A34

\bibitem[{{Kasliwal}(2009)}]{2009CBET.2015....3K}
{Kasliwal}, M.~M. 2009, CBET, 2015, 3

\bibitem[{{Kasliwal} {et~al.}(2011){Kasliwal}, {Cenko}, {Kulkarni}, {Ofek},
  {Quimby}, \& {Rau}}]{2011ApJ...735...94K}
{Kasliwal}, M.~M., {Cenko}, S.~B., {Kulkarni}, S.~R., {et~al.} 2011, \apj, 735,
  94

\bibitem[{{Kasliwal} {et~al.}(2009){Kasliwal}, {Rau}, {Salvato}, {Cenko},
  {Ofek}, {Quimby}, \& {Kulkarni}}]{2009ATel.1886....1K}
{Kasliwal}, M.~M., {Rau}, A., {Salvato}, M., {et~al.} 2009, ATel, 1886

\bibitem[{{Kent}(1987)}]{1987AJ.....94..306K}
{Kent}, S.~M. 1987, \aj, 94, 306

\bibitem[{{Kenyon} \& {Fernandez-Castro}(1987)}]{1987AJ.....93..938K}
{Kenyon}, S.~J., \& {Fernandez-Castro}, T. 1987, \aj, 93, 938

\bibitem[{{Lee} {et~al.}(2012){Lee}, {Riffeser}, {Seitz}, {Bender}, {Fliri},
  {Hopp}, {Ries}, {B{\"a}rnbantner}, \& {G{\"o}ssl}}]{2012A&A...537A..43L}
{Lee}, C.-H., {Riffeser}, A., {Seitz}, S., {et~al.} 2012, \aap, 537, A43

\bibitem[{{Li} {et~al.}(2011){Li}, {Bloom}, {Podsiadlowski}, {Miller}, {Cenko},
  {Jha}, {Sullivan}, {Howell}, {Nugent}, {Butler}, {Ofek}, {Kasliwal},
  {Richards}, {Stockton}, {Shih}, {Bildsten}, {Shara}, {Bibby}, {Filippenko},
  {Ganeshalingam}, {Silverman}, {Kulkarni}, {Law}, {Poznanski}, {Quimby},
  {McCully}, {Patel}, {Maguire}, \& {Shen}}]{2011Natur.480..348L}
{Li}, W., {Bloom}, J.~S., {Podsiadlowski}, P., {et~al.} 2011, \nat, 480, 348

\bibitem[{{Magnier} {et~al.}(1992){Magnier}, {Lewin}, {van Paradijs},
  {Hasinger}, {Jain}, {Pietsch}, \& {Truemper}}]{1992A&AS...96..379M}
{Magnier}, E.~A., {Lewin}, W.~H.~G., {van Paradijs}, J., {et~al.} 1992, \aaps,
  96, 379

\bibitem[{{Massey} {et~al.}(2006){Massey}, {Olsen}, {Hodge}, {Strong},
  {Jacoby}, {Schlingman}, \& {Smith}}]{2006AJ....131.2478M}
{Massey}, P., {Olsen}, K.~A.~G., {Hodge}, P.~W., {et~al.} 2006, \aj, 131, 2478

\bibitem[{{Newsham} {et~al.}(2013){Newsham}, {Starrfield}, \&
  {Timmes}}]{2013arXiv1303.3642N}
{Newsham}, G., {Starrfield}, S., \& {Timmes}, F. 2013, ArXiv e-prints,
  arXiv:1303.3642

\bibitem[{{Nishiyama} \& {Kabashima}(2010)}]{2010CBET.2305....2N}
{Nishiyama}, K., \& {Kabashima}, F. 2010, CBET,
  2305, 2

\bibitem[{{Nishiyama} {et~al.}(2010){Nishiyama}, {Kabashima}, \&
  {Yusa}}]{2010CBET.2483....1N}
{Nishiyama}, K., {Kabashima}, F., \& {Yusa}, T. 2010, CBET, 2483, 1

\bibitem[{{Nugent} {et~al.}(2011){Nugent}, {Sullivan}, {Cenko}, {Thomas},
  {Kasen}, {Howell}, {Bersier}, {Bloom}, {Kulkarni}, {Kandrashoff},
  {Filippenko}, {Silverman}, {Marcy}, {Howard}, {Isaacson}, {Maguire},
  {Suzuki}, {Tarlton}, {Pan}, {Bildsten}, {Fulton}, {Parrent}, {Sand},
  {Podsiadlowski}, {Bianco}, {Dilday}, {Graham}, {Lyman}, {James}, {Kasliwal},
  {Law}, {Quimby}, {Hook}, {Walker}, {Mazzali}, {Pian}, {Ofek}, {Gal-Yam}, \&
  {Poznanski}}]{2011Natur.480..344N}
{Nugent}, P.~E., {Sullivan}, M., {Cenko}, S.~B., {et~al.} 2011, \nat, 480, 344

\bibitem[Osborne et al.(2011)]{2011ApJ...727..124O} Osborne, J.~P., Page, 
K.~L., Beardmore, A.~P., et al.\ 2011, \apj, 727, 124 

\bibitem[{{Page} {et~al.}(2010){Page}, {Osborne}, {Evans}, {Wynn}, {Beardmore},
  {Starling}, {Bode}, {Ibarra}, {Kuulkers}, {Ness}, \&
  {Schwarz}}]{2010MNRAS.401..121P}
{Page}, K.~L., {Osborne}, J.~P., {Evans}, P.~A., {et~al.} 2010, \mnras, 401,
  121

\bibitem[{{Perryman} \& {ESA}(1997)}]{1997ESASP1200.....P}
{Perryman}, M.~A.~C., \& {ESA}, eds. 1997,
  {The HIPPARCOS and TYCHO catalogues. Astrometric and photometric star
  catalogues derived from the ESA HIPPARCOS Space Astrometry Mission} (ESA Special Publication, Vol. 1200; Noordwijk: ESA)

\bibitem[{{Pietsch}(2010)}]{2010AN....331..187P}
{Pietsch}, W. 2010, Astronomische Nachrichten, 331, 187

\bibitem[{{Pietsch} {et~al.}(2010{\natexlab{a}}){Pietsch}, {Henze}, {Burwitz},
  {Kaur}, {Hartmann}, {Milne}, \& {Williams}}]{2010ATel.3001....1P}
{Pietsch}, W., {Henze}, M., {Burwitz}, V., {et~al.} 2010{\natexlab{a}}, ATel, 3001

\bibitem[{{Pietsch} {et~al.}(2007{\natexlab{a}}){Pietsch}, {Burwitz},
  {Greiner}, {Barsukova}, {Fabrika}, {Moiseev}, {Valeev}, {Goranskij}, \&
  {Hornoch}}]{2007ATel.1009....1P}
{Pietsch}, W., {Burwitz}, V., {Greiner}, J., {et~al.} 2007{\natexlab{a}}, ATel, 1009

\bibitem[{{Pietsch} {et~al.}(2007{\natexlab{b}}){Pietsch}, {Haberl}, {Sala},
  {Stiele}, {Hornoch}, {Riffeser}, {Fliri}, {Bender}, {B{\"u}hler}, {Burwitz},
  {Greiner}, \& {Seitz}}]{2007A&A...465..375P}
{Pietsch}, W., {Haberl}, F., {Sala}, G., {et~al.} 2007{\natexlab{b}}, \aap,
  465, 375

\bibitem[{{Pietsch} {et~al.}(2010{\natexlab{b}}){Pietsch}, {Lloyd}, {Henze},
  {Burwitz}, {Knaur}, {Hartmann}, {Milne}, {Williams}, {Liakos},
  {Hatzidimitriou}, \& {Niarchos}}]{2010ATel.2896....1P}
{Pietsch}, W., {Lloyd}, J., {Henze}, M., {et~al.} 2010{\natexlab{b}}, ATel, 2896

\bibitem[{{Rau}(2007)}]{2007ATel.1276....1R}
{Rau}, A. 2007, ATel, 1276

\bibitem[{{Rau} {et~al.}(2007){Rau}, {Burwitz}, {Cenko}, {Updike}, {Hartmann},
  {Milne}, \& {Williams}}]{2007ATel.1242....1R}
{Rau}, A., {Burwitz}, V., {Cenko}, S.~B., {et~al.} 2007, ATel, 1242

\bibitem[{{Rodr{\'{\i}}guez-Gil} {et~al.}(2009){Rodr{\'{\i}}guez-Gil},
  {Ferrando}, {Rodr{\'{\i}}guez}, {Bode}, {Huxor}, {Giles}, \&
  {Mackey}}]{2009ATel.2166....1R}
{Rodr{\'{\i}}guez-Gil}, P., {Ferrando}, R., {Rodr{\'{\i}}guez}, D., {et~al.}
  2009, ATel, 2166

\bibitem[{{Rosino}(1973)}]{1973A&AS....9..347R}
{Rosino}, L. 1973, \aaps, 9, 347

\bibitem[{{Schaefer}(2010)}]{2010ApJS..187..275S}
{Schaefer}, B.~E. 2010, \apjs, 187, 275

\bibitem[{{Schaefer} \& {Pagnotta}(2012)}]{2012Natur.481..164S}
{Schaefer}, B.~E., \& {Pagnotta}, A. 2012, \nat, 481, 164

\bibitem[{{Shafter}(1997)}]{1997ApJ...487..226S}
{Shafter}, A.~W. 1997, \apj, 487, 226

\bibitem[{{Shafter} {et~al.}(2011{\natexlab{a}}){Shafter}, {Bode}, {Darnley},
  {Ciardullo}, {Cao}, \& {Hornoch}}]{2011ATel.3825....1S}
{Shafter}, A.~W., {Bode}, M.~F., {Darnley}, M.~J., {et~al.} 2011{\natexlab{a}},
  ATel, 3825

\bibitem[{{Shafter} {et~al.}(2010{\natexlab{a}}){Shafter}, {Bode}, {Darnley},
  {Ciardullo}, \& {Misselt}}]{2010ATel.3006....1S}
{Shafter}, A.~W., {Bode}, M.~F., {Darnley}, M.~J., {Ciardullo}, R., \&
  {Misselt}, K.~A. 2010{\natexlab{a}}, ATel, 3006

\bibitem[{{Shafter} {et~al.}(2011{\natexlab{b}}){Shafter}, {Bode}, {Darnley},
  {Misselt}, {Rubin}, \& {Hornoch}}]{2011ApJ...727...50S}
{Shafter}, A.~W., {Bode}, M.~F., {Darnley}, M.~J., {et~al.} 2011{\natexlab{b}},
  \apj, 727, 50

\bibitem[{{Shafter} {et~al.}(2011{\natexlab{c}}){Shafter}, {Ciardullo}, {Bode},
  \& {Darnley}}]{2011ATel.3699....1S}
{Shafter}, A.~W., {Ciardullo}, R., {Bode}, M.~F., \& {Darnley}, M.~J.
  2011{\natexlab{c}}, ATel, 3699

\bibitem[{{Shafter} {et~al.}(2010{\natexlab{b}}){Shafter}, {Ciardullo}, {Bode},
  {Darnley}, \& {Misselt}}]{2010ATel.2987....1S}
{Shafter}, A.~W., {Ciardullo}, R., {Bode}, M.~F., {Darnley}, M.~J., \&
  {Misselt}, K.~A. 2010{\natexlab{b}}, ATel, 2987

\bibitem[{{Shafter} {et~al.}(2008{\natexlab{a}}){Shafter}, {Ciardullo}, {Bode},
  {Darnley}, {Misselt}, {Nishiyama}, \& {Kabashima}}]{2008ATel.1834....1S}
{Shafter}, A.~W., {Ciardullo}, R., {Bode}, M.~F., {et~al.} 2008{\natexlab{a}},
  ATel, 1834

\bibitem[{{Shafter} {et~al.}(2011{\natexlab{d}}){Shafter}, {Ciardullo},
  {Darnley}, \& {Bode}}]{2011ATel.3727....1S}
{Shafter}, A.~W., {Ciardullo}, R., {Darnley}, M.~J., \& {Bode}, M.~F.
  2011{\natexlab{d}}, ATel, 3727

\bibitem[{{Shafter} {et~al.}(2010{\natexlab{c}}){Shafter}, {Ciardullo},
  {Darnley}, {Bode}, \& {Misselt}}]{2010ATel.2898....1S}
{Shafter}, A.~W., {Ciardullo}, R., {Darnley}, M.~J., {Bode}, M.~F., \&
  {Misselt}, K.~A. 2010{\natexlab{c}}, ATel, 2898

\bibitem[{{Shafter} {et~al.}(2010{\natexlab{d}}){Shafter}, {Ciardullo},
  {Darnley}, {Bode}, \& {Misselt}}]{2010ATel.2949....1S}
---. 2010{\natexlab{d}}, ATel, 2949

\bibitem[{{Shafter} {et~al.}(2010{\natexlab{e}}){Shafter}, {Ciardullo},
  {Darnley}, {Bode}, \& {Misselt}}]{2010ATel.2909....1S}
---. 2010{\natexlab{e}}, ATel, 2909

\bibitem[{{Shafter} {et~al.}(2013){Shafter}, {Curtin}, {Pritchet}, {Bode}, \&
  {Darnley}}]{2013arXiv1307.2296S}
{Shafter}, A.~W., {Curtin}, C., {Pritchet}, C.~J., {Bode}, M.~F., \& {Darnley},
  M.~J. 2013, to appear in Stella Novae: Future and Past Decades,
  P. A. Woudt \& V. A. R. M. Ribeiro (eds), ASPCS, ArXiv e-prints, arXiv:1307.2296

\bibitem[{{Shafter} {et~al.}(2012{\natexlab{a}}){Shafter}, {Darnley}, {Bode},
  {Ciardullo}, \& {Hornoch}}]{2012ATel.3850....1S}
{Shafter}, A.~W., {Darnley}, M.~J., {Bode}, M.~F., {Ciardullo}, R., \&
  {Hornoch}, K. 2012{\natexlab{a}}, ATel, 3850

\bibitem[{{Shafter} {et~al.}(2012{\natexlab{b}}){Shafter}, {Hornoch},
  {Ciardullo}, {Darnley}, \& {Bode}}]{2012ATel.4503....1S}
{Shafter}, A.~W., {Hornoch}, K., {Ciardullo}, J.~V.~R., {Darnley}, M.~J., \&
  {Bode}, M.~F. 2012{\natexlab{b}}, ATel, 4503

\bibitem[{{Shafter} {et~al.}(2012{\natexlab{c}}){Shafter}, {Hornoch},
  {Ciardullo}, {Darnley}, \& {Bode}}]{2012ATel.4368....1S}
{Shafter}, A.~W., {Hornoch}, K., {Ciardullo}, R., {Darnley}, M.~J., \& {Bode},
  M.~F. 2012{\natexlab{c}}, ATel, 4368

\bibitem[{{Shafter} {et~al.}(2012{\natexlab{d}}){Shafter}, {Hornoch},
  {Ciardullo}, {Darnley}, \& {Bode}}]{2012ATel.4658....1S}
---. 2012{\natexlab{d}}, ATel, 4658

\bibitem[{{Shafter} {et~al.}(2012{\natexlab{e}}){Shafter}, {Hornoch},
  {Ciardullo}, {Darnley}, \& {Bode}}]{2012ATel.4391....1S}
---. 2012{\natexlab{e}}, ATel, 4391

\bibitem[{{Shafter} {et~al.}(2010{\natexlab{f}}){Shafter}, {Hornoch},
  {Darnley}, {Bode}, {Ciardullo}, \& {Misselt}}]{2010ATel.3039....1S}
{Shafter}, A.~W., {Hornoch}, K., {Darnley}, M.~J., {et~al.} 2010{\natexlab{f}},
  ATel, 3039

\bibitem[{{Shafter} {et~al.}(2009){Shafter}, {Rau}, {Quimby}, {Kasliwal},
  {Bode}, {Darnley}, \& {Misselt}}]{2009ApJ...690.1148S}
{Shafter}, A.~W., {Rau}, A., {Quimby}, R.~M., {et~al.} 2009, \apj, 690, 1148

\bibitem[{{Shafter} {et~al.}(2008{\natexlab{b}}){Shafter}, {Ciardullo},
  {Burwitz}, {Henze}, {Pietsch}, {Milne}, {Misselt}, {Williams}, {Hartmann},
  {Updike}, {Bode}, \& {Darnley}}]{2008ATel.1851....1S}
{Shafter}, A.~W., {Ciardullo}, R., {Burwitz}, V., {et~al.} 2008{\natexlab{b}},
  ATel, 1851

\bibitem[{{Shafter} {et~al.}(2011{\natexlab{e}}){Shafter}, {Darnley},
  {Hornoch}, {Filippenko}, {Bode}, {Ciardullo}, {Misselt}, {Hounsell},
  {Chornock}, \& {Matheson}}]{2011ApJ...734...12S}
{Shafter}, A.~W., {Darnley}, M.~J., {Hornoch}, K., {et~al.} 2011{\natexlab{e}},
  \apj, 734, 12

\bibitem[{{Sirianni} {et~al.}(2005){Sirianni}, {Jee}, {Ben{\'{\i}}tez},
  {Blakeslee}, {Martel}, {Meurer}, {Clampin}, {De Marchi}, {Ford}, {Gilliland},
  {Hartig}, {Illingworth}, {Mack}, \& {McCann}}]{2005PASP..117.1049S}
{Sirianni}, M., {Jee}, M.~J., {Ben{\'{\i}}tez}, N., {et~al.} 2005, \pasp, 117,
  1049

\bibitem[Skrutskie et al.(2006)]{2006AJ....131.1163S} Skrutskie, M.~F., 
Cutri, R.~M., Stiening, R., et al.\ 2006, \aj, 131, 1163 

\bibitem[{{Stark} {et~al.}(1992){Stark}, {Gammie}, {Wilson}, {Bally}, {Linke},
  {Heiles}, \& {Hurwitz}}]{1992ApJS...79...77S}
{Stark}, A.~A., {Gammie}, C.~F., {Wilson}, R.~W., {et~al.} 1992, \apjs, 79, 77

\bibitem[{{Starrfield} {et~al.}(2012){Starrfield}, {Timmes}, {Iliadis}, {Hix},
  {Arnett}, {Meakin}, \& {Sparks}}]{2012BaltA..21...76S}
{Starrfield}, S., {Timmes}, F.~X., {Iliadis}, C., {et~al.} 2012, Baltic
  Astronomy, 21, 76

\bibitem[{{Steele} {et~al.}(2004){Steele}, {Smith}, {Rees}, {Baker}, {Bates},
  {Bode}, {Bowman}, {Carter}, {Etherton}, {Ford}, {Fraser}, {Gomboc}, {Lett},
  {Mansfield}, {Marchant}, {Medrano-Cerda}, {Mottram}, {Raback}, {Scott},
  {Tomlinson}, \& {Zamanov}}]{2004SPIE.5489..679S}
{Steele}, I.~A., {Smith}, R.~J., {Rees}, P.~C., {et~al.} 2004, in Society of
  Photo-Optical Instrumentation Engineers (SPIE) Conference Series, Vol. 5489,
  Society of Photo-Optical Instrumentation Engineers (SPIE) Conference Series,
  ed. J.~M. {Oschmann}, Jr., 679

\bibitem[{{Tang} {et~al.}(2013){Tang}, {Cao}, \&
  {Kasliwal}}]{2013ATel.5607....1T}
{Tang}, S., {Cao}, Y., \& {Kasliwal}. 2013, ATel, 5607

\bibitem[Tang et al.(2014)]{2014ApJ...786...61T} Tang, S., Bildsten, L., 
Wolf, W.~M., et al.\ 2014, \apj, 786, 61 

\bibitem[Tody(1986)]{1986SPIE..627..733T} Tody, D.\ 1986, \procspie, 627, 
733 

\bibitem[Tody(1993)]{1993ASPC...52..173T} Tody, D.\ 1993, Astronomical Data 
Analysis Software and Systems II, 52, 173 

\bibitem[{{Truran} \& {Livio}(1986)}]{1986ApJ...308..721T}
{Truran}, J.~W., \& {Livio}, M. 1986, \apj, 308, 721

\bibitem[{{Valeev} {et~al.}(2009){Valeev}, {Barsukova}, {Sholukhova},
  {Medvedev}, {Hornoch}, {Kusnirak}, {Pietsch}, \&
  {Fabrika}}]{2009ATel.2208....1V}
{Valeev}, A., {Barsukova}, E., {Sholukhova}, O., {et~al.} 2009, ATel, 2208

\bibitem[{{Whelan} \& {Iben}(1973)}]{1973ApJ...186.1007W}
{Whelan}, J., \& {Iben}, Jr., I. 1973, \apj, 186, 1007

\bibitem[{{White} {et~al.}(1995){White}, {Giommi}, {Heise}, {Angelini}, \&
  {Fantasia}}]{1995ApJ...445L.125W}
{White}, N.~E., {Giommi}, P., {Heise}, J., {Angelini}, L., \& {Fantasia}, S.
  1995, \apjl, 445, L125

\bibitem[{{Williams} {et~al.}(2004){Williams}, {Garcia}, {Kong}, {Primini},
  {King}, {Di Stefano}, \& {Murray}}]{2004ApJ...609..735W}
{Williams}, B.~F., {Garcia}, M.~R., {Kong}, A.~K.~H., {et~al.} 2004, \apj, 609,
  735

\bibitem[{{Williams}(1992)}]{1992AJ....104..725W}
{Williams}, R.~E. 1992, \aj, 104, 725

\bibitem[{{Williams} {et~al.}(2013{\natexlab{a}}){Williams}, {Darnley}, {Bode},
  \& {Shafter}}]{2013ATel.5611....1W}
{Williams}, S.~C., {Darnley}, M.~J., {Bode}, M.~F., \& {Shafter}, A.~W.
  2013{\natexlab{a}}, ATel, 5611

\bibitem[{{Williams} {et~al.}(2013{\natexlab{b}}){Williams}, {Darnley}, {Bode},
  \& {Shafter}}]{2013arXiv1303.1980W}
---. 2013{\natexlab{b}}, to appear in Stella Novae: Future and Past
Decades, P. A. Woudt \& V. A. R. M. Ribeiro (eds), ASPCS, ArXiv e-prints, arXiv:1303.1980

%\bibliography{paper}
\end{thebibliography}

\end{document}